\documentclass[epj, final]{svjour}
\usepackage{latexsym}
\usepackage{graphicx}
\usepackage{caption}
\usepackage{subcaption}
\usepackage{amsmath}
 \usepackage{color}
 \usepackage[table]{xcolor}
 \usepackage{gensymb}
\usepackage{hyperref}
\usepackage[numbers,sort&compress]{natbib}
\usepackage{placeins}
\usepackage {float}
\begin{document}
\title{Spallation reactions. A successful interplay between modeling and applications.}
\author{J.-C. David\inst{} 
}                     
%
%
\institute{CEA/Saclay, Irfu/SPhN, 91191 Gif-sur-Yvette, Cedex, France}
\date{Received: date / Revised version: date}
%
\abstract{
The spallation reactions are a type of nuclear reaction which occur in space by interaction of the cosmic rays with interstellar bodies. The first spallation reactions induced with an accelerator took place in 1947 at the Berkeley cyclotron (University of California) with 200 MeV deuterons and 400 MeV alpha beams. They highlighted the multiple emission of neutrons and charged particles and the production of a large number of residual nuclei far different from the target nuclei. The same year R. Serber describes the reaction in two steps: a first and fast one with high-energy particle emission leading to an excited remnant nucleus, and a second one, much slower, the de-excitation of the remnant. In 2010 IAEA organized a worskhop to present the results of the most widely used spallation codes within a benchmark of spallation models. If one of the goals was to understand the deficiencies, if any, in each code, one remarkable outcome points out the overall high-quality level of some models and so the great improvements achieved since Serber. Particle transport codes can then rely on such spallation models to treat the reactions between a light particle and an atomic nucleus with energies spanning from few tens of MeV up to some GeV. An overview of the spallation reactions modeling is presented in order to point out the incomparable contribution of models based on basic physics to numerous applications where such reactions occur. Validations or benchmarks, which are necessary steps in the improvement process, are also addressed, as well as the potential future domains of development. Spallation reactions modeling is a representative case of continuous studies aiming at understanding a reaction mechanism and which end up in a powerful tool.
} 
\maketitle
\section{Introduction}
\label{Intro}





The spallation reactions are one branch of the nuclear reactions. Usually they are defined as the interaction of a particle, most of the time a nucleon, with a nucleus. The projectile energy lies more or less between 100 MeV and 3000 MeV. During this interaction many nucleons are emitted and the residual nucleus can be very different from the target nucleus. 

Early studies of spallation reactions took place in 1947 in the Berkeley cyclotron at the University of California \cite{BRO47}. Several experiments (items A8 and A9 of \cite{KAP47}) with beams of deuterons and alpha of energy 200 MeV and 400 MeV, respectively, were able to demonstrate the multiple emission of neutrons and charged particles resulting in the production of a large spectrum of residual nuclei. Measurements of neutron spectra were also taken and compared, quite successfully, to the theory of R. Serber who was the first to describe, also in 1947, the process of the spallation reaction in two stages \cite{SER47}.

From a historical and semantics point of view, the spallation term seems to be first used by G. T. Seaborg in his thesis in 1937. The first publications mentioning the spallation term date back to 1948: O'Connor and Seaborg \cite{OCO48} (in the title) and Seaborg and Perlman (\cite{SEA48}, page 587). For the record and as anecdote Serber didn't mention the spallation term in his aforementioned article, but he titled it {\it Nuclear Reactions at High Energies}. The meaning of the word {\it spallation} is given below to avoid ambiguity later.

It is usual to define the spallation reactions as nuclear reactions between a fast particle and an atomic nucleus. In recent decades physicists prefer to speak of protons of a few hundred MeV up to a few GeV on a nucleus, perhaps because during this time the spallation has been closely associated with the ADS (Accelerator-Driven Systems), for which the beam is made of protons. Actually the particle type and the energy limits can be slightly different according to the researcher and/or to the period. Moreover regarding the modeling we will see that the spallation reactions are a set of successive or competing nuclear processes : binary collisions, pre-equilibrium, evaporation, fission, binary decay, multifragmentation. In addition in all directions some extensions can be done, towards lower and higher projectile energies and using more complex projectiles. Thus since the goal is to describe nuclear reactions, like the spallation reactions, within the largest ranges of energy and of projectile type, the needed assumptions and approximations in the modeling must not only be kept in mind in order to know the validity domain, but they define this domain. In other words assumptions and approximations define the reactions we are studying. Moreover, for sake of clarity, the definitions of the generic terms {\it particle} and {\it energetic particle} adopted in this paper, when they are used alone, are the following. Particle : lepton or hadron with a mass less than or equal to the $\alpha$ particle. Therefore all nuclei are heavier than the $\alpha$ particle. Energetic particle : particle having an energy such that it sees the atomic nucleus like a set of free nucleons. The lowest energy of the incident particle is thus limited to around 150 MeV . The upper limit is more difficult to determine, but it makes sense not to go far beyond the heaviest baryonic resonances implemented.

The reasons and motivations for studying these reactions are manifold. Spallation reactions exist in several places and are linked to several domains. In space, Galactic and Solar Cosmic Rays (GCR and SCR) induce such reactions on spacecrafts and interstellar bodies like the meteorites. In the first case the knowledge of the reactions are mandatory for safety reasons. Radiation damage to the electronics and radioprotection of the astronauts must be under control in order to safely achieve the mission \cite{MAN11}. In the second case the spallation reactions occuring in the meteorites can allow us to obtain information on their history via the residual isotopes produced inside provided we have sufficient knowledge of their production mechanism (\cite{AMM09} and references therein).  In addition, GCR and SCR are partly responsible, via spallation reactions, of the abundance of Li, Be and B \cite{MAS00}. In accelerators, the spallation reactions are used most of the time either as neutron sources or as radioactive isotope beams. The term ''spallation neutron source'' is usually used to describe facilities that provide neutron beams (which are eventually cooled to get the right wavelength) to probe matter. Such facilities have existed for a long time, for example SINQ (in Switzerland) \cite{SINQ} and ISIS (in UK) \cite{ISIS}, and new ones have been recently built in the USA and in Japan, respectively SNS (operational in 2006) \cite{SNS} and JSNS (in 2008) \cite{JSNS}. The next planned high-power neutron facility in Europe, the European Spallation Source (ESS) \cite{ESS} will be built in Lund (Sweden) and should deliver first neutrons around 2020, and more or less at the same time the China Spallation Neutron Source (CSNS) \cite{CSNS} should also be operational. However the neutron production via spallation reactions can also be used to control  hybrid reactors. Such facilities, named ADS (Accelerator-Driven System), consist of a nuclear reactor, with a spallation target surrounded by the fuel, and of an accelerator. The added spallation neutrons enable control of the reactor even with fuels containing actinides delivering only a low fraction of delayed neutrons. Since the reactor is subcritical by the prompt neutrons and  the accelerator can be shut down quickly, such a facilty could help to transmutate nuclear wastes safely. The use of spallation neutrons in order to fission a nucleus can also be a way to deliver radioactive ion beams. A target, made generally of actinides and surrounding a spallation target, undergoes fission reactions via the spallation neutrons and the fission products, once extracted and selected, can be used as a secondary beam of exotic nuclei. A complementary way is to directly use the radioactive nuclei produced by spallation. According to the expected beam, one or the other of the methods has to be chosen as well as the material of the spallation target, to get the highest flux. This type of radioactive ion beam facility is called ISOL, for Isotope Separation On-Line, and we can mention three of them that are operating today:  CERN-ISOLDE and GANIL-SPIRAL in Europe and TRIUMF in North America. In the more remote future a high intensity ISOL facility is planned in Europe: EURISOL. In the meantime, new facilities will be built and operated in the near future, for example HIE-ISOLDE (CERN), SPES (Legnaro) and SPIRAL2 (GANIL). Finally we can also cite another application, proton (or hadron) therapy, where the reaction destroys the cancerous cells, and the proton, unlike the usual radiotherapy, delivers only low-dose side-effects to surrounding tissue. 

In addition to those motivations driven by application of the reaction, the spallation reaction study is obviously interesting in itself. Understanding a reaction mechanism involving a particle, with an energy between 100 MeV and 2-3 GeV, and any atomic nucleus (light and heavy), i.e. being capable of reproducing emitted particle spectra in energy and in angle, as well as the mass, charge and isotopic distributions of the residual nuclei, is a real challenge. This challenge in modeling is nevertheless necessary, because, due to the broad projectile energy range, to the large target nucleus spectrum and to the numerous output channels, each experimental data set can only be used to validate a part of the reaction mechanism. It is impossible to launch a program aiming at covering all types of spallation reactions.

This paper aims at presenting the spallation reactions from modeling to application in parallel with the improvements over the last few years. It is divided into four parts: modeling, validation, application and future development. Most of the time publications dealing with spallation reactions address only one of the first three parts, and when two of them are addressed, validation is included, but with a minor and ancillary role. The first section, modeling, is focussed on the outlines. A review paper could be written only on spallation reaction modeling. We then aim at covering only the main aspect of the modeling from the basic ideas up to the computational codes. Details are sometimes given to show either the various domains involved in the modeling or the difficulties met and the assumptions needed. For example two models will be explained more in depth, INCL4 for the first stage and Abla for the second stage of the reaction, because,  first, these two codes have been extensively used in the part dedicated to applications and, second, they are good examples of modeling evolution. 
Nevertheless, as far as possible, numerous other spallation codes are discussed in the second section devoted to validation. Whether combined to modeling or to application, validations are usually addressed in a specific purpose. Here validation is discussed in a more general way with a brief history of various benchmarks carried out in the two last decades, with an overview of the benchmarks done by the transport code developers  and with the results obtained within the Benchmark of Spallation Models, an IAEA project where the best known spallation models were involved, comparing their calculation results to an experimental data set chosen in order to cover all aspects of the spallation reaction, in input  channels and output channels.
The third section gives four examples of the use of spallation reaction modeling: the EURISOL project,  the MEGAPIE target, the ESS facility and the cosmogenic nuclide production in meteorites. The goal is to discuss respectively designing and optimisation, predictive power and reliability, feasibility and uncertainty estimate, and the use of modeling to mitigate lack of experimental data.
Finally the fourth section is devoted to the topics less studied up to now, like e.g. the pion and strangeness production, especially towards the high energies. 

This paper does not aim at being complete, giving all details in all topics. Such an objective is out the scope of a review article, and that has been successfully addressed in the book entitled {\it Handbook of Spallation Research} \cite{FIL09}, written by Filges and Goldenbaum. The goal of this paper is rather to point out the reasons for the successes of the simulations without hiding the remaining deficiencies and to give references to help the reader go to into further detail. This is also the opportunity to show the results achieved by physicists involved in spallation reaction study over several decades.


\section{Modeling and associated codes}
\label{modeling}

The basics of modeling spallation reactions have been stated by R. Serber in 1947 \cite{SER47}. Unlike the low-energy reactions where there is formation of a compound nucleus, which then de-excites, the high-energy reactions, say above 100 A.MeV, cannot form directly in general a compound nucleus due to the large mean free path of the incident nucleon. This large mean free path, thus associated with a small de Broglie wavelength, allows the incident nucleon to see the nucleons in the nucleus as if they were free and to start a cascade of binary collisions. Different scenarios for this cascade are possible depending on the energy and the impact parameter of the incident particle going from ejection of a single nucleon, taking with it all of the incident energy, to the capture of the projectile leaving the nucleus in a state of strong excitation. Once excited, the nucleus enters a second and slower phase, the de-excitation. Here again different scenarios compete according to the mass, excitation energy and angular momentum of the remnant nucleus. The first and rapid phase is of about $10^{-22}$ s and the second in the order of $10^{-18}$ s. In addition to these two phases sometimes included is a third one named pre-equilibrium. This step is actually an intermediate step since it deals with the transition between cascade and de-excitation and more precisely how the cascade is stopped. The need of this additional phase is then strongly connected with the cascade modeling.

The codes associated with the modeling are of course the codes handling each phase of the spallation reactions, but there exists also others associated codes: the particle transport codes. These codes aim at simulating interactions of particles with a massive target, i.e. which has a spatial extension and therefore is the seat of successive reactions due to the secondary particles. They usually use libraries to handle the transport of particles, but, according to the type of the particle and its energy, reaction models are needed when the experimental or evaluated data are unavailable. This is typically the case of spallation reactions. Although some libraries have emerged \cite{KOR10,WAT11}, they have been built thanks to models.

This section is divided into three parts. The first two ones deal with the two above mentioned phases of the spallation reaction (the pre-equilibrium being addressed with the cascade). We will give the main aspects, the basic ideas, and details only when necessary to understand the topics discussed in the following sections of this paper. The last subsection is devoted to transport codes related to the spallation regime, because these codes are most of the time the link between modeling and applications. The transport codes need models, but in return the models had to be upgraded to be used optimally in transport codes.

\subsection{Intranuclear cascade (INC) - First phase}
\label{cascade}






Modeling of the first phase consists of simulating the time evolution of a system of nucleons in a nucleus possibly in interaction with each other.
Several articles have been published on the basic foundations of spallation reactions and include among others Aichelin \cite{AIC91}, Bunakov \cite{BUN85} and Cugnon \cite{CUG12}. Aichelin and Bunakov demonstrate in two different ways how the problem moves from quantum-mechanical equations to transport equations. Bunakov in addition explains that the intranuclear cascade (INC) modeling is nothing else than a Monte-Carlo method to solve the transport equations. This point is taken up in \cite{CUG12} by Cugnon to explain the good results obtained by some codes using this modeling. In the following we summarize the very main steps to get the master equations. In this way we will show what are the difficulties to be solved and the approximations to be made to be able to simulate spallation reactions.

The time evolution of a quantum system is given by the Schr\"odinger equation, using the wave function, or by the Liouville-Von Neumann equation (eq. \ref{eq:dens}), using the density matrix $\rho$,
\begin{equation}
\label{eq:dens}
\imath\hbar\frac{\partial \rho}{\partial t} = [H,\rho] ,
\end{equation}
where $H$ is the Hamiltonian.

The goal being to solve this equation,  a first step is to use reduced density matrices (eq. \ref{eq:densred}), which are the density matrices of a subsystem of $k$  nucleons,
\begin{equation}
\label{eq:densred}
\rho_{1,...,k} = \frac{A!}{(A-k)!} Tr_{(k+1,...,A)} \rho ,
\end{equation}
such that $\rho_{1,...,A} = \rho$, with $A$ the total number of nucleons. Doing so the Liouville-Von Neumann equation is replaced by a system of $A$ coupled equations (eq. \ref{eq:bbgky}), so-called BBGKY\footnote{BBGKY: Bogolioubov-Born-Green-Kirkwood-Yvon} hierarchy,
\begin{equation}
\label{eq:bbgky}
\begin{split}
\imath\hbar\frac{\partial}{\partial t}\rho_{1,...,k}(t) = 
\frac{A!}{(A-k)!} Tr_{(k+1,...,A)} \hspace{4cm}\\
\left(
 \sum \limits_{\underset{}{i=1}}^k\left[T_{i},\rho(t)\right] +  
 \frac{1}{2} \sum \limits_{\underset{i \neq j}{i,j}}^k\left[V_{ij},\rho(t)\right]+ \right. \hspace{2.3cm}\\
\left. \frac{1}{2} (A-k) \sum \limits_{\underset{}{i=1}}^k\left[V_{i(k+1)},\rho(t)\right] 
\right). \hspace{2.3cm}
\end{split}
\end{equation}
Such a system does not simplify the solution of the $A$-nucleons system, however it allows to reduce the chain through physical approximations and so make the system solvable.  In our case collisions between two particles have to be taken into account and they occur in a nucleus that can be characterized by its potential. Thus only the first two equations (eqs. \ref{eq:bbgky1} and \ref{eq:bbgky2}) have to be considered, with the two interacting particles (1 and 2) and a third partner (3), which represents the nuclear potential,
\begin{equation}
\label{eq:bbgky1}
\imath\hbar\frac{\partial}{\partial t}\rho_{1}(t)
 = 
\left[T_{1},\rho_{1}(t)\right] +
\frac{1}{2} Tr_{(2)}\left[V_{12},\rho_{1,2}(t)\right], \hspace{1.2cm}
\end{equation}
\begin{equation}
\label{eq:bbgky2}
\begin{split}
\imath\hbar\frac{\partial}{\partial t}\rho_{1,2}(t)
 = 
\left[T_{1}+T_{2},\rho_{1,2}(t)\right] +\hspace{3.cm}\\
\frac{1}{2} \left[V_{12},\rho_{1,2}(t)\right]+
\frac{1}{2} Tr_{(3)}\left[V_{13}+V_{23},\rho_{1,2,3}(t)\right].
\end{split}
\end{equation}
$T_{i}$ and $V_{jk}$ are respectively the kinetic energy and interacting potential operators of the particles $i$, $j$ and $k$.

Through the use of the Wigner transformation, $f$, of the density matrices,
\begin{equation}
f(\vec{r},\vec{p}) = \int d^3\vec{r'} \hspace{0.3cm} e^{i\frac{\vec{p}.\vec{r'}}{\hbar}} \hspace{0.3cm} 
\rho({\vec{r}+\frac{\vec{r'}}{2},\vec{r}-\frac{\vec{r'}}{2}}) , \nonumber
\end{equation}
and the use of three main assumptions, which are (as mentioned in \cite{CUG12}),
\begin{list}{$\bullet$}{}
\item the two-body correlations are unimportant in the collision term,
\item the mean potential is a smooth function on the spatial extension of the particle wave functions, and
\item the collisions are independent, that is the scattering wave function becomes asymptotic before the next collision takes place,
\end{list}
the transport equation, also called BUU or VUU\footnote{BUU: Boltzmann-\"{U}hling-Uhlenbeck; VUU: Vlasov-\"{U}hling-Uhlenbeck} equation \cite{UU33}, can be written as
\begin{equation}
\label{eq:equu}
\begin{split}
\frac{\partial f_1}{\partial t} + \vec{v}.\vec{\nabla}_rf_1 - \vec{\nabla}_rU.\vec{\nabla}_pf_1 = - \int \frac{d^3\vec{p}_2d^3\vec{p}_3d^3\vec{p}_4}{(2\pi)^6} \hspace{0.1cm}
\sigma v_{12}  \hspace{4cm}\\
 \delta^3(\vec{p}_1+\vec{p}_2-\vec{p}_3-\vec{p}_4) \delta(\epsilon_1+\epsilon_2-\epsilon_3-\epsilon_4) \hspace{4cm}\\
\left[f_1f_2\left(1-f_3\right)\left(1-f_4\right)-f_3f_4\left(1-f_1\right)\left(1-f_2\right)\right].\hspace{3.8cm}\nonumber
\end{split}
\end{equation}
$f_i$ are the distribution functions (1 and 2 for the two colliding particles, 3 and 4 for the two outgoing particles), $\sigma v_{12}$ the collision term (most of the time based on experimental data), the $f_if_j$ and $(1-f_i)$ terms are respectively the overlapping of input/output distributions and the Pauli principle, and finally the delta functions are the momentum and energy conservations.

Doing so Bunakov \cite{BUN85} not only explained how it was possible to move the problem from quantum-mechanical equations to transport equations, through several assumptions matching spallation reactions, but also demonstrated the basic foundations of the intranuclear cascade  (INC) modeling.

From the practical point of view the different steps encountered during the intranuclear cascade in all modeling codes are: initialization, transport, event treatment and stopping. Since all these details are explained for each available model in articles and/or user's manuals (a list of references is given in section \ref{validations}), only the main features are described below. Nevertheless three papers must be cited here. The two articles of Metropolis {\it et al.} \cite{MET58a,MET58b}, where the first calculation results about spallation reactions using a computer\footnote{The first calculations by hand were done by M. L. Golberger in \cite{GOL48}.} were published, and the one of Bertini \cite{BER63}, where an improved version of the previous model is used. Those works are the basis of modern codes.

Both projectile and nucleus have to be initialized. Regarding the projectile, type and energy are known, but its impact parameter is taken randomly and coulomb deviation considered. If the projectile is a composite particle, its structure must be given in the same way as for the target. The target nucleus is defined by its mass, its charge, the potentials felt by the particles, the momentum of each nucleon (most of the time a Fermi gas distribution is used), and the spatial distribution of the nucleons. Two ways exist to define this distribution. Either the distribution is continuous, often several concentrical density regions, or discrete, i.e. positions are sampled in a Wood-Saxon distribution, for example.

Two different methods are applied to move and follow the particles participating to the cascade. With the time-like transport all the particles are followed at the same time, while with the space-like transport the patrticles are followed ones after the others.

Three events exist during the intranuclear cascade: collision, resonance decay and reflection/transmission at the nucleus surface.
Collisions can be elastic or inelastic. Most of the time experimental data (cross-sections) are used to define interaction probabilities first and second what are the output products and their characteristics (types, energies, momenta), each selection being done randomly. When necessary Pauli blocking is taken into account. Cross-section parameterizations, number and type of collisions and of resonances taken into account, and the way to apply Pauli blocking are different from one model to another; this can already  explained the broad spectrum of results that will be shown in section \ref{validations}. Decay of the resonances is not as simple as it seems. It is not only the production of two particles (Pauli blocking included), but also the right behavior of the resonance. This particle has a lifetime that can be more complicated than the inverse of the decay width. For example, a phase-space factor has been added to the $\Delta(1232)$ lifetime, because this resonance is  a $\pi N$ system at low energy. More details are given in \cite{CUG97}, where the $\Delta(1232)$ has been studied. Moreover, experimental data for the other resonances are sometimes scarce, and, due to their widths and proximity (in energy), some can overlap and make difficult a clear treatment. This explains, in addition to the energy domain, why sometimes only the $\Delta(1232)$ is implemented. The third event happens when the particle reaches the surface; it can be emitted or reflected. To be emitted the particle must be energetic enough, i.e. be able to overcome nuclear and coulomb potential. Experimental results on composite particle spectra display high energy tails, which indicate emission of composite particles during the intranuclear cascade phase. However the basis of INC are clearly the treatment of the transport of nucleons with their two-body interactions, i.e.  without clusterization. So, up to now, the only way to produce composite particles during the cascade is to add a coalescence model. Before leaving the nucleus, a nucleon can drag and aggregate one or more nucleons, close enough to it in space and momentum. This procedure extends the INC applicability to a satisfactory level.

\begin{figure}[hbt]
\begin{center}
\resizebox{.5\textwidth}{!}{
\includegraphics{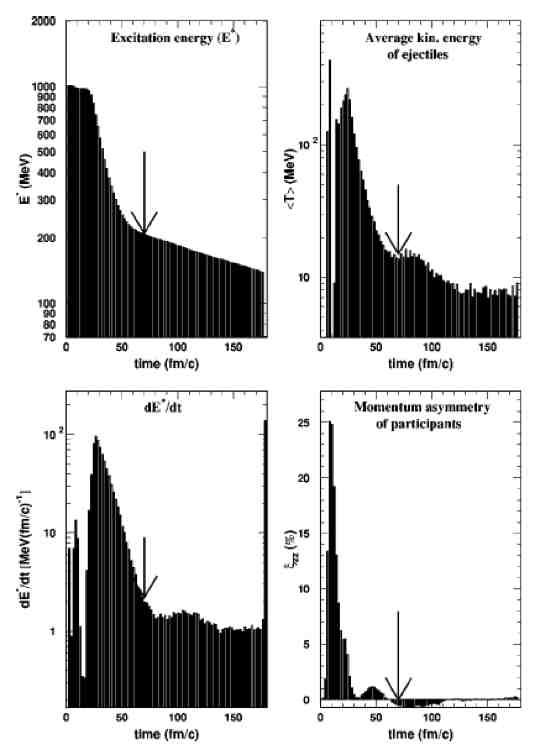}}
\end{center}
\caption{\label{Fig_tfin}
Time evolution of four observables during the intranuclear cascade INCL4. Nucleus excitation energy is plotted at the top-left, mean kinetic energy of the emitted particles at the top-right, time derivative of the excitation energy at the bottom-left and momentum asymmetry of the participants at the bottom-right. The arrows point out the time that can be considered as the end of the cascade. Reprinted figure with permission from A. Boudard \cite{BOU02}. \href{http://link.aps.org/abstract/PRC/v66/e044615} {Copyright (2014) by the American Physical Society.}}
\end{figure}

Finally different criteria are used to stop the cascade and to start the de-excitation phase of the remnant nucleus. We can mention three of them: cutoff energy, stopping time and deviation from an optical absorptive potential. The first one is the simplest. When its kinetic energy falls below a given cutoff energy (depending on the type of the particle or not) the particle no longer contributes to the cascade and, once there are no more participants, the cascade is stopped. The stopping time criterion, defined and used in INCL4 \cite{BOU02}, is based on physical features. The two stages, cascade and de-excitation, are two different mechanisms (duration, characteristics of the initial and final products) and the hot remnant nucleus, de-exciting during the second phase, can be seen as a compound nucleus, i.e. with no memory. In \cite{BOU02} the authors of INCL4 describe the result of their study that ends up in an equation giving this stopping time versus the target mass. Fig.\ref{Fig_tfin} shows the time evolutions of four observables used to determine this time. The last and third criterion, used in CEM03 \cite{MAS08}, is based on optical potentials. The cascade is stopped when the deviation between the imaginary parts of the calculated optical potential (from interaction cross-sections and Pauli blocking effects) and of a phenomenological optical potential exceeds a chosen value. The cascade formalism is then considered inappropriate. This criterion is described in \cite{MAS08}. In this paper the authors explain the difficulties to find the right value to assess this deviation, which governs the switch to a pre-equilibrium model. 

Such a pre-equilibrium model is sometimes used between the cascade and the de-excitation phases. Several versions exist, but almost all are based on the exciton model developed by Griffin \cite{GRI66}. According to the use or not of this intermediate phase, the duration of the cascade is obviously different or, maybe more correctly, mass, charge and excitation of the remnant nucleus are larger, if this phase is called. While some intranuclear cascade models need such pre-equilibrium models to improve their capability, this is not the case of some others. This will be shown in section \ref{validations} and this topic is related to the low energy interaction during the cascade.

After mentioning in \cite{CUG12} the three main assumptions  needed to draw the master transport equation, J. Cugnon points out also that the INC models do more than solving this equation. In particular the first assumption, {\it the two-body correlations are unimportant in the collision term}, is not made, since the previous collisions influence the next ones. Thus collision after collision n-body correlations are propagated leading to reliable prediction of the spectral characteristics of secondary particles. Bunakov already discussed the domain of validity of INC model from its basic assumptions. He considered that INC method can be applied with a  target mass above A$\approx$10 and with an incident projectile energy above about ten MeV, and added that {\it from the point of view of its basic principles the INC model is not inferior to the other reaction models, say Glauber's model, the optical model or DWBA}. Nevertheless the INC method has obviously its own limits, such as the need of a smooth function for the potential that makes the use of INC difficult to treat Nucleus-Nucleus reactions and some justifications are at least necessary. Still from the basic assumptions, Bunakov suggests some modification to extend or improve the INC capabilities. We can cite the treatment of the nucleon mean free path, which should be calculated carefully for the low energy particles ($\le$ 100 MeV), and the use of DWBA calculations for the first collisions to improve the threshold behavior.  Since then INC models have been refined and their reliability significantly improved, thus we mention hereafter two studies related to Bunakov suggestions. 

In their paper \cite{CUG05}, J. Cugnon and P. Henrotte investigate the low-energy limit validity of the intranuclear cascade on a purely empirical basis. They compare their INCL4.3 model to experimental data with  projectile energies from $\sim$25 MeV to $\sim$250 MeV and sometimes to other models dedicated to this energy domain. The first observation was the rather good results well below the often quoted lowest limit and the second was that INCL4.3 could compete with more dedicated models, even if there was room for improvement. They then study two main features: the Pauli blocking implementation and the number of collisions. Pauli blocking can be accounted for via different ways. In this study they used a "strict" and a "statistical" version. "Strict" means that collisions are allowed only if the momenta of the outgoing particles are larger than the Fermi momentum. "Statistical" version is based on the phase-space occupation probabilities around the outgoing particles and a comparison with a random number. The conclusion was the best results were obtained  using both of them, the "strict" Pauli blocking for the first collision and the "statistical" one for the others. According to the way positions and momenta are assigned to the nucleons to prepare the target nucleus, and taking into account the population evolution, the best implementation was a priori not so easy to define. Nevertheless it is clear that the Pauli blocking plays a significant role in the mean free path. Regarding the number of collisions, it is defined as the level of the collision in the cascade, in other words as the incrementation of the number of collisions undergone previously. This study combined with a comparison of other models, most of them using exciton states, shows first that INCL gave fairly good results compared to these more elaborate models in this low-energy domain, but also that the collision number was closely related to the exciton number. In addition to the comparisons with experimental data, this latter observation is a clue that a pre-equilibrium phase is not mandatory in the spallation modeling, provided the INC model treat carefully the low energy collision. 

\begin{figure}[hbt]
\begin{center}
\resizebox{.3\textwidth}{!}{
\includegraphics{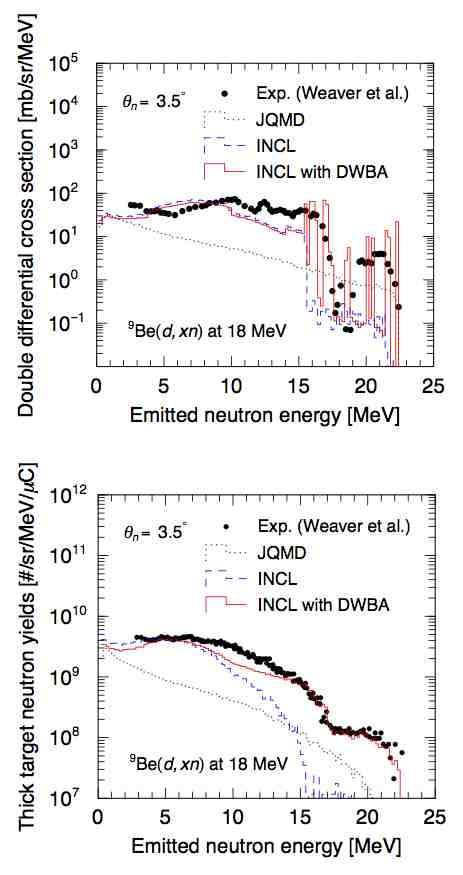}}
\end{center}
\caption{\label{Fig_dwba}
Double differential cross-sections and thick target yields (respectively top and bottom panel) at $\theta_n = 3.5^\circ$ in $^{\text 9}$Be($d,xn$) for 18 MeV. Experimental data  \cite{WEA73} are represented by closed circles, and results calculated by Particle and Heavy Ion Transport code System (PHITS) with JQMD, INCL, and INCL with DWBA are denoted by dotted, dashed, and solid lines, respectively. Figures drawn from \cite{HAS14}. \href{http://www.sciencedirect.com/science/article/pii/S0168583X14005096} {Copyright {\copyright} 2014 Elsevier B.V.}}
\end{figure}
The use of a DWBA model to improve behavior at the low-energy limits has been investigated recently. This study combines a version of INCL, INCL4.5, to a DWBA method \cite{HAS14}, because INCL treats well the incoherent processes and DWBA the coherent ones. Details on the two models can be found in the references, but what  has to be mentioned here is the need for fitted parameters in DWBA, especially for the optical potential used. Thus this study focused on ($d,xn$) spectra (and also ($d,xp$)) on $^{\text nat}$Li, $^{\text 9}$Be, and $^{\text nat}$C targets at incident energies ranging from 10 to 40 MeV. From the technical point of view the use of INCL or DWBA is chosen randomly according to the ratio $\frac{\sigma_{DWBA}}{\sigma_{total}}$ of the reaction cross-sections. We plot results in Fig. \ref{Fig_dwba}, one is the elementary production of neutrons at a forward angle (upper part) and the second one a neutron thick target (2.5 mm) yield (lower part), INCL and DWBA being implemented in the PHITS transport code. We clearly see the two domains: INCL4.5 reproduces well the neutrons at half the incident energy of a deuteron and DWBA the high energy part (the first colision having then a heavy weight).

\subsection{De-excitation - Second phase}

As previously stated, spallation reactions are divided in two mechanisms: the intranuclear cascade and the de-excitation.
\begin{figure}[hbt]
\begin{center}
\resizebox{.35\textwidth}{!}{
\includegraphics{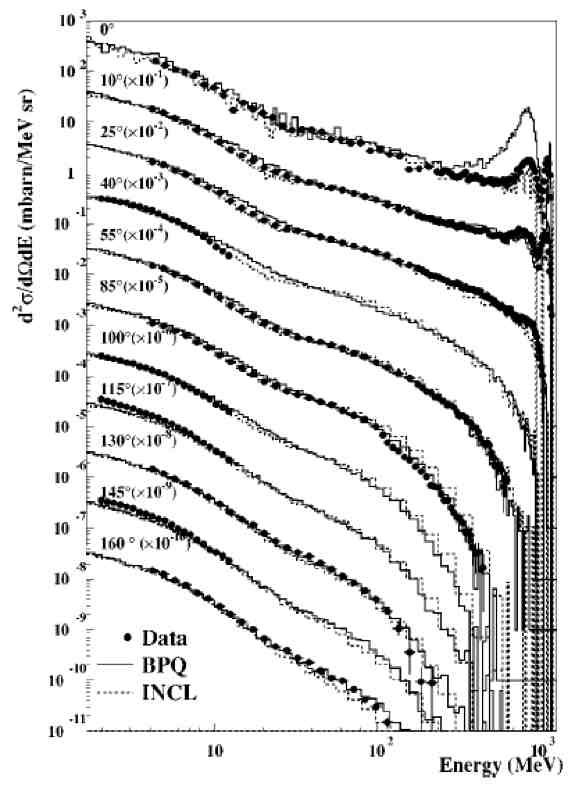}}
\end{center}
\caption{\label{Fig_spn} Neutron double differential cross-sections for the reaction p(1200 MeV) + W. Reprinted figure with permission from S. Leray \cite{LER02} where details can be read. \href{http://link.aps.org/abstract/PRC/v65/e044621} {Copyright (2014) by the American Physical Society.}}
\end{figure}
This can be illustrated with Fig. \ref{Fig_spn} where neutron spectra are plotted for several emission angles.  The high-energy part (cascade) is strongly focused in the forward direction and characterized by two peaks, the elastic one (sharp) and the quasi-elastic one (broader) due to the $\Delta(1232)$ resonance, whereas the low energy part (de-excitation), much more important, is isotropic.
The large amount of emitted nucleons during the de-excitation stage can also be seen in mass and charge distributions of the residues, as shown in Fig. \ref{Fig_femass} for the reaction $^{56}$Fe+p, where residue mass distributions measured at five projectile energies, from 300 A.MeV up to 1500 A.MeV (here inverse kinematics has been used), are plotted. The upper part of Fig. \ref{Fig_femass} shows also that the production of the light nuclei increases with the projectile energy, in such a way that, as seen on the lower part, the yield of some very light residual nuclei becomes significant. This latter observation already indicates that the emission of nucleons is not the only way to de-excite.
\begin{figure}[hbt]
\begin{center}
\hspace*{-.3cm}
\resizebox{.4\textwidth}{!}{
\includegraphics{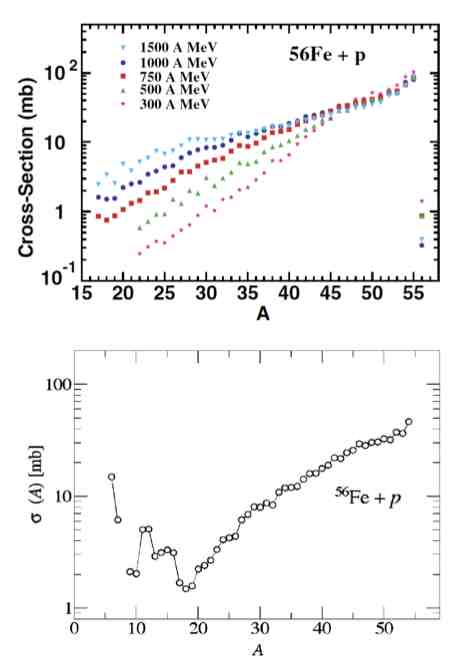}}
\end{center}
\caption{\label{Fig_femass}
Mass distribution of the residual nuclei in the spallation reaction $^{56}$Fe+p. Above at five different beam energies (\href{http://link.aps.org/abstract/PRC/v75/e044603} {Reprinted figure with permission from C. Villagrasa \cite{VIL07}}) and below at 1000 A.MeV (\href{http://link.aps.org/abstract/PRC/v70/e054607} {Reprinted figure with permission from P. Napolitani \cite{NAP04}}). Copyright (2014) by the American Physical Society.}
\end{figure}
Another complexity of the de-excitation phase arises when the target mass increases as shown with Fig. \ref{Fig_pbcharge}.
\begin{figure}[hbt]
\begin{center}
\hspace*{-.3cm}
\resizebox{.39\textwidth}{!}{
\includegraphics{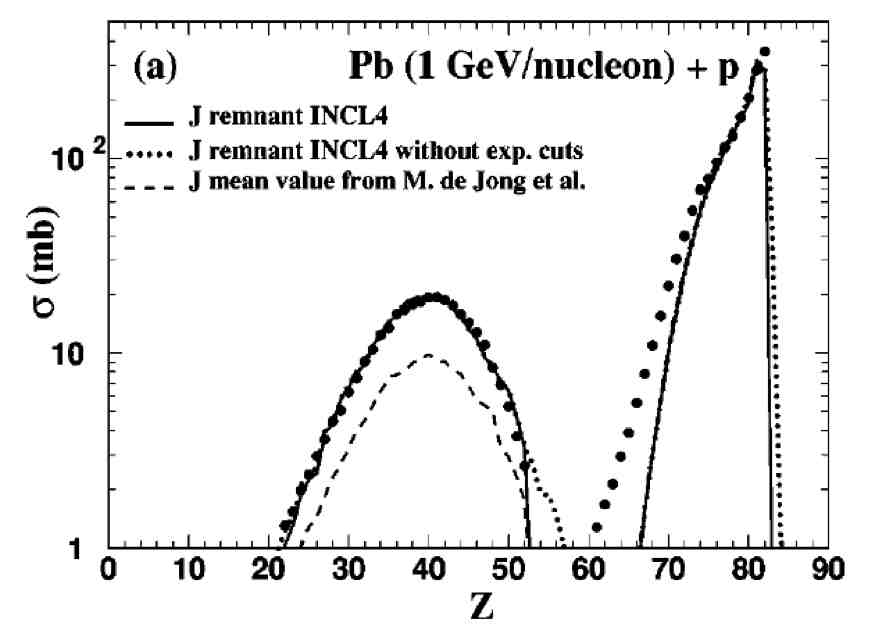}}
\end{center}
\caption{\label{Fig_pbcharge}
Charge distribution of the residual nuclei in the spallation reaction $^{208}$Pb+p at 1000 A.MeV (Reprinted figure with permission from A. Boudard \cite{BOU02}). \href{http://link.aps.org/abstract/PRC/v66/e044615} {Copyright (2014) by the American Physical Society}. Data \cite{ENQ01} are the black points.}
\end{figure}
On this charge distribution, in addition to the decreasing cross-section with decreasing residual charge, on the high Z part,  a bump centered around the half of the target charge appears. This bump indicates a high-energy fission process.

Once the nucleus target reaches an equilibrium state, the intranuclear cascade vanishes and is followed by the de-excitation of this compound nucleus (remnant) characterized by its mass, charge, excitation energy and angular momentum. The most common mechanisms used to describe this phase are the particle evaporation and the fission process, but others have been proposed either as alternative ways or as additional channels to cover the whole mass (and charge) distributions. Thus we report hereafter the five mechanisms possibly encountered in the spallation modeling for the de-excitation step: evaporation, fission, binary breakup, multifragmentation and Fermi breakup. The same approach is used in all these processes, that is the statistical approach. The number of microstates of a given macrostate (here the remnant nucleus) determines the probability of a channel and the final channel is chosen randomly.

\subsubsection{Evaporation}
\label{evaporation}

Understanding the evaporation of  particles means to be able to determine: i) the type of the emitted particle or more precisely the ratios between all types , ii) how many of these particles are emitted for each type, and iii) the energy spectra of the emitted particles. To answer these questions the emission probability of each type of particles as a function of the kinetic energy must be calculated. This is done starting with the following assumptions:
\begin{list}{$\bullet$}{}
\item the nucleus is sufficiently excited to apply a statistical description, and
\item the emission rate is equal to the rate of absorption (detailed balance principle).
\end{list}
Doing so one obtains the emission probability per units of time and energy of a particle ($part$) with a kinetic energy $\varepsilon$.
\begin{equation} \label{eq:probev}
P_{part}(\varepsilon) = \frac{\rho_f(E^*_f)}{\rho_i(E^*_i)} (2s+1) (4\pi p^2dp/h^3) \sigma_c(\varepsilon),
\end{equation}
where $\rho_i(E^*_i)$($\rho_f(E^*_f)$) is the state density of the initial(final) nucleus with $E^*_i$($E^*_f$) the excitation energy of the initial(final) nucleus, $(4\pi p^2dp/h^3)$ the phase-space of the emitted particle with a momentum $p$, $s$ its spin, and $\sigma_c(\varepsilon)$  the capture cross-section of the particle by the final nucleus. The two main ingredients are the state density and the capture cross-section. Their choices largely explain the differences between the codes.

Usually the state density is defined with the Fermi gas model using a canonical ensemble rather than a micro-canonical one to simply the formalism. Then a new parameter arises, the level density parameter $a$, which relates the excitation energy and the temperature: $E^* = a T^2$.
Thus, considering the nucleons distributed on equidistant levels at the Fermi energy (place of the excitation energy), the state density, with no angular momentum dependence, can be written 
\begin{equation}\label{eq:weisskopf}
\rho(E^*)= \frac{\sqrt{\pi}}{12}\frac{e^{2\sqrt{aE^*}}}{a^{1/4}{{E^*}^{5/4}}},
\end{equation}
and with angular momentum dependence $J$ 
\begin{equation}
\rho(E^*,J)= (2J+1) \left( \frac{\hbar^2}{2I}  \right)^{3/2} \frac{\sqrt{a}}{12}\frac{e^{2\sqrt{aU}}}{U^2},
\end{equation}
where $I$ is the nucleus inertia moment and $U = E^* - E_{rot}$, with $E_{rot}$ the rotational energy. The point is then the expression of the level density parameter $a$. Numerous works deal with $a$ and it is impossible to draw here a list and even a summary of the different values and/or formul{\ae} used. Nevertheless the knowledge of this parameter has been improved since the first estimates. In \cite{WEI37} Wei\ss kopf gives values for $a$, in \cite{GIL65} Gilbert and Cameron take into account the pairing and the shell effects, and later these effects have been refined in \cite{IGN75} and \cite{SHM82}. The deformation of the nucleus and the diffusivity of its surface have been investigated in \cite{IGN76,TOK81,GUE88}, even if the temperature dependence can be neglected in the spallation domain, this has been studied in \cite{SHL91} and the collective effects are included in the level density parameter used in the Abla code \cite{JUN98}.

The second basic ingredient for determining the width or probability of emission of a particle is the capture cross-section as a function of its kinetic energy $\varepsilon$. Various expressions can be found in literature, from the simplest one to the most sophisticated.
The simplest formula is the one used in 1937 by Wei\ss kopf \cite{WEI37}, but taking into account the Coulomb barrier $B_{coul}$ of charged particles
\begin{equation} \label{eq:secinv}
\sigma_c(\varepsilon) = \pi R^2 (1-\frac{B_{coul}}{\varepsilon}),
\end{equation}
with R, radius of the nucleus.
In 1940 Wei\ss kopf and Ewing improve the formula in \cite{WEI40}. It takes into account the angular momentum by a decomposition into partial waves and the cross-section becomes
\begin{equation}
\sigma_c(\varepsilon) = \frac{\pi}{k^2} \sum_l (2l+1)Q_l(\varepsilon),
\end{equation}
where $k$ is the wave number and $l$ the angular momentum. Wei\ss kopf-Ewing split their $Q_l(\varepsilon)$ factor into two parts. The first one is related to the incoming wave, $P_l(\varepsilon)$, which is the penetration probability into the surface of the nucleus, and the second one related to the energy exchange probability between the particle and the nucleus, $\xi_l(\varepsilon)$, called sticking probability. $P_l(\varepsilon)$ contains the barrier for charged particles (actually the Coulomb barrier and the centrifugal barrier) and, regarding $\xi_l(\varepsilon)$, only the average value on $l$ was available.
A very similar formula is given by Hauser-Feshbach \cite{HAU52},
\begin{equation}  \label{eq:xstl}
\sigma_c(\varepsilon) = \frac{\pi}{k^2} \sum_l (2l+1)T_l(\varepsilon).
\end{equation}
The $Q_l(\varepsilon)$ factor is replaced by the penetrability or transmission factor $T_l(\varepsilon)$, which has become more sophisticated. For example it can be written as
\begin{equation}
T_l(\varepsilon) = \frac{1}{1-t_l(\varepsilon)},
\end{equation}
with 
\begin{equation}
t_l(\varepsilon) =e^{\left( 2\int\sqrt{2\frac{\mu}{\hbar}[V(l,r)-\varepsilon]} dr \right)} ,
\end{equation}
where $\mu$ is the reduced mass of the system and $V (l,r)$ the potential taking into account the Coulomb potential, the centrifugal potential and the nuclear potential. While the centrifugal potential is clearly known and Coulomb potential also, although refinements exist, the nuclear potential is more complicated and different expressions exist such as those of Huizenga \cite{HUI62} or Bass \cite{BAS74}. Whatever the formula used for the capture cross-section, it amounts mainly to calculating the $V (l,r)$ potential.

When using experimental data to fit the formula of the inverse or capture cross-section, one has to pay attention to the fact that the compound nucleus is not in its ground state, but heated. This is generally not a problem, except at low energy. Some other aspects must be taken into account, like the tunneling effect. This point is obviously important for the low-energy light charged particles, as explained in the section describing Abla07 in \cite{FIL08}.

The formalism of evaporation is well posed, especially for the neutron, but the expressions of its ingredients may vary and explain the differences in results according to the models, mostly for charged particles. For these latter particles it is worth mentioning that evaporation models often also emit light nuclei. While typically the same formalism is used for nucleons up to alpha, the strategy can change when the number of channels increases (emission of any type of nuclei), because the number of emission widths or probabilities increases significantly and the computation time also. The need to evaporate nuclei with a charge higher or equal than 3 has been seen in Fig. \ref{Fig_femass}. This is however not the only way to generate such nuclei. Other mechanisms, such as binary decay or multifragmentation, are also used to do the job.

Finally the gamma evaporation is a special case. Gammas do not compete with the other channels, except at the end of the de-excitation stage, when the nucleus is much colder. Then neutrons and gammas compete. Most of the codes include the statistical emission of gammas through the giant dipole resonance. However this treatment ends up not always with the splitting into the ground state and the metastable states. Nevertheless we can mention a recent model, using in particular the Evaluated Nuclear Structure Data File \cite{ENSDF}, that has been developped within the code PHITS and gives interesting results on isomer production \cite{PHI13}. A comparison between the codes providing the isomer production should be interesting, because this domain has not been studied very much up to now.

\subsubsection{Fission}

When excited a heavy nucleus can also undergo fission. This process competes with all evaporation channels and even with all de-excitation channels. The treatment of the fission  process is indeed divided in two separate parts: calculation of the fission width (or probability), which governs the competition, and, if fission occurs, determination of the characteristics of the two fission products (mass, charge and excitation energy).

\begin{figure}[hbt]
\begin{center}
\hspace*{-.3cm}
\resizebox{.35\textwidth}{!}{
\includegraphics{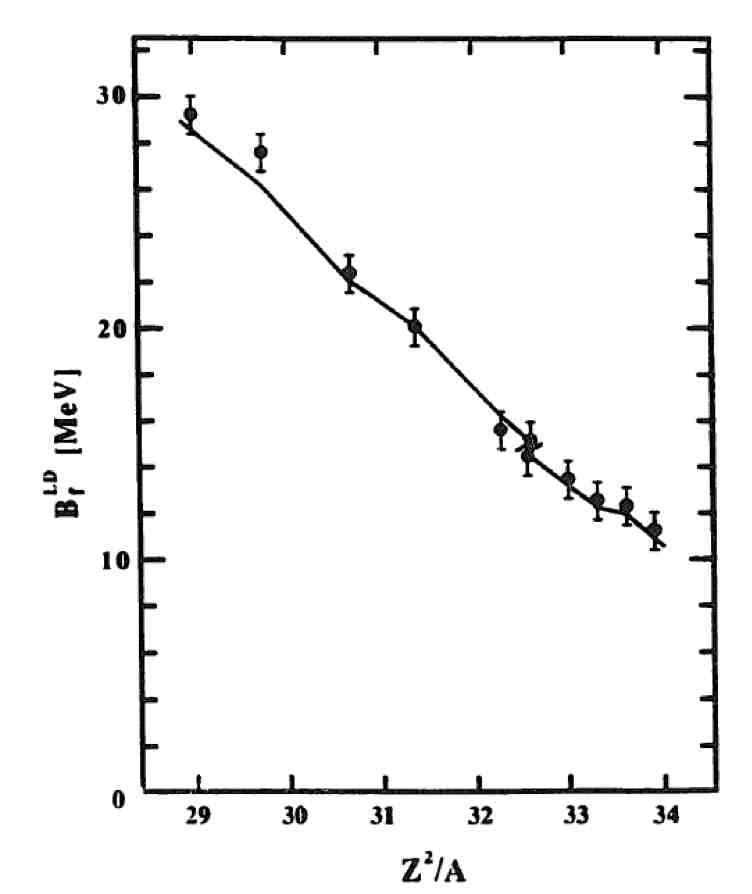}}
\end{center}
\caption{\label{Fig_fissility} Dependence of the barrier heights on $Z^2/A$. The points are extracted from the analysis of nuclear fissility done in \cite{ILJ92}. The curve is liquid drop model calculation results. Figure from \cite{ILJ92} (Courtesy of V. Muccifora). \href{http://www.sciencedirect.com/science/article/pii/037594749290278R} {Copyright {\copyright} 2014 Elsevier B.V.}}
\end{figure}
Two parameters explain the instability of excited nuclei against fission: the fissility parameter $Z^2/A$ and the angular momentum $J$. The fissility parameter describes the competition between the repulsive Coulomb energy ($\propto Z^2/A^{1/3}$) and the attractive surface energy ($\propto A^{2/3}$) (see Fig.\ref{Fig_fissility}). The higher the fissility, the lower the fission barrier.
The angular momentum increases the deformation of the nucleus and then decreases the fission barrier. The results obtained in the work of A. Sierk on a macroscopic model for  rotating nuclei \cite{SIE86} are used by most of the fission models in the spallation domain.

Fission, unlike evaporation, is a collective phenomenon. So passing from a state of individual degrees of freedom to a state of collective degrees of freedom takes time, the transition time to reach the highest fission  probability. The random movement of particles in a dissipative medium is managed by the Langevin or Fokker- Planck equations. If the stationary equations are solvable analytically, this is not the case for the dynamical ones and long computation times are needed. As is often done to save time, approximations are required, but the results are convincing \cite{JUR05}. Two consequences of what has been stated previously: taking into account the viscosity via the stationary equations, the fission width is lower than expected, and second the maximum fission width value needs time to be reached (dynamical effects). The width can then be expressed as: 
\begin{eqnarray}
\Gamma_f(t) = \Gamma_{BW} . K . f_\beta(t) \nonumber ,
\end{eqnarray}
where 
\begin{list}{$\bullet$}{}
\item $\Gamma_{BW}$ is the Bohr-Wheeler fission width,
\begin{equation}
= \frac{1}{2\pi\rho_i(E^*,J)}  \int_0^{E^*-B_f(J)} \hspace{-1.4cm}\rho_{saddle}(E^*-B_f(J)-\varepsilon) d\varepsilon ,
\end{equation}
with $\rho_{saddle}(E^*-B_f(J)-\varepsilon) $ the level density at the saddle-point and $B_f(J)$ the fission barrier,
\item $K$ the Kramers factor ($ = \sqrt{1+\left(\frac{\beta}{2\omega_0}\right)^2}-\left(\frac{\beta}{2\omega_0}\right)$, where $\beta$ is the reduced dissipation coefficient and $\omega_0$  the frequency of the harmonic oscillator describing the potential at the saddle-point deformation) , and
\item $f_\beta(t)$ a function simulating the dynamical dissipation effects.
\end{list}

Once the fission channel is chosen, characteristics of the fission fragments must be determined. In a few words we can say that experimental data are usually used for the determination of masses and charges, often assuming Gaussian distributions, and taking care to preserve the initial mass and charge. The excitation energy is shared in respect to the mass ratios, and, if the kinetic energy of each fragment is based on their Coulomb repulsion, adjustments on experimental data is often needed also. One of the most reliable models developed on this topic is the one of J. Benlliure \cite{BEN98}. Once again, all models used to model a part of the spallation process must be fast enough to give results in a reasonable time, thus, combining this constraint with the goal of building a predictive model, J. Benlliure worked out a semi-empirical nuclear-fission model.

\begin{figure}[hbt]
\begin{center}
\resizebox{.4\textwidth}{!}{
\includegraphics{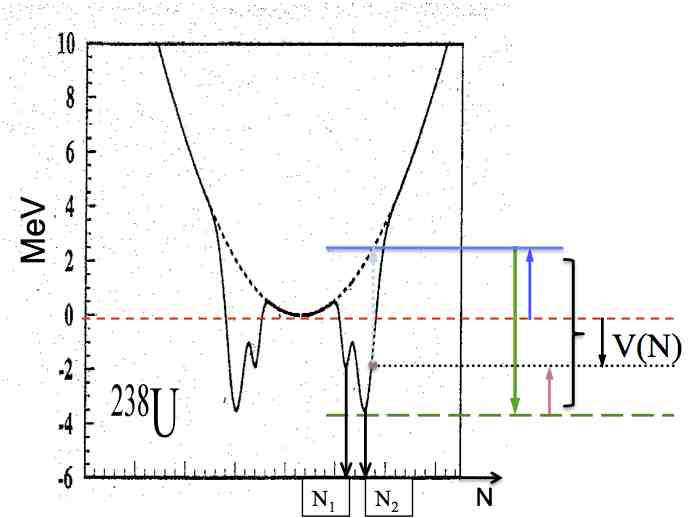}}
\end{center}
\caption{\label{Fig_potfis}
Potential energy at the fission barrier for $^{238}U$ versus the neutron number of one of the fragments. Colors of arrows match the contributions in equation \ref{eq:pot-fis}. The original figure comes from \cite{BEN98} (Courtesy of J. Benlliure). \href{http://www.sciencedirect.com/science/article/pii/S0375947497006076} {Copyright {\copyright} 2014 Elsevier B.V.}}
\end{figure}

The yield, $Y(E,N)$, of a given fission fragment, here characterized by its number of neutrons, is the ratio of its number of transition states above the potential, at the fission barrier, to the sum of these numbers for all possible fragments:
\begin{eqnarray}
Y(E,N) = \frac{\int_0^{E-V(N)}\rho_N(U)dU}{\sum_{N=0}^{N_{\text{fis}}}\int_0^{E-V(N)}\rho_N(U)dU} \nonumber .
\end{eqnarray}
$\rho$ is the density of state, $E$ the excitation energy, $V(N)$ the potential energy at the fission barrier and $N_{\text{fis}}$ the number of neutrons in the fissioning nucleus. It turns out that this yield can also be written in an approximate way,
\begin{eqnarray}
Y(E,N) \sim e^{2\sqrt{aE^*}} \nonumber ,
\end{eqnarray}
with $a$ a level density parameter simply taken equal to $A/8$ in this case and $E^* = E - V(N)$ the available energy above the potential. Therefore determining the distribution of fission fragments amounts to define the potential $V(N)$. The formula used is the following (colors related to Fig.~\ref{Fig_potfis}):
\begin{align}\label{eq:pot-fis}
 V(N) = &  \textcolor{blue}{C_{sym} (N_{\text{fis}}/2 - N)^2} + \nonumber \\
 &  \textcolor{green}{\delta U_i} +  \\
 &  \textcolor{magenta}{C_{asym(i)}(N_i - N)^2} \nonumber
\end{align}
and an example is given for $^{238}U$ in Fig. \ref{Fig_potfis} to illustrate the different contributions.

The macroscopic part of the potential is represented by a symmetrical parabola with the \textcolor{blue}{$C_{sym}$} parameter and the microscopic part by the shell effects characterized by the two positions $N_i$ ($i=1,2$), the respective strengths \textcolor{green}{$\delta U_i$} and the two curvatures  \textcolor{magenta}{$C_{asym(i)}$}. These parameters are adjusted on experimental data. Finally a random sample is used to determine the $N$ of the fragment. Other characteristics (charge, excitation energy) are obtained by including considerations of conservation, like the ratio of the numbers of neutrons and protons in the fissioning nucleus, but may take into account some effects seen experimentally as the polarization charge in the case of the asymmetric fission. To reproduce at best the experimental facts, features are defined at the most appropriate time. Thus the mass asymmetry is defined at the saddle point, low variation between the saddle point and the scission point, while the $N/Z$ ratio is defined at the scission point, because it can vary even after the saddle point.

Most of the other fission models used in spallation modeling are based on the Atchison's fission model \cite{ATC80} with refinements and new parameterizations, as the model implemented in PHITS  by Furihata \cite{FUR01a}.

\subsubsection{Binary breakup}

Evaporation of particles and high energy fission can explain almost all the charge and mass distributions, except for intermediate mass nuclei, as previously mentioned with the bottom part of Fig.\ref{Fig_femass}. Several means can be used to complete this description, like the extension of evaporation to heaviest nuclei than $\alpha$, taking into account their excitation energy, or the multifragmentation process addressed in the next subsection. Another idea for the creation of these nuclei is the extension of the binary breakup. Evaporation can be seen as a highly asymmetric binary decay and high energy fission as a symmetric binary decay. For Moretto \cite{MOR88}, {\it[...] there is no need to consider the two extremes of this distribution as two independent processes. Rather, one would conclude, fission and evaporation are the two, particularly (but accidentally) obvious extremes of a single statistical decay process, the connection being provided in a very natural way by the mass asymmetry coordinate.} He then generalized the model of Bohr-Wheeler fission by adding an extra dimension: the mass asymmetry. The decay width of the extended binary breakup is then
\begin{eqnarray}
\begin{array}{l}
\Gamma_{y}(E^*,J)dy = \\
\frac{1}{2\pi\rho_i}  \int_0^{E^*-B_f(J)}\hspace{-.1cm}\int \frac{dydp_y}{h} \rho_{sad}(E^*-B(J,y)-\frac{p^2_y}{2m_y}\varepsilon) d\varepsilon ,
\end{array}
\end{eqnarray}
where $y$ is the mass asymmetry, $p_y$ its conjugate momentum and $m_y$ the associated inertia.
A similar, but simpler, formalism is used in the de-excitation code GEMINI, described in \cite{FIL08}. Charity applied a simplified formula for the width where the mass asymmetry is contained in the barrier
\begin{eqnarray}
\begin{array}{l}
\Gamma_{Z,A}(E^*,J) = \\
\frac{1}{2\pi\rho_i}  \int_0^{E^*-B_f(J)} \rho_{saddle}(E^*-B_{Z,A}(J)-\varepsilon) d\varepsilon ,
\end{array}
\end{eqnarray}
where $B_{Z,A}(J) = B_{A}(J) + \Delta M +\Delta E_{coul} -\delta W -\delta P$, with $\Delta M$ and $\Delta E_{coul}$ the mass and Coulomb energy corrections due to the mass asymmetry, $\delta W$ and $\delta P$ respectively the shell and pairing effects.

In practice, this binary decay process, useful in some codes to produce the intermediate mass fragments, are still competing with the evaporation and fission channels, whose the well established formalisms give good results.

\subsubsection{Multifragmentation}

We previously mentioned that intermediate mass isotopes appear when the excitation energy increases. Their production can be taken into account by an extensive evaporation, i.e. beyond the $\alpha$ particle or by a generalized binary breakup. Another physical process is suggested, the multifragmentation \cite{BON76}, i.e. the breakup of the target nucleus in several nuclei prior to its thermal equilibrium. The idea underlying this formalism comes from the time equilibration of the excited nucleus between two successive emissions. This relaxation time becomes comparable to the time between two emissions of particle when excitation is greater than about 3 MeV, and thus the models based on a thermal equilibrium nucleus would no longer be justified. This relaxation time is equal to $2R/c_s$, where $R$ is the radius of the nucleus and $c_s$ the speed of sound in the nucleus. Estimates of the latter parameter can be found in reference \cite{SU88}. To treat this multifragmentation different formalisms have been developed and we can cite, according to a ranking of J. P. Bondorf \cite{BON95}, TDHF\footnote{TDHF: Time-Dependent Hartree-Fock} and QMD\footnote{QMD: Quantum Molecular Dynamics} models for the microscopic dynamics and BUU/VUU for kinetic models. Another formalism is preferred in the case of spallation reactions with high excitation energies, i.e. with numerous degrees of freedom: the statistical approach.

Briefly, the statistical model seeks to give a weight to each partition of the system and then one is randomly selected. Microcanonical, canonical and grand canonical ensembles can be used according to the knowledge of the system, to the possible approximations and to the necessary time to achieve reliable results. Thus, taking the case of the SMM\footnote{SMM: Statistical Multifragmentation Model} model, described in \cite{FIL08}, the number of partitions quickly becomes huge (for 100 nucleons, $\sim 2. 10^8$, and for 200, $\sim 4. 10^{12}$ \cite{BOT00}), and simplifications based on physical findings are made. For rather low excitation energies the microcanonical ensemble is used, but considering only multiplicities lower than or equal to 3. At high excitation energies the grand canonical ensemble is considered, as the number of partitions becomes too large. The choice of the partition is based on the statistical weights $\propto e^{S_p}$ in the first case, with the entropy $S_p$ given by A. Botvina in \cite{FIL08}
\begin{equation}
\begin{split}
S_p = ln\left( \prod_{A,Z}g_{A,Z} \right) + ln\left( \prod_{A,Z}A^{3/2} \right) - ln\left( A_0^{3/2} \right) - \\
ln\left( \prod_{A,Z}n_{A,Z}! \right) + \left(M-1\right)ln\left(\frac{V_f}{\lambda_{T_p}^3}\right) + \\
1.5\left(M-1\right) + \sum_{A,Z}\left( \frac{2T_p}{E_0} - \frac{\partial F_{A,Z}^S(T_p)}{\partial T_p} \right)  ,
\end{split}
\end{equation}
and the average multiplicities of fragments of mass A and charge Z in the
second case, 
\begin{equation}
\langle n_{A,Z}\rangle = g_{A,Z} \frac{V_f}{\lambda_{T_p}^3} A^{3/2} e^{\left[-\frac{1}{T}\left(F_{A,Z}(T,V)\right)-\mu A - \nu Z\right]} .
\end{equation}

These statistical weights and average multiplicities are based on several variables\footnote{$g_{A,Z}$ is the spin degeneracy factor, $n_{A,Z}$ the number of fragments with mass A and charge Z in the partition, $A_0$ ($E_0$) the mass (excitation energy) of the system, $M$ the multiplicity of fragments, $\lambda_{T_p}$ the thermal nucleon wavelength, and $\mu$ and $\nu$ the chemical potentials.} like the free energy $F_{A,Z}$, and a volume $V_f$ related to the freeze-out volume $V$ by $V=V_0+V_f$, where $V_0$ is the volume of the system with a  regular nuclear density. This volume $V_f$ corresponds to the nucleus configuration where it is no longer influenced by the Coulomb forces and thus is no longer bound by the nuclear forces. Then the fragments are dispersed by Coulomb repulsion from this freeze-out volume, after having been randomly positioned.

Another approach has been studied in the de-excitation code Abla07 \cite{FIL08}. K.-H. Schmidt follows a practical way relied on experimental observations \cite{SCH02}. Even though, to determine the partition, it is not enough to take into account the phase space only, knowledge of the dynamics are still poor. So, first of all, this multifragmentation process is not competing with the other de-excitation channels and the remnant excited nucleus, derived from the intranuclear cascade, can undergo a multifragmentation if its temperature is above a threshold. In this case two options are offered: a fixed value for this temperature, or the use of a value depending on the mass of the excited nucleus \cite{NAT02}.
The mass and charge of the emitted nuclei are obtained in two steps.
The first step is the determination of the amount of mass to be released (and thus the energy as well) to achieve a nucleus temperature below the multifragmentation threshold. This method is based on three points: i) When a unit mass is ejected, the nucleus loses energy  ii) according to some experiments \cite{ENQ99} this energy goes from 10 MeV for an excitation energy of 2.9 A.MeV to 5 MeV for an energy of 11.8 A.MeV, and iii) the $N/Z$ ratio of the initial nucleus is retained for the multifragmentation process, allowing to define the charge that must be emitted.
The second step defines the masses $A_i$, charges $Z_i$  and excitation energies of the emitted nuclei $i$. The masses are determined from a power law of the production cross-section observed experimentally
\begin{eqnarray}
\frac{d\sigma}{dA} \propto A^{-\tau} \nonumber ,
\end{eqnarray}
where the $\tau$ parameter is taken in Abla07 linearly dependent of the excitation energy per nucleon. The masses are then defined iteratively until their sum matches exactly the ejected mass calculated in the first phase. Charges $Z_i$ related to the masses $A_i$ are sampled in Gaussian distributions in such a way that the mean charge $Z_{mean}$ respects the $A/Z$ of the hot nucleus and that the width $\sigma_Z$ depends on the temperature and on the symmetric term of the equation of state of the nuclear matter $C_{sym}$, whose studies can be found in \cite{BUY05,SOU07,BOT06},
\begin{eqnarray}
\sigma_Z^2 = \frac{T}{C_{sym}} \nonumber  .
\end{eqnarray}
Finally, the excitation energy of the emitted nuclei ($Z > 2$) is based on the threshold  temperature and on their density parameter level, and their kinetic energy is due to the Coulomb repulsion and thermal agitation. For $Z \leq 2$ particles kinetic energy is drawn from a Maxwellian distribution.

As for the fission products, the various excited fragments from multifragmentation can in turn undergo evaporation and/or fission for the heaviest ones and, for the lightest, in some models, a Fermi breakup. This latter process is described hereafter.

\subsubsection{Fermi breakup}

The Fermi breakup has been originally developed by E. Fermi to explain the multiple emission of $\pi$ in high energies\footnote{at that time, a few hundreds MeV} nuclear reactions \cite{FER50}. He considered that the large number of possible states of the system leading to multiple pion production justify a statistical approach. The probability of occurrence of a given channel was equal to its statistical weight (eq. \ref{eq:pstat-fb}), 
\begin{equation}\label{eq:pstat-fb}
\text{statistical weight} = \left( \frac{V}{8\pi^3\hbar^3} \right)^{n-1} \rho_n(E)  ,
\end{equation}
with $n$ the number of components (particles) in the final state, $\rho_n(E)$ the density of final states depending on the final total kinetic energy and $V$ the volume of decay (often taken as the volume corresponding to normal nuclear density).
This method was then used for highly excited light nuclei \cite{EPH67}, since the hypotheses were the same.
Once a decay channel is selected, it remains to assign to each particle its kinetic energy from the Coulomb repulsion, as in the case of multifragmentation.
It is clear that the Fermi breakup and the statistical description of the multifragmentation are based on the same principle, a calculation of the probabilities of all possible partitions and selection by random sampling. Moreover, they involve both the volume where the explosion happens, even if in the original Fermi case it was not a volume of freeze-out. Recently a study \cite{CAR11}, aiming at merging the Fermi-breakup model and the statistical multifragmentation model based on the microcanonical ensemble, showed that it was possible if the source volume was equivalent, which seems unfortunately not always the case.

\subsection{Transport codes}

\begin{figure}[hbt]
\begin{center}
\resizebox{.35\textwidth}{!}{
\includegraphics{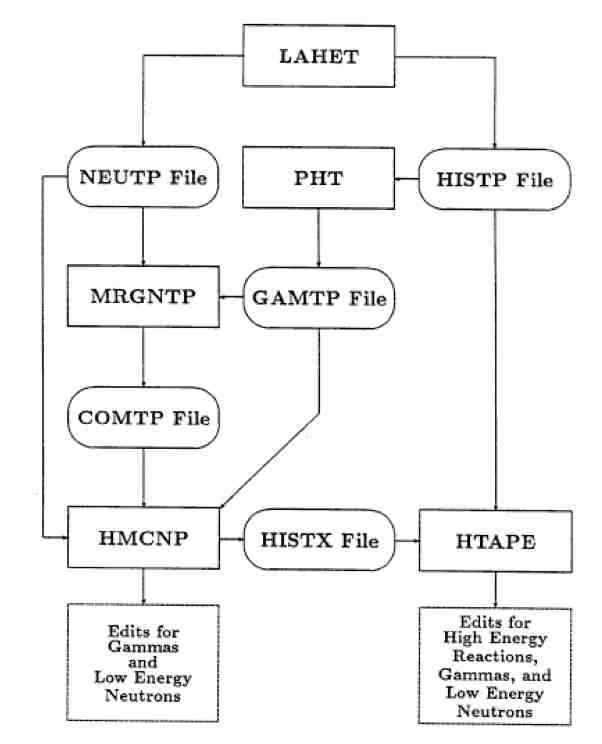}}
\end{center}
\caption{\label{Fig_lcs}
The LAHET Code System (LCS) developed in Los Alamos, with the LAHET code for the high (spallation) part and the HMCNP code for the low energy. Figure from \cite{PRA89}.}
\end{figure}

Most people dealing with spallation reactions are not model developers, but users. For their projects involving such reactions in massive targets, they need computational tools to get results. Moreover, they must also account for the particle transport in the given targets, as well as the slowdown, because the secondary particles can in turn induce other reactions. Thus the spallation codes must be implemented in a particle transport code, if the goal is not only to understand the elementary process, but also to offer a reliable tool to the scientific community. The transport codes, which manage the energy sector of spallation reactions must, to be complete, handle all the processes encountered by a particle of energies ranging from a few GeV down to a few meV (for neutrons). For questions of knowledge of physics and for technical matters, two choices are possible to treat particle interactions: databases or modeling. Thus, regarding the interactions of low-energy particles including neutrons, the choice of the database has been made. No model is able to reproduce the numerous resonances observed in the cross-sections, but, fortunately, the not too broad energy range (from a few MeV to 20/150 MeV) allows experimental determination of the needed data, with the use of evaluations when necessary (see e.g. \cite{SAL01}). In the spallation domain the situation is exactly the opposite. The energy range, from a few hundred MeV up to a few GeV, opens so many output channels that building a database fed by a battery of experiments is unimaginable, whereas our knowledge of the physics of spallation reactions provides an acceptable and now reliable modeling of the interactions. This separation, or difference, between low and high energies in the codes or systems of transport codes is  illustrated in Fig. \ref{Fig_lcs} with the LAHET Code System (LCS) \cite{PRA89}. The high-energy part is handled by LAHET (Los Alamos High Energy Transport), based on the previous code HETC \cite{ARM72,CHA72,GAB85}, itself an extension of NMTC \cite{COL71}, whereas for the low-energy part the code HMCNP, modified version of MCNP \cite{BRI86}, is used.

Another common point shared by the transport codes is the necessary reliability. They not only need to give the best results, which sometimes leads to a compromise between sophisticated models and reasonable running time, but also cover an energy range broader and broader, in addition to different operating systems evolving. The maintenance is crucial to keep a transport code alive. This aspect can have repercussions on the development of spallation models. 

As already said, a spallation model can be fully recognized only if it can be used via a transport code, this means, if it is able to give results in all cases (projectile, energy, target) and is reliable enough. Another point, rarely mentioned, is that its implementation in a transport code ensures in return the continued existence of such a model. Thus, for example, the INCL intranuclear cascade code has been implemented in several transport codes. Early in 2000 a collaboration with D. Prael allowed the use of INCL4.2 \cite{BOU02} and Abla \cite{JUN98} in the LAHET3.16 \cite{PRA01} code, and so in the LCS (see Fig.\ref{Fig_lcs}). 
The goal was to be able to apply INCL combined with Abla on thick targets. This work done, and the first results being encouraging \cite{DAV03},  the implementation in MCNPX2.5.0 (Monte Carlo N-Particle eXtended) was straighforward \cite{HEN05}. The particle transport codes handling spallation reaction have two origins: reactor physics and high-energy particle physics. Nowadays both families tend to join. Thus, willing to  provide INCL to most of the people, it has been decide to implement it in GEANT4 (GEometry ANd Tracking) \cite{GEA03}. This code is a toolkit used by the high-energy physicists. The point was that only the codes written in C++ could be implemented. Then A. Boudard, co-developer of INCL, initiated the translation of the Fortran77 version of INCL in C++. The work, started with P. Kaitaniemi \cite{KAI11}, has been continued and finalized by D. Mancusi \cite{MAN14}. This huge work has been the opportunity to re-think the INCL code and has made its maintenance easier. In the mean time developers of the PHITS (Particle and Heavy Ion Transport code System) transport code \cite{SAT13} decided to implement and use the INCL4.6  \cite{BOU13} version as their default intranuclear cascade model. This example is mentioned to stress the point that the high quality of the physics model, if absolutely necessary, is unfortunately not enough to make a computational code useful for the community. The coding part becomes then an essential activity.

Coming back to the origin of the transport codes, the studies of meteorites clearly illustrate their convergence regarding the spallation reactions. Meteorites are interstellar bodies irradiated by the cosmic rays, mainly protons, whose energy spectrum, going from a few MeV up to several TeV, peaks around 1 GeV, i.e. the spallation  domain. We will come back later to such studies, but the energy range is so broad that people involved in this topic were interested in many types of codes. This is clearly seen by the use first of LCS by Reedy and Masarik \cite{REE94} in 1994 and then by the use of GEANT3\cite{GEA87} combined to MCNP \cite{BRI93} by Masarik and Beer \cite{MAS99} five years later. Finally Tab. \ref{Tab_trnsp} gives an overview of the main particle transport codes dealing with spallation reactions with their characteristics. The FLUKA (FLUktuierende KAskade) \cite{FLU07a,FLU07b} and MARS (Math Analysis by Random Sampling) \cite{MOK04a,MOK04b,MOK04c,MOK04d} codes comes from the high energy physics. The former is developed mainly at CERN and the latter at Fermilab.

\begin{table*}
\begin{minipage}[c]{17cm}
\center
\begin{tabular}{|c|l|c|c|c|c|c|} \hline
Code  & INC (Pre-equilibrium)       & De-excitation   & Projectiles  & Upper energy limit\footnote{In some cases the limit can be changed by the user.}  & language \\ \hline\hline
                    &                              &                                                 
& n, p & 3.5 GeV & \\ 
                    &  Bertini (MPM\footnote{The multistage multistep pre-equilibrium exciton model \cite{PRA88}.})      &   
& $\pi$ & 2.5 GeV & \\ 
                    &                &                                                               
&  & & \\ \cline{2-2}\cline{4-5}
                    &                              &                                                 
&  n, p & 0.8 GeV\footnote{Above Bertini is used or FLUKA. It has to be mentioned that this is an old version of the FLUKA hadron generator, contained in FLUKA87\cite{FLU87}. That version has very little in common with the modern FLUKA\cite{FLU07a,FLU07b}.} & \\ 
                    &  Isabel  (MPM)     &   Dresner\footnote{Dresner is always combined with RAL\cite{ATC80} or ORNL\cite{BAR81} model for fission.}  or Abla                    
&  $\pi$ & 1.0 GeV\footnote{Like nucleons.} & \\  
                    &                &                                                              
& d, t, $^3$He, $\alpha$ & 1.0 GeV/nucleon\footnote{Above the LAQGSM  \cite{MAS06} model is invoked.} & Fortran 90\\ \cline{2-2}\cline{4-5}
MCNPX2.7  &                &                                                              
& n, p & $\sim$3 GeV& \\ 
MCNP6       &  INCL4.2               &                    
&  $\pi$ & $\sim$2.5 GeV & \\ 
                    &                &                                                              
&  d, t, $^3$He, $\alpha$ & $\sim$3 GeV/nucleon & \\ \cline{2-5}
                    &  \multicolumn{2}{c|}{}                                              
&  n, p & 5 GeV\footnote{Above the LAQGSM code is used.} & \\ 
                   &   \multicolumn{2}{c|}{CEM03\cite{MAS05}  + GEM\cite{FUR00}}                   
&  $\pi$ & 2.5 GeV & \\ 
                   &  \multicolumn{2}{c|}{}    
&  & & \\ \hline\hline
                    &                     &                                                 
& n, p &   3 GeV\footnote{Above the JAM code \cite{JAM00} is used.} & \\ 
PHITS2. 64 &  INCL4.6      &    GEM                  
&  $\pi$ &    3 GeV\footnote{Like nucleons.} & Fortran 77\\ 
                    &                     &                                                 
&  d, t, $^3$He, $\alpha$ &  3 GeV/nucleon\footnote{Above the JQMD code \cite{JQM95} is used.} & \\
                   &  &  
&  & & \\ \hline\hline
                    &                              &                                                  
& n, p & 10 GeV & \\ 
                    &  Bertini Intranuclear      &  \multicolumn{1}{l|}{Own Evaporation\footnote{Based on the work of Dostrovsky \cite{DOS59}.} (or GEM), }  
& $\pi$ & 10 GeV & \\ 
                    &  Cascade (+Preeq.\footnote{GEANT4 has its own pre-equilibrium model based on \cite{GUD83}.})    &  \multicolumn{1}{l|}{Fission,}                                               
&  & & \\ \cline{2-2}\cline{4-5}
                    &                              &     \multicolumn{1}{l|}{Multifragmentation\footnote{Based on \cite{BON95} and \cite{BOT87}.},}                           
&  n, p & 10 GeV & \\ 
GEANT4                    &  Binary cascade (+Preeq.)     &  \multicolumn{1}{l|}{and Fermi breakup}                 
&  $\pi$ & 10 GeV &C++  \\  
                    &                &            \multicolumn{1}{l|}{models }                                         
& d, t, $^3$He, $\alpha$ & few GeV/nucleon & \\ \cline{2-2}\cline{4-5}
                   &                &                                                        
& n, p & $\sim$3 GeV& \\ 
                   &  INCL++\cite{MAN14}               &         \multicolumn{1}{l|}{or AblaV3\footnote{Which is the Abla model previously mentioned, based on \cite{JUN98} and \cite{BEN98}.}}                               
&  $\pi$ & $\sim$3 GeV & \\ 
                    &                &                                                   
&  d, t, $^3$He, $\alpha$ & $\sim$3 GeV/nucleon & \\ \hline\hline
                    &                     &                                                 
& n, p &   5 GeV\footnote{Above the Glauber-Gribov model \cite{BER72} is used.} & \\ 
          &  PEANUT(GINC+preeq.\footnote{Based on the work of M. Blann \cite{BLA72}.})      &     \multicolumn{1}{l|}{Own Evaporation,}                  
&  $\pi$ &  5 GeV & \\  \cline{2-2}\cline{4-5}
FLUKA & & \multicolumn{1}{l|}{Fission} & & & Fortran 77\\
                    &   r-QMD-2.4\footnote{This model handle nucleus-nucleus interaction below 5 GeV/nucleon \cite{SOR95}.}                  &       \multicolumn{1}{l|}{and Fermi breakup models\footnote{Evaporation based on the Weisskopf-Ewing \cite{WEI40} formalism and fission on the Atchison model \cite{ATC80}.}}                                           
&  d, t, $^3$He, $\alpha$ &  5 GeV/nucleon\footnote{Above the DPMJET model (based on \cite{RAN95}) is used.} & \\
                   &  &  
&  & & \\ \hline\hline
                    &  &                                         
& n, p &   5 GeV\footnote{Above the LAQGSM code is used.} & \\ 
MARS &  CEM03  & GEM                   
&  $\pi$ &    5 GeV & \\ 
                   & & 
&  & & Fortran 77\\  \cline{2-5}
                    & LAQGSM  & GEM                                                   
&  d, t, $^3$He, $\alpha$ &  800 GeV/nucleon & \\
                   &  &  
&  & & \\ \hline\hline
\end{tabular} 
\caption{\label{Tab_trnsp} Overview of the main particle transport codes dealing with spallation reactions. Sometimes, according to the codes, the multi choices (the physics models, their reliability in a given energy range according to a type of projectile) make it difficult to summarize in a table, thus for more details, readers are invited to look at the user's manual of each code. While some codes allow the use of other projectiles than the nucleon, pion and composite particles up to the $\alpha$ particle, the list is limited to the projectiles that match the definition of spallation reaction given and addressed in this paper.}
\end{minipage}
\end{table*}

Most of the time the lower energy limit is determined by the availability of data. For neutron projectile, databases exist for all nuclei below 20 MeV and sometimes also from 20 MeV to 150 MeV, whereas for light charged particles the spallation model must be invoked down to 1 MeV.
Many more details on these transport codes can be found in the given references. These codes deal with energy ranges broader than the spallation domain and so use either databases or other kinds of models. Some offer the possibility to mix data and codes when both exist (e.g. MCNPX).

\subsection{Conclusions on the modeling}
\label{concmodel}

Most of the time the models describing the spallation reactions, or at least each phase, rely on the same basics, but the differences are in the details of the implementation, the way to do it and to put all the necessary mechanisms, ingredients, and {\it only} the necessary ones to be predictive and efficient. 
The knowledge of the spallation reactions is also driven by the need for simulations where they occur, in space or in accelerators, which gives another reason for getting the best description in terms of reliability and performance.
The two next sections report on these aspects, with the validation of the spallation models, where their predictive powers prove their basic foundations, and their use and usefulness in various topics.

\section{Validations}
\label{validations}

Model validation is an important step to know the value of a model. If everybody agrees with this preliminary remark, one must nevertheless, at first, define more precisely what is meant by {\it validation}. Validating can be understood in two slightly different ways depending on whether it is an end or a beginning. In other words, according to whether you are a model developer or a user, the goal of the validation is not the same and nor is the way to do it. The developer of a model sets goals to be achieved, he/she seeks to replicate such and such observable, and once he/she has built the model, he/she tests it on experimental data to judge its performance, to see if the mechanisms implemented fit the reality, and in this case the overall qualitative approach is important. The user of a model wants to know before making calculations, which are the models the most worthy of trust for the topic of interest and what are the margins of uncertainty. The approach is more targeted and quantitative.

Trying to summarize the works on benchmarking of spallation models would be audacious, since all users did their own particular validations at a given time for a given project with one or several models, which makes the job too laborious and the information too specific. Therefore, in order to give an idea of the efforts regarding the validation process of the spallation models and to tentatively draw some useful conclusions, in the following we draw up a brief history of the recent benchmarking campaigns, we address the work done by some transport codes teams, show the results of the most recent and most complete benchmark on elementary processes done under the auspices of the IAEA, and finally we mention some examples of validations in thick targets needing the use of transport codes.

\subsection{Overview of benchmark campaigns in the last two decades}
\label{benchhisto}

In the 1990s the Nuclear Energy Agency (NEA) of the OECD carried out a benchmark for three years, comparing codes in the field of spallation reactions. The goal was to get an inventory of modeling tools for spallation reactions in view of designing of accelerator-driven systems (ADS), dedicated to the transmutation of nuclear wastes. Two exercises were related to thin targets \cite{BLA93,MIC97}, involving microscopic spallation reactions, and one exercise was dedicated to thick targets \cite{FIL95}, thus including secondary reactions and therefore requiring the use of a transport code. The exercises on thin targets have attracted about twenty participants each. Chronologically the first \cite{BLA93} dealt with neutrons and protons emitted, while the second \cite{MIC97} dealt with the residual nuclei. A dozen models were implied in the thick target exercise. After these three years several conclusions could be drawn. The first observation was that there was room for improvement in all models. Used nuclear masses and binding energies were not always the correct ones, a rough value being used instead of the measured value. To alleviate this problem, details concerning  level densities and shell effects had to be taken into account in order to correctly reproduce some reaction thresholds. Light composite particle yields were often badly reproduced and pre-equilibrium emission seemed to be needed to remove this deficiency. A more recent benchmark, discussed later in section \ref{benchiaea}, will show more precisely that this type of emission is necessary all along the intranuclear phase, but also that the role of the pre-equilibrium is not so clear. Modeling of light fragmentation products yields have also to be improved. The same is true for  fission products as well. But in these cases the effort to remove deficiencies had to be done in the frame of de-excitation models. Finally models showed different behavior with low-energy projectiles. This is again due to the details in the models, as for the previous problems with reaction thresholds. A conclusion, not mentioned in the reports, implies that the models are globally good, and that it is time to turn to details.
Moreover an evaluation of several models concerning a number of different observables with different targets and energies is always difficult. First, very rarely, a model turns out to be better than another one for any of the observables; second,  experimental data are not always sufficiently numerous to be representative of the whole set of possible reactions. Hopefully, the lack of experimental data will trigger new experiments that are expected to enlarge data sets. 

After these NEA benchmarks, knowledge and simulation of spallation reactions have shown significant progress in Europe, thanks to successive projects dealing with transmutation of nuclear waste and ADS. These projects included experiments intended to increase the amount of data and then to help improve the models of spallation. There first was the HINDAS \cite{MEU05} (High and Intermediate Energy Nuclear Data for Accelerator-Driven System) project from 2000 to 2003, which benefited from a preliminary previous work \cite{MEU00}, then NUDATRA \cite{NUD11} (NUclear DAta for TRAnsmutation) from 2005 to 2010 followed by ANDES \cite{AND13} (Accurate Nuclear Data for Nuclear Energy Sustainability) from 2010 to 2013.

In HINDAS, as in the other two projects, a report on the performance of the models was done at the beginning, firstly pointing out the difficulty in reproducing some existing experimental data or the dispersion of predictions of different models on observables where little or no experimental data exist, and secondly allowing defining some specific domains to be improved. At the end of the project, models were tested especially on the new data obtained. If several models were tested at the beginning, like Bertini, Isabel (both implemented in the LAHET Code System) and INCL  for the intranuclear cascade and Dresner-Atchison and GEMINI for the de-excitation part, the spallation codes that were improved and combined were INCL for the intranuclear cascade, ending up in INCL4.2 \cite{BOU02}, and Abla \cite{JUN98} for the de-excitation. This combination of codes has shown good progress in the simulation of the spallation. It may be mentioned, among others, the ability of INCL4.2 to give the right reaction cross-section thanks to a diffuse nuclear surface, and a good reproduction of fission products for Abla through a new analytical approximation to the solution of the Fokker-Planck equation for the time-dependent fission width and a semi-phenomenological model \cite{BEN98} for the fission product yields. A coalescence model implemented in INCL, allowing cluster emission, showed already encouraging results in a previous version \cite{LET02}. This idea will be applied later in an improved version of INCL \cite{BOU03}. This last point is also related to the excitation energy of the remnant, given by the intranuclear cascade model to the de-excitation model. This energy being a crucial input for the de-excitation, the intranuclear model must include the main channels in the right way to get it. It will be seen then that the pion emission still have then to be improved in INCL. The new experimental data obtained during this project will greatly help to validate the models, but also to show the way to improve them, and are still now considered as essential. We can cite the data on neutrons at SATURNE \cite{LER02} and COSY \cite{HER01}, the data on light charged particles at COSY \cite{HER00} and the isotopic distributions for the residue production at GSI in inverse kinematics \cite{ENQ01,TAI03,BER03}, complementary to the excitation functions. More details can be found in references \cite{BOU02,MEU05}.

After HINDAS, INCL4.2 and Abla give generally rather good results, without free parameters, i.e. parameters defined once and for all, however both could still be improved. Thus some defects were to be corrected: production of light charged particles (up to $\alpha$), production of intermediate mass fragments, called IMF (with A $\le$ $\sim$20), production of residues close to the target and handling of low-energy projectiles. This work was conducted within NUDATRA \cite{NUD11}. Although not mentioned in the list, the pion production in INCL has been improved by the works on nucleon and pion potentials \cite{AOU04,AOU06}, which is also a required step to get the right excitation energy. The establishment of a process of coalesence allowed the production of composite particles up to alpha in a first step (INCL4.3 \cite{BOU03}) and up to nuclei with A$\le$10 \cite{CUG11} later (but reduced in INCL4.5 \cite{BOU13} to A$\le$8, for a question of computational time) and regarding Abla (become Abla07) the evaporation of nucleons and $\alpha$ was extended to any type of particles and nuclei in addition of a multifragmentation or breakup phase, open before the other de-excitation channels and depending on a threshold temperature. Regarding the projectiles with energies below 200 MeV, great improvements have been obtained with INCL by taking into account the Coulomb deflection and by handling correctly the soft collisions. This latter point is also mentioned as a good example demonstrating that a good prescription can prevent from improving a model if it is used outside of its original goal. In INCL the low-energy reactions, called soft collisions, are neglected, because they occurred often with no changes in the  progress of the cascade, since most of the time they are Pauli-blocked. However this prescription must be removed, or at least modified, for the first collision of a low-energy projectile with a nucleon of the target nucleus, otherwise the number of events without interaction will be too high and the reaction cross-section significantly underestimated. Doing so, the INCL4.5 version was able then to take care correctly of the low-energy reactions. Finally, and unfortunately, the problem of the residues close to the target remained an open question. The experiments performed and used during this project were on light charged particle emission \cite{BUB07,BUD08} and on the evolution of the fission cross-sections according to the projectile energy \cite{AYY14}.

\begin{figure}[hbt]
\begin{center}
\resizebox{.3\textwidth}{!}{
\includegraphics{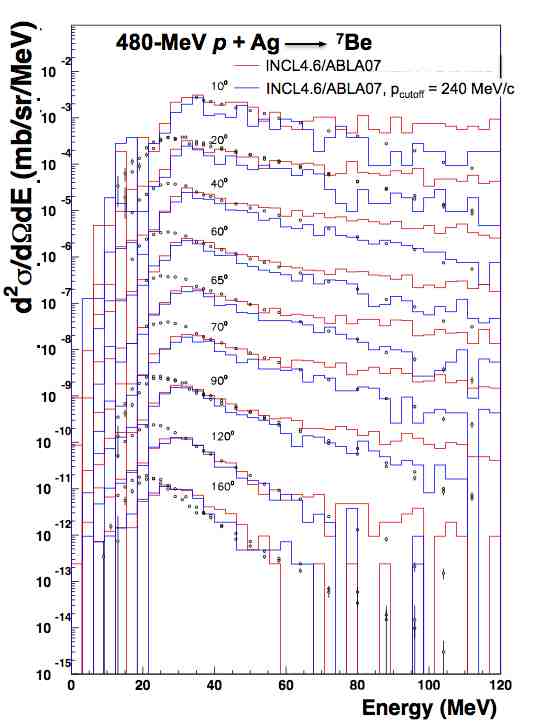}}
\end{center}
\caption{\label{Fig_coal-be7}
Double differential cross-sections for the production of $^7$Be for the reaction p(480 MeV)+Ag, as function of the produced $^7$Be kinetic energy. Calculations with INCL4.6 (updated version of INCL4.5) combined to ABLA07 (red lines) are compared with the version including a momentum cut-off (blue lines) and with the experimental data (black symbols) \cite{GRE80}. Figure from \cite{CUG13}.}
\end{figure}

A little after the NUDATRA project it was decided to test the codes on a specific energy domain, i.e. in the range 150-600 MeV. This has been done in the ANDES project \cite{AND13}. This energy range corresponds to the future prototype of the ADS Myrrha planned in Mol, Belgium. Data collected for the {\it Benchmark of Spallation Models}, discussed later in section \ref{benchiaea}, were used with new experimental data for the energy in question. The details of this study are published in an ANDES report \cite{DAV11}. Findings indicated good results, even if some defects were to be corrected and which tended to increase as the incident energy decreased. Moreover, since on one hand the particles in a spallation target for ADS will be slowed down below 150 MeV and the secondary low-energy particles can play a role, and, on the another hand, databases are missing for some types of particles in this energy range, the low energy interaction modeling will be also studied. Regarding intermediate mass fragment emission, if the INCL4.5-Abla07 gave rather good results, double differential spectra showed a high energy tail overestimated by INCL4.5. The coalescence model, where an outgoing particle aggregates with other nucleons, is based on a space-phase proximity criterion, and the use of a momentum cut-off seems to be needed as indicated by Fig. \ref{Fig_coal-be7}.
Improvement of clusters at low energy, together with residues close to the target, has been done considering the Fermi sea definition. In the cascade there are two types of nucleons: those which underwent a collision, the participants, and the others, the spectators. However, a participant that is slowed down below the Fermi energy is considered back to spectator. Actually the energy criterion is the Fermi energy plus a small quantity. In a first attempt this value was a constant value, 18 MeV, and the results were improved, but with a spurious hole in the neutron spectra. This was cured using a softening "back-to-spectator" condition, i.e. requiring that the energy of the participant falls below the emission threshold. This example showed once again that a careful implementation is required to take full advantage of a good idea. Regarding the secondary particles, INCL was able to emit clusters and handled interactions with particles up to $\alpha$, but no special attention was paid to the low-energy light cluster collisions. Two aspects are important in this case: the scenarios (or entrance channels) and some specific ingredients. The scenarios that have to be taken into account are: formation of an (almost) undisturbed projectile fragment by non-interacting nucleons, formation of a compound nucleus (fusion) and cascade of interacting projectile nucleons; the two ingredients that become crucial at low energies are: Coulomb distortion and Q-values. This way the new version of INCL was  made much more reliable in this field. More details are given in \cite{DAV13} and in section \ref{toolmegapie}. A last example of deficiency studied during ANDES was the overestimation of yields of some nuclei close to the target and more precisely a typical case : the (p,2p) process. The overestimation of the cross-section could be improved by increasing the excitation energy, which facilitates the neutron evaporation. These (p,2p) events are probably peripheral and the nucleons involved are the ones with the larger momenta, because usually nucleons are treated as classical particles, as far as their momentum is concerned, and sampled in a Fermi sphere distribution. Now, nucleons are quantum particles and the wave function of the less energetic nucleons in the nucleus can extend up to the surface. Thus, accounting for this quantum property, the removed peripheral nucleons could take away less energy, and so the excitation energy of the remnant nucleus could be increased. Encouraging results has been obtained in a recent study published by D. Mancusi \cite{MAN15}, where the quantum-mechanical features of the wave functions are mimicked by using a simple shell-model that allows deep orbitals to contribute to the surface energy content.

In these three European projects the models, and more precisely INCL4 and Abla, were benchmarked and improved with some specific goals, but not used to design an experiment or a facility. In other words it was done mainly by the code developers. As mentioned in the introduction, one can also test models with the first aim not to improve them, but to use them in the best way, that is to choose the most suitable model, knowing its uncertainties concerning a studied subject. For example we can cite the work done within the EURISOL-DS (EURopean Isotope Separation On-Line Radioactive Ion Beam Facility - Design Study) project, from 2005 to 2009.
The project aimed at designing a facility delivering radioactive ion beams using two types of spallation targets: one producing the ions directly from spallation reactions and the other using the spallation neutrons to induce fission in a uranium target. Several types of materials were to be studied and at first a benchmark of the tools, spallation codes, was done. As already said, there is no use here to discuss all results, but it is worth saying that for this project three of the five transport codes listed in Tab. \ref{Tab_trnsp} were studied as well as all model combinations in MCNPX2.5.0. The reader will find information, concerning MCNPX2.5.0, in the references \cite{RAP06,DAV07a,DAV07b}, where it was concluded that the two model combinations Isabel-Abla and INCL-Abla, as well as the CEM2k model, were the three best options for this purpose. Regarding FLUKA, it was first benchmarked on the elementary residue production cross-sections, reproducing the experimental data within a factor of two in most of the cases \cite{FEL06}, like the best options of MCNPX2.5.0. Then it was compared to neutron multiplicities \cite{PIE06} with more or less good results and finally validated on neutron attenuation in shielding materials \cite{ENE06} where the transport of the particles is important. In \cite{ENE09} a comparison of PHITS with MCNPX2.5.0\footnote{Using the LAQGSM model in this case} for some ion-ion interactions was performed. The question of considering these latter reactions as spallation reactions is still open. The numerous possibilities concerning the target materials forced researchers to search for a large set of experimental data, which will be efficiently reused for the {\it Benchmark of Spallation Models}. However, the benchmark of elementary processes in the transport codes within EURISOL-DS was obviously focussed on the projectile energy, i.e. around 1 GeV, and more benchmarks can be found on the websites of the developers of these transport codes as discussed in the next section.

Regarding the experimental data, used for such validations, great efforts have been made to collect and offer high quality data. We can cite for the elementary reactions the Experimental Nuclear Reaction Data base EXFOR \cite{EXFOR}. This database originally dedicated to low-energy neutron cross-sections has been extended to light charged particles and to higher energies, say around 1 GeV, but sometimes beyond. EXFOR aims at providing experimental data with all necessary information, when available, to make them usable. EXFOR is a network of Nuclear Reaction Data Centres (NRDC\footnote{https://www-nds.iaea.org/nrdc/}) coordinated by the International Atomic Energy Agency (IAEA). Exchange of information between developers, users and experimentalists are promoted by the Working Party on international nuclear data Evaluation Co-operation (WPEC) established by the NEA Data Bank  in close co-operation with the Nuclear Data Section of the IAEA. For example a subgroup dealing with {\it Quality Improvement of the EXFOR database} \cite{KON11} gave the opportunity to improve the data related to spallation reactions (some errors were corrected) and the way to find them\footnote{Some data come from experiments done in inverse kinematics, this means that the heavy nucleus is the beam and the light particle the target. These data must be also found when considering the light particle as the beam.}. On the side of thick, massive or complex targets, each project is a new one, however some typical cases exist and a joint project was launched in 1996  by the NEA Data Bank and ORNL/RSICC (Oak Ridge National Laboratory, Radiation Safety Information Computational Center): SINBAD (Shielding Integral Benchmark Archive Database). The objective was to produce a database containing an internationally established set of radiation shielding and dosimetry data. For example, the experimental data used within EURISOL-DS for the validation of FLUKA on neutron attenuation came from the HIMAC experiments included in SINBAD. More information can be found in \cite{KOD06}.

Validation is a long and complex process, starting with numerous benchmarks in laboratories done {\it privately} or within collaborations on international projects followed by the work of the transport code developers, and, from time to time, with international benchmarks as the ones of NEA and IAEA making the point on the spallation reaction modeling. The following sections give more details on some examples to show the difficulties and the outcomes of the different types of benchmarks.

\subsection{Benchmarks within transport codes}
\label{benchtrans}

Particle transport codes dealing with spallation reactions are tools supposed to work with all kinds of material (or target), all types of projectiles, with a module handling the lower energy domain when the particles are slowed down to a threshold depending on the spallation models. Moreover most of these codes also manage higher energy reactions. Thus a benchmarking or validation activity cannot aim at covering all possible studies, as it is written in the MCNPX2.6.0 User's Manual \cite{MCN08}: {\it All computer simulation codes must be validated for specific uses, and the needs of one project may not overlap completely with the needs of other projects. It is the responsibility of the user to ensure that his or her needs are adequately identified, and that benchmarking activities are performed to ascertain how accurately the code will perform. The benchmarking done for code developments for the MCNPX sponsors may or may not be adequate for the needs of the user's particular program. We make our benchmarking efforts public as they are completed, but the user must also develop a rigorous benchmarking program for his/her own application.}
Nevertheless great efforts have been carried out by the transport code developers to give indications on the reliability of their code, often motivated by particular applications or current topics. We give hereafter some examples of the work performed by three of them: PHITS, MCNPX/MCNP6 and GEANT4.

Around forty references related to benchmark studies can be found on the website of the PHITS code \cite{PHIXX}. PHITS covers a large energy scale from say 0 to 200 GeV with not only light hadrons as projectiles, but also heavy ions, leptons and gammas, and the work is almost always motivated by a specific goal, e.g. safety, medicine, cosmic rays, accelerators. Thus, if more than ten papers deal with benchmarks with thick target (shieldings studies) and more or less the same amount for heavy ion reactions, the number of papers on the validation of the elementary processes of the spallation models in PHITS is far from being the majority. However these studies deal with difficult points and experiments are sometimes carried out to provide the needed experimental data, which are then useful for the whole community. Hereafter some examples are cited, first to mention what has been done by the PHITS developers, but also to address some points raised by these benchmarks and discussed later. It must be added that the PHITS team was involved in the {\it Benchmark of Spallation Models} developed in section \ref{benchiaea}.

Several spallation models are available in the PHITS code and benchmarks are often used to know which one is the best in a particular case. It is not the place here to summarize the history which ends up in the choice of the default options or default models in PHITS, and all information is in \cite{PHIXX}. However three interesting points, at least, can be drawn from this validation work. The first one concerns the libraries, the second one the neutron-induced spallation reactions and the last one the pions and resonances. Below 20 MeV or 150 MeV it is common to use data libraries instead of models, the former ones being much faster and validated. Beyond, the problem is the large energy range and the numerous output channels, which makes the building of such libraries tedious. Nevertheless this topic is developed later in section \ref{toollibraries}, and we can mention here that, at least in two papers, PHITS developers validated their own JENDL-HE libraries. Briefly the GNASH \cite{YOU92} code has been used below 150 MeV and the JQMD model above, and up to 3 GeV, to construct this library. In the paper dealing with forward neutron emission from the reaction p(100-200 MeV)+C \cite{IWA12} the JENDL-HE library shows its ability to reproduce the Quasi-Free peak contrary to the spallation models. This is due to the choice of the right and dedicated models to build the library in the specific energy ranges. However this asset can become a drawback: a gap at the frontiers of the models used. This appears in the neutron activation cross-sections study \cite{YAS13}, where a discontinuity is seen around 250 MeV. The experimental data used for this study were measured on purpose, because, while below 20 MeV numerous experimental data are available, they are scarce between 20 MeV and 200 MeV and nothing exists above. Thus several experiments had been carried out at the Research Center for Nuclear Physics (RCNP) at Osaka using a quasi-monoenergetic neutron beam of around 270 MeV and around 380 MeV (references in \cite{YAS13}). This kind of data is strongly needed in particular for extraterrestrial matter irradiation by cosmic rays, and some other techniques have been developed to extract such information \cite{LEY11}. The spallation model used by PHITS in this study \cite{YAS13} is INCL4.6 \cite{BOU13} and the results are reasonable, with obviously no discontinuity in energy. This specific point on excitation functions with neutron projectiles is discussed in section \ref{toolcosmogeny}. A third point, discussed in another validation of PHITS on neutron production via the pion interactions \cite{IWA04}, is on the better way to take into account the resonances. The place where the $\Delta(1232)$ is produced in the nucleus is invoked to explain the production of negative and positive pions in iron and lead, as well as the nucleon resonances taken into account in JQMD to understand the differences between the two types of pions. This last point is an example of what must be included to get the best results in a simpler and faster way. The production of pions around GeV, like in this paper, and at higher energy can be handled in two ways: either by the decay of the resonances considered as participants of the intranuclear cascade or forgetting this intermediate step and taking into account the direct production of pions. So, for example, the first way is implemented in JQMD, whereas the INCL code used the second method. This way to proceed is explained in more details in sections \ref{spalhe} and \ref{spalstrange}.

Like PHITS, MCNPX/MCNP6 put in its website the references of the works done about validations. Around fifty papers are cited and the half dedicated to the spallation reactions. The rest is mainly on low energy neutron and criticality. The validations of the spallation models in MCNPX/MCNP6 have been carried out most of the time by S. Mashnik, focussing on the CEM code, but also LAQGSM at higher energies, with comparison to experimental data and other codes implemented in MCNPX/MCNP6 or not. So, through these benchmarks, we can follow the development of the CEM code, from CEM95 \cite{MAS95} to CEM03.03 \cite{MAS08} via CEM2k \cite{MAS00b}. There is no doubt that this continuous work of validation to test and to know the points to improve is the reason why in MCNP6 the CEM03.03 has become the default option, replacing the famous Bertini-Dresner combination. Among all these references \cite{MCNXX}, we can cite the one on validation and verification of MCNP6 \cite{MAS11}, where eighteen tests are presented in details with the input files to help the users. Light and heavy targets with mainly protons, but also neutrons and pions as projectiles are considered in a wide energy range. Three tests are dedicated to gamma-induced reactions around GeV. Photonuclear reactions are not usually considered as spallation reactions, but this interesting topic is addressed in the last section \ref{front}. The fifteen other benchmarks were mainly dedicated to the elementary processes with thirteen tests, two tests dealing with thick targets. In addition, considering the spallation reactions as the interaction of  a light particle with a nucleus, we can mention two other tests described in another report dedicated to the LAQGSM code \cite{MAS11b}. One on backward neutron emission from a thick lead target, a difficult task for the spallation models, and another one induced by alphas, not handled by CEM03.03. Doing so the codes have been verified, i.e. can work with no bugs, and validated, i.e. give reasonable results and can be considered reliable.

The website of GEANT4 offers also results of benchmarking works. However, some differences exist with the two previous transport codes. If in the three cases the references of the experimental data used are given, only comments on the results are available in the GEANT4 case (one article can nevertheless be mentioned \cite{BAN11}), probably because, in return, an impressive number of plots from the calculation results have been obtained and are put in the website, which allows anybody to make its own analysis. We see clearly with the GEANT4 validation procedure, compared to the other two, that a compromise has to be made between the number of tests and the detail of the analysis. Once again the validations can be undertaken with different goals: user-oriented,  developer-oriented or in-between. GEANT4 seems the most user-oriented regarding the amount of information provided, whereas PHITS and MCNPX/MCNP6 give maybe less results\footnote{difficult to compare, because the cases studied are not always the same ones}, but with most of the time a critical analysis. There is however one difficulty that each has to face, which is the regular updating of such benchmarks. Nevertheless the work done and the information made available by the transport codes concerning validation is a great database for the user as well as for the developer.

Finally we can mention two papers dealing with an inter-comparison of transport codes, in particular those in Tab. \ref{Tab_trnsp}, but at the frontiers of the spallation  reactions. One benchmarked heavy ion reactions \cite{SIH08} and the other one high energy reactions \cite{MOK07}, i.e. the lowest projectile energy being 12 GeV/c. In those works the number of reactions were limited to draw clear conclusions.

\subsection{IAEA {\it\bf Benchmark of Spallation Models}}
\label{benchiaea}

Around fifteen years after the benchmarks carried out by the NEA and followed by great improvements in the spallation reactions modeling, three institutes, CEA (France), Soreq (Israel) and FZJ (Germany), decided to launch another validation exercise under the auspices of the IAEA. In order to get the most discerning conclusions the goal was to focus on the elementary processes only. Nevertheless, trying to make an assessment of the spallation models used around the world can be an ambitious task for many reasons:
\begin{list}{$\bullet$}{}
\item scanning all areas of spallation reactions (type of projectile, projectile energy, type of nucleus target, observable) 
\item involving the most of existing models, 
\item asking all participants to perform all calculations with identical output formats, 
\item creating the tools for the analysis of results, 
\item making the most comprehensive analysis, 
\item making this exercise valuable to developers {\it and} users, and 
\item making available the information and tools used for future next exercises. 
\end{list}
This IAEA {\it Benchmark of Spallation Models} is developed hereafter for several reasons. First, up to now,  it is the most complete benchmark in this field, and second, if some papers have been written to summarize the results obtained \cite{LER11,DAV11b}, some information provided during the exercise can still be analyzed and some points are underlined here. The author apologizes in advance for the resolution of the plots shown in this section, most of the time taken from the website of the benchmark. The goal of these plots is here often to show the trend, not to discuss a particular model, then if the reader requires more information, all information is available on the website.


A first workshop was held in Trieste in February 2008 to define the content of this benchmark. It brought together developers presenting their models and experimentalists who made a review of existing data. The first result of this workshop is a report \cite{FIL08} with, in particular, interesting descriptions and details of many models. Another important point was the compromise on the experimental data that should be used for the comparisons. The aim was to cover the broadest spectrum of spallation reactions taking into account two constraints: the availability of data and the amount of work that would lead to the participants and the organizers. Taking into account these objectives and constraints, it was decided to consider as projectiles only nucleons, mainly protons with incident energy from 20 MeV to 3000 MeV, but, as targets, heavy, intermediate and light nuclei. The observables were spectra of particles (neutrons, protons, deuterons, tritons, $^3$He, $\alpha$ and pions), neutron multiplicities, isotopic (mass and charge) distributions of the residual nuclei, and isotope production function of the energy of the projectile (excitation functions). 

\begin{table*}
\begin{minipage}[c]{17cm}
\center
\begin{tabular}{|c|c|c|c|c|} \hline
Projectile  & Target   & Energy (MeV) & Observable  & Reference \\ \hline\hline
n &  $^{nat}$Fe                      &    65                          & DDXS - neutron & \cite{HJO96} \\ \hline
p &  $^{208}$Pb                     &    63                          & DDXS - neutron & \cite{GUE05} \\ \hline
p &  $^{nat}$Pb                      &  256                          & DDXS - neutron & \cite{MEI92} \\ \hline
p &  $^{nat}$Fe, $^{nat}$Pb  &   800                         & DDXS - neutron & \cite{AMI92} \\ \hline
p &  $^{nat}$Fe, $^{nat}$Pb  &   800, 1200, 1600     & DDXS - neutron & \cite{LER02} \\
   &                                         &                                  & Multiplicity - neutron &  \\ \hline
p &  $^{nat}$Fe, $^{nat}$Pb  & 3000                         & DDXS - neutron & \cite{ISH97} \\ \hline
p &  $^{nat}$Fe                     &     1200                     & Multiplicity distribution - neutron & \cite{HER01} \\ \hline
p &  $^{nat}$Pb                     &     1200                     & Multiplicity distribution - neutron & \cite{LET00} \\ \hline\hline
n &  $^{209}$Bi                      &    542                          & DDXS - proton, deuton, triton & \cite{FRA90} \\ \hline
p &  $^{56}$Fe, $^{209}$Bi    &    62                          & DDXS - proton, deuton, triton, $^3$He, $\alpha$ & \cite{BER73} \\ \hline
p &  $^{208}$Pb                    &    63                          & DDXS - proton, deuton, triton, $^3$He, $\alpha$ & \cite{GUE05} \\ \hline
p &  $^{27}$Al, $^{197}$Au   &    160                          & DDXS - $\alpha$ & \cite{COW96} \\ \hline
p &  $^{nat}$Ni                      &    175                          & DDXS - proton, deuton, triton, $^3$He, $\alpha$ & \cite{BUD09} \\ \hline
p &  $^{nat}$Ni                      &    175                          & DDXS - proton                                                      & \cite{FOR91} \\ \hline
p &  $^{208}$Pb                    &    800                         & DDXS - proton                                                       & \cite{CHR80,MCG84} \\ \hline
p &  $^{181}$Ta                    &    1200                         & DDXS - proton, deuton, triton, $^3$He, $\alpha$ & \cite{HER06} \\ \hline
p &  $^{197}$Au                    &    1200                         & DDXS - proton, deuton, triton, $^3$He, $\alpha$ & \cite{BUD08} \\ \hline
p &  $^{197}$Au                    &    2500                         & DDXS - proton, deuton, triton, $^3$He, $\alpha$ & \cite{LET02,BUB07} \\ \hline\hline
p &  $^{nat}$C, $^{27}$Al, $^{,nat}$Cu and $^{nat}$Pb &    730 & DDXS - $\pi^{\pm}$ & \cite{COC72} \\ \hline
p &  $^{27}$Al                     &    2205 & DDXS - $\pi^{-}$ & \cite{ENY85} \\ \hline\hline
p &  $^{56}$Fe                    &    300 & Isotopic distribution\footnote{Experiments done in inverse kinematics}  & \cite{VIL07} \\ \hline
p &  $^{56}$Fe                    &    1000 & Isotopic distribution  & \cite{VIL07,NAP04} \\ \hline
p &  $^{208}$Pb                  &    500 & Isotopic distribution  & \cite{AUD06} \\ \hline
p &  $^{208}$Pb                  &    1000 & Isotopic distribution  & \cite{ENQ01} \\ \hline
p &  $^{238}$U                   &    1000 & Isotopic distribution  & \cite{TAI03,BER03,RIC06,BER06} \\ \hline\hline
p &  $^{nat}$Fe                   &             & Excitation function  & \cite{MIC97,MIC95,SCH96,MIC02,AMM08,TIT08} \\ \hline
p &  $^{nat}$Pb                   &             & Excitation function  & \cite{GLO01,LEY05,TIT06} \\ \hline\hline
\end{tabular} 
\caption{\label{Tab_IAEA} Experimental data sets considered in the IAEA  {\it Benchmark of Spallation Models}.}
\end{minipage}
\end{table*}
Table \ref{Tab_IAEA} lists the experimental data sets. These data were then also put in the ASCII format to be easy to use during this benchmark, but also by anyone, since the files have been made available via a dedicated website (http://www-nds.iaea.org/spallations/). Each of these files has been associated with another one summarizing the necessary information on the experimental conditions. As regards the energy range under consideration, the domain below say 150 MeV has been validated also, because, when used in a transport code, the models are called when no database exists, and especially at low energies. Also, it is interesting to know at least the behavior of the models in these domains, where they are not supposed to give good results. 
\begin{table*}
\begin{minipage}[c]{17cm}
\center
\begin{tabular}{|c|c||c|c|c|c|c|c|} \hline
Code  & Representative   & Neutron & Neutron    & p, d, t,  & Pions & Residue & Excitation  \\ 
          &                             &  DDXS  & Multiplicity & $^3$He, $\alpha$          &           &  distributions & functions \\ \hline\hline
cem03-02 &  A. Sierk, S. Mashnik, & \cellcolor{green} & \cellcolor{green} & \cellcolor{green} & \cellcolor{green} & \cellcolor{green} & \cellcolor{green}   \\ 
                 &  K. Gudima, M. Baznat  & \cellcolor{green} & \cellcolor{green} & \cellcolor{green} & \cellcolor{green} & \cellcolor{green} & \cellcolor{green}  \\ \hline
cem03-03 &  K. Gudima, M. Baznat, & \cellcolor{green} & \cellcolor{green} & \cellcolor{green} & \cellcolor{green} & \cellcolor{green} & \cellcolor{green}   \\ 
                 &  A. Sierk, S. Mashnik & \cellcolor{green} & \cellcolor{green} & \cellcolor{green} & \cellcolor{green} & \cellcolor{green} & \cellcolor{green}  \\ \hline
cascade04 &  H. Kumawat & \cellcolor{green} & \cellcolor{green} & \cellcolor{green} & \cellcolor{green} & \cellcolor{green} & \cellcolor{green}   \\ \hline
cascade-asf & A. Konobeyev & \cellcolor{green} & \cellcolor{green} & \cellcolor{green} & \cellcolor{green} & \cellcolor{green} & \cellcolor{green}   \\ \hline
cascadex &  Yu. Korovin & \cellcolor{green} & \cellcolor{red} & \cellcolor{green} & \cellcolor{red} & \cellcolor{red} & \cellcolor{green}   \\ \hline
phits-bertini &   & \cellcolor{green} & \cellcolor{green} & \cellcolor{green} & \cellcolor{green} & \cellcolor{green} & \cellcolor{green}   \\  \cline{1-1}\cline{3-8}
phits-jam &  N. Matsuda & \cellcolor{green} & \cellcolor{green} & \cellcolor{green} & \cellcolor{green} & \cellcolor{green} & \cellcolor{green}   \\ 
\cline{1-1}\cline{3-8}
phits-jqmd &   & \cellcolor{green} & \cellcolor{green} & \cellcolor{green} & \cellcolor{green} & \cellcolor{green} & \cellcolor{green}   \\ \hline
Geant4-bertini & Geant4 Hadronic  & \cellcolor{green} & \cellcolor{green} & \cellcolor{green} & \cellcolor{green} & \cellcolor{green} & \cellcolor{green}   \\\cline{1-1}\cline{3-8}
Geant4-binary & Group  & \cellcolor{green} & \cellcolor{green} & \cellcolor{green} & \cellcolor{green} & \cellcolor{green} & \cellcolor{green}   \\ \hline
isabel-abla07 &   & \cellcolor{green} & \cellcolor{green} & \cellcolor{green} & \cellcolor{green} & \cellcolor{green} & \cellcolor{green}   \\ \cline{1-1}\cline{3-8}
isabel-gemini++ & D. Mancusi  & \cellcolor{green} & \cellcolor{green} & \cellcolor{green} & \cellcolor{green} & \cellcolor{green} & \cellcolor{green}   \\ \cline{1-1}\cline{3-8}
isabel-smm &    & \cellcolor{green} & \cellcolor{green} & \cellcolor{green} & \cellcolor{green} & \cellcolor{green} & \cellcolor{green}   \\ \hline
incl4.5-abla07 &   & \cellcolor{green} & \cellcolor{green} & \cellcolor{green} & \cellcolor{green} & \cellcolor{green} & \cellcolor{green}   \\ \cline{1-1}\cline{3-8}
incl4.5-gemini++ & D. Mancusi  & \cellcolor{green} & \cellcolor{green} & \cellcolor{green} & \cellcolor{green} & \cellcolor{green} & \cellcolor{green}   \\ \cline{1-1}\cline{3-8}
incl4.5-smm &    & \cellcolor{green} & \cellcolor{green} & \cellcolor{green} & \cellcolor{green} & \cellcolor{green} & \cellcolor{green}   \\ \hline
MCNPX  &  F. Gallmeier  & \cellcolor{green} & \cellcolor{green} \textcolor{red}{No multiplicity} & \cellcolor{green} & \cellcolor{green} & \cellcolor{green} & \cellcolor{green}   \\ 
(Bertini-Dresner)  &    & \cellcolor{green} & \cellcolor{green}\textcolor{red}{distributions}  & \cellcolor{green} & \cellcolor{green} & \cellcolor{green} & \cellcolor{green}   \\ \hline
\end{tabular} 
\caption{\label{Tab_Part} Spallation models or model combinations involved in the IAEA {\it Benchmark of Spallation Models}. More details on the models (versions, ingredients, ...) at https://www-nds.iaea.org/spallations/2010ws/}
\end{minipage}
\end{table*}

Any person involved had to commit to do the calculations for all the reactions mentioned in the exercise, in order to get a complete and consistent computational results database. Regarding the excitation functions it was agreed that the participants gave the direct (or independent) production cross-sections, while the experimental data also included cumulative productions, which take into account the radioactive progenitors. Actually only one model for the calculation of the cumulative cross-sections was used by the organizers to avoid bias in the analysis of these observables. Details on this cumulative model can be found in \cite{SCH11}. Table \ref{Tab_Part} shows the models or model combinations involved in this benchmark and the calculation results provided (in green) or not (in red). The first result was twofold: the most widely used spallation models (except the one in the FLUKA transport code) were involved and a complete picture of the spallation reactions modeling could be drawn.

In order to analyze the calculation results a huge number of figures (several thousands) have been built. Most of the figures are the results of each model compared to the experimental data, but the results of all models were also plotted on a common figure, for a given experimental data set, to make a global inter-comparison. We will see hereafter that this type of figure allows the trends to be seen, if the experimental data are collectively well reproduced or not, if the models give more or less the same results, with the same quality or shortcomings, or if the results are very different. In addition to the {\it usual} figures, Figures of Merit (FoM), or deviation factors, were also drawn for each model and observable. Those tools deserve some attention, because they are symptomatic of the goal of the validation. Classical figures comparing the calculation results with experimental data of an observable can be seen as a global and qualitative approach. The FoM are quantitative, but focused on one characteristic. The choice of the tools is driven by the aim of the validation. The FoM that were used are the following factors:
\begin{eqnarray}\label{eq:forme-pos}
          &R =&   \frac{1}{N}\sum_{i=1}^{N}\frac{\sigma_{i}^{calc}}{\sigma_{i}^{exp}} \nonumber \\
          &H =&   \left(\frac{1}{N}\sum_{i=1}^{N}\left( \frac{\sigma_{i}^{exp}-\sigma_{i}^{calc}}{\Delta\sigma_{i}^{exp}}\right)^2\right)^{1/2}  \nonumber \\
          &F =&    10^{\left( \frac{1}{N}\sum_{i=1}^{N}\left[ log\left(\sigma_{i}^{exp}\right)-log\left(\sigma_{i}^{calc}\right) \right]^2 \right)^{1/2}}          \\
          &S =&    10^{\left\{   \frac{\sum_{i=1}^{N}\left[ \frac{log\left(\sigma_{i}^{calc}\right)-log\left(\sigma_{i}^{exp}\right)}{\left( \Delta\sigma_{i}^{exp}/\sigma_{i}^{exp} \right)} \right]^2}{\sum_{i=1}^{N}\left[ \frac{\sigma_{i}^{exp}}{\Delta\sigma_{i}^{exp}}\right)^2} \right\}^{1/2} }         \nonumber  \\
          &M =&  min\left\{ \sum_{i=1}^{N}\sigma_{exp} \hspace{0.2cm}log\left(\frac{\sigma_{exp}}{\sigma_{calc}}\right), \sum_{i=1}^{N}\sigma_{calc} \hspace{0.2cm}log\left(\frac{\sigma_{calc}}{\sigma_{exp}}\right)\right\}  \nonumber \\
          &P_{x} =&   \frac{N_{x}}{N}  \text{; ($N_{x}$: number of points $i$ such as $\frac{1}{x} < \frac{\sigma_i^{calc}}{\sigma_i^{exp}} < x$)} \nonumber
\end{eqnarray}
where $\sigma_{i}^{exp}$ ($\sigma_{i}^{calc}$) is the measured  (calculated) cross-section, $\Delta\sigma_{i}^{exp}$ the experimental uncertainty and $N$ the number of experimental data points. Each of these coefficients shows a particular characteristic of the calculation result. $R$ is the crude mean ratio, and is not symmetrical, since it is dominated by the overestimation, $H$ is the mean $\chi^2$, $F$ is a symmetric version of $R$, i.e. overestimation and underestimate treated on an equal footing, $S$ is similar to $F$, but taking into account the weight of the experimental errors, $M$, called intrinsic discrepancy \cite{MIC09}, can be seen as the $F$ factor, but assessing the shape of the distribution, and finally $P_x $, which allows to get a rough idea of the accuracy of a model for a given calculation.

Two types of analysis have been carried out and presented during the "Second Advanced Workshop on Model Codes for Spallation Reactions" at Saclay (France) in 2010. On one hand each participant was asked to explain the successes and deficiencies of their models and, on the other hand, a global analysis showed the main trends and a status of the spallation  modeling. This last task has been divided in three subtasks, neutron, light charged particles and residues,  taken care of respectively by J.-C. David (CEA-France), F. Gallmeier (ORNL-USA) and R. Michel (ZSR-Germany). A summary of this work can be read in \cite{LER11,DAV11b}, but the great deal of information built up in the website dedicated to this benchmark can be still used to study more details. Here is a picture we can paint of the spallation modeling based on this benchmark.

Regarding the global analysis, two methods were used to rate the models. The first one was suggested and used by F. Gallmeier for the light charged particles. It is based on a quantitative procedure. The more the calculation results deviate from the experimental data, the lower is its rating. The second one was suggested by R. Michel and based on a more qualitative method. The ratings for both methods are given in Table \ref{tab:Tab_iaea-rating}.
\begin{table}
\center
\begin{tabular}{cc} \hline\hline
\vspace{-0.25 cm}    
  &  \\ 
  Acceptance band [eval/x ; eval*x]  & Points \\  \hline\vspace{-0.25 cm}
  &  \\ 
  x=5     & 1 \\
  x=3     & 2 \\
  x=2     & 3 \\
  \hspace{0.3 cm}x=1.4  & \hspace{0.1 cm}4 \vspace{-0.25 cm}\\
  &  \\  \hline\hline
\end{tabular} 
\medskip

\begin{tabular}{cc} \hline\hline
\vspace{-0.25 cm}    
  &  \\ 
  Quality & Points \\  \hline\vspace{-0.25 cm}
  &  \\ 
  Good                                                          & 2 \\
  Moderately good, minor problems             & 1 \\
  Moderately bad, particular problems         & -1 \\
  Unacceptably bad, systematically wrong  & -2\\
  &  \\  \hline\hline
\end{tabular} 
\caption{\label{tab:Tab_iaea-rating} Ratings used to analyze the benchmark results: upper part for light charged particles and lower part for neutrons and residues.}
\end{table}
Obviously the plots, and sometimes the FoM, were studied to go into the details (ingredients, mechanisms, ...).
\medskip 

\begin{figure}[hbt]
\begin{center}
\resizebox{.5\textwidth}{!}{
\includegraphics{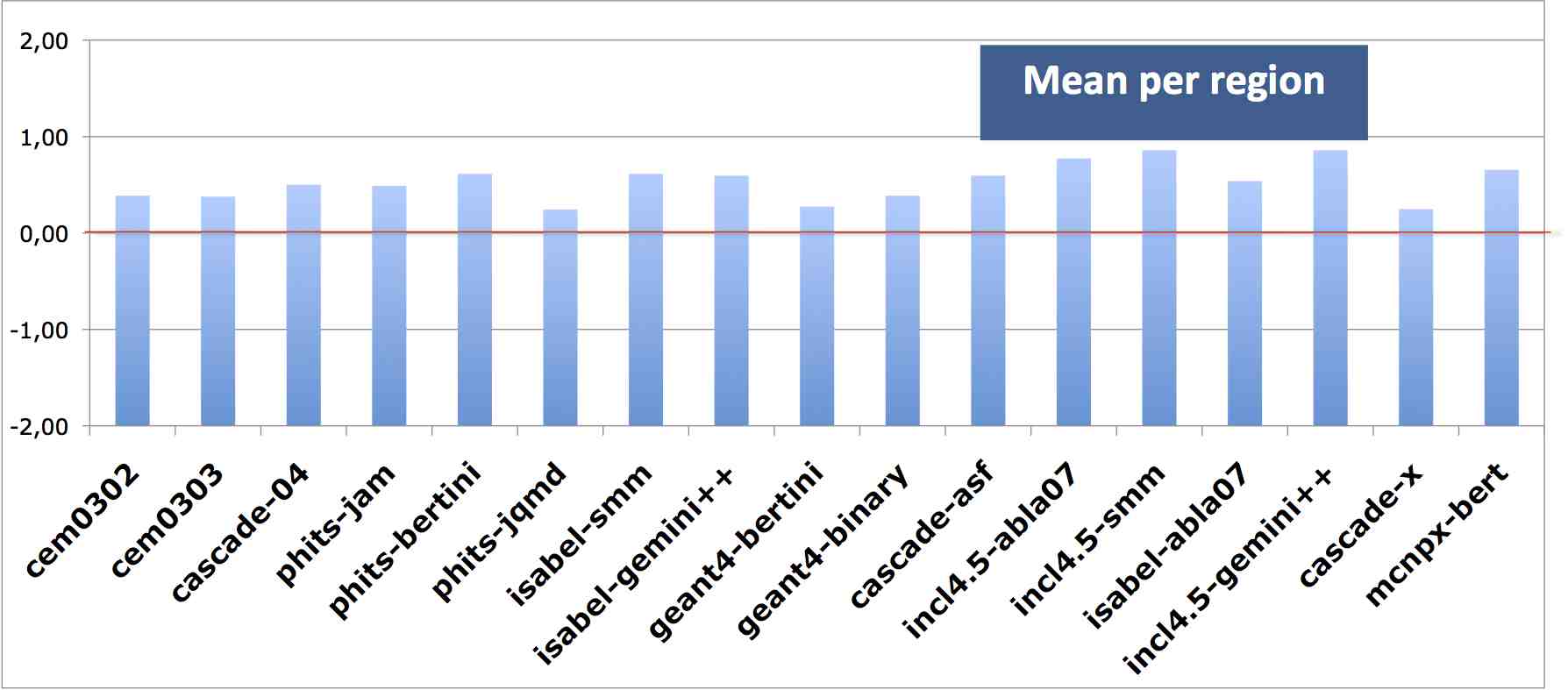}}
\end{center}
\caption{\label{Fig_neutron-rate}
Rating results obtained using the method given in the lower part of Table \ref{tab:Tab_iaea-rating} for neutron double differential cross-sections.}
\end{figure}

The data studied for the neutrons are mainly the double differential cross-sections, to which the rating procedure was applied, but also the multiplicities and multiplicity distributions. In order to separate the different mechanisms, each spectrum was divided into energy (0-20 MeV related to the de-excitation step; 20-150 MeV for the intermediate phase sometimes called pre-equilibrium; beyond 150 MeV for the intranuclear cascade) and into angle (forward angles ($\le$ 45\degree); intermediate angles (45\degree $\le \theta \le$ 135\degree), and backward angles ($\ge$ 135\degree)). Figure \ref{Fig_neutron-rate} shows the result of ratings averaged over these regions and over all the reactions studied. One can clearly see that the dispersion between the models is small and the overall results are correct. However, a more detailed study of the spectra can point to deficiencies and differences.

In the low-energy part the models give good results. This is because evaporation of neutrons is the preferred route of the de-excitation and the formalism of evaporation is well known. Nevertheless two features may be noted. On the one hand the border between the neutrons from the intranuclear cascade and the neutrons from the de-excitation is very difficult to locate precisely, thus towards 20 MeV the differences between the models are more pronounced, and on the other hand, the other side of the spectrum to very low energies, the models can differ a lot. This last point is possibly linked to the capture cross-section of neutrons used in evaporation models. The values used for these cross-sections rest most of the time on experimental measurements obviously made with cold nuclei, while here the nuclei are hot, therefore the energy levels are different and the probability of low-energy neutron capture modified.

\begin{figure}[hbt]
\begin{center}
\hspace*{-.6cm}
\resizebox{.45\textwidth}{!}{
\includegraphics[trim=0.3cm 0.7cm 1.05cm 0.5cm,clip]{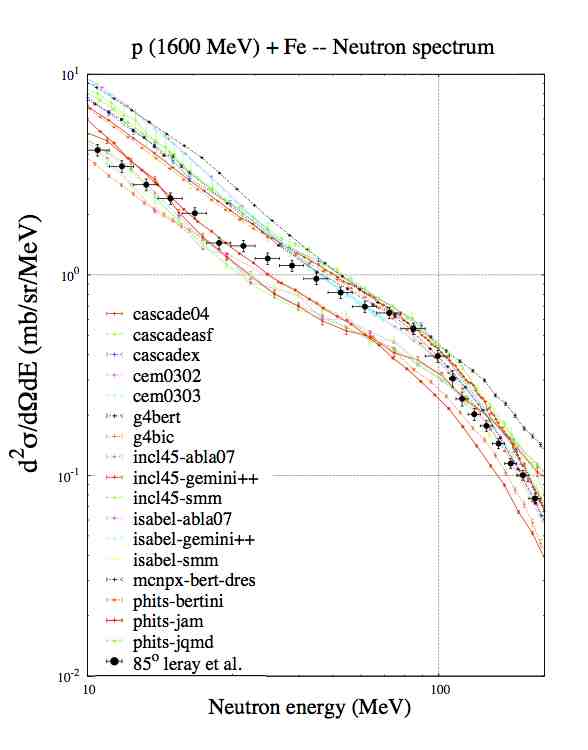}}
\end{center}
\caption{\label{Fig_neut-pre1}
Double differential cross-section production of neutrons at an angle of 85{\degree} from the reaction p (1600MeV) + Fe. This plot focuses on the intermediate energy part (20 MeV - 150 MeV). The seventeen models (listed in Tab. \ref{Tab_Part}) are plotted and the experimental data taken from \cite{LER02}.}
\end{figure}

\begin{figure}[!hbt]
\begin{center}
\hspace*{-.5cm}
\resizebox{.49\textwidth}{!}{
\includegraphics{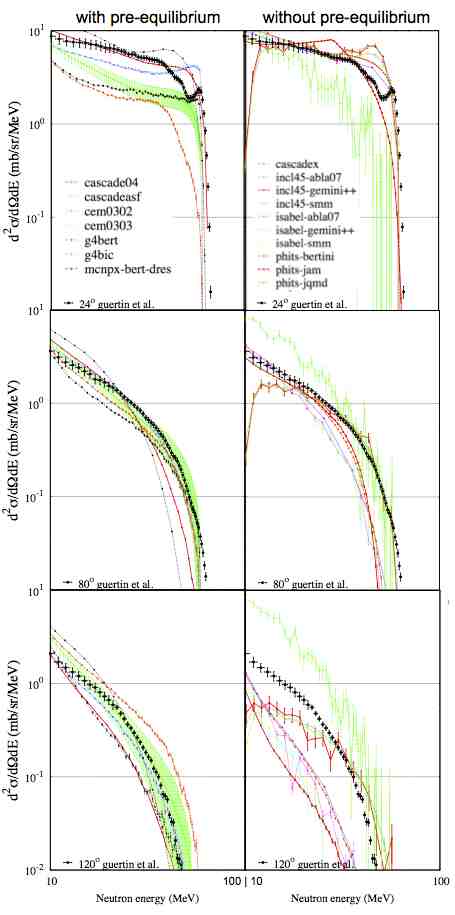}}
\end{center}
\caption{\label{Fig_neut-pre2}
Double differential cross-section production of neutrons at three angles (24{\degree} on the upper part, 80{\degree} in the middle and 120{\degree} on the lower part) from the reaction p(63MeV) + $^{208}$Pb. The left part deals with the models including a pre-equilibrium step and the right part the models without pre-equilibrium. Experimental data taken from \cite{GUE05}.}
\end{figure}
\begin{figure}[htb]
\begin{center}
\vspace*{-.3cm}
\resizebox{0.52\textwidth}{!}{
\includegraphics[trim=1cm 3.5cm 1cm 1.9cm,clip]{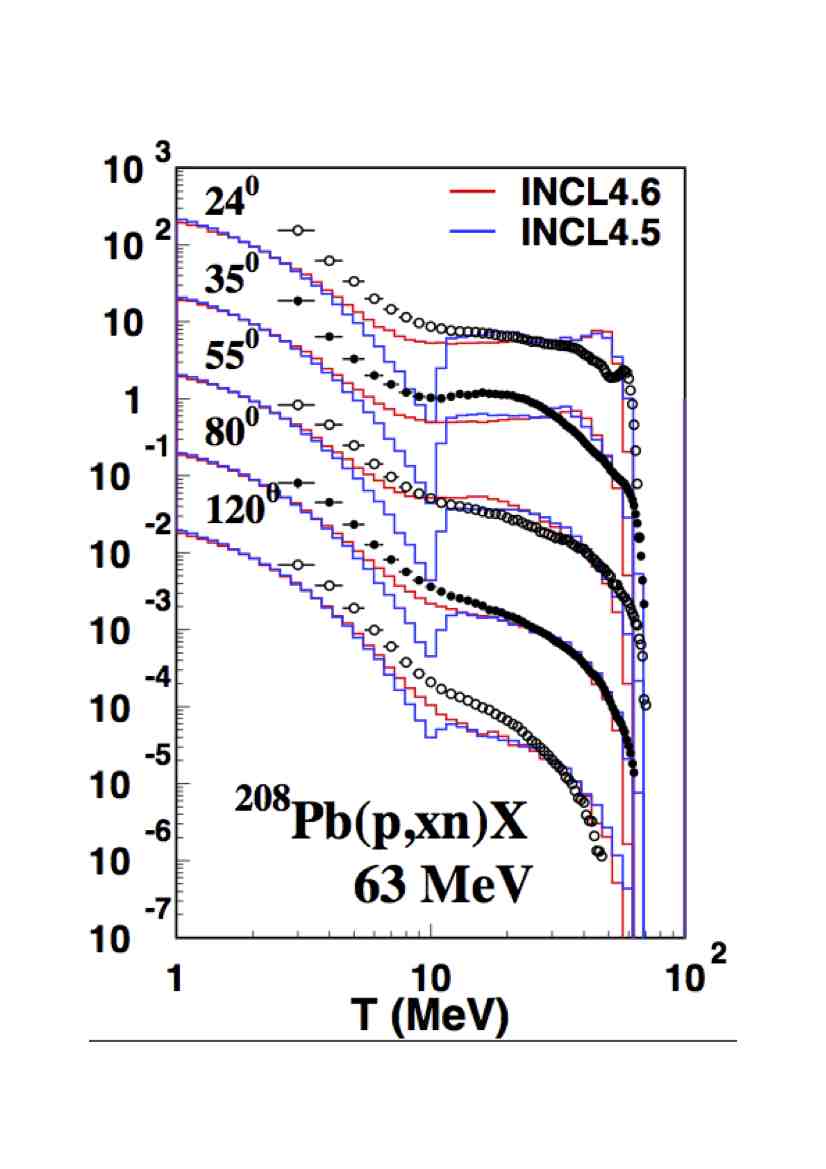}}
\end{center}
\caption{\label{Fig_neut-pre-ab}
Comparison of INCL4.5 (blue) and INCL4.6 (red) predictions for the neutron double differential cross-sections in p(63MeV) + Pb collisions. For the sake of clarity, the various spectra have been multiplied by $10^0$, $10^{−1}$, $10^{−2}$, etc., for increasing angles, starting at the lowest angle. Experimental data taken from \cite{GUE05}. Figure from \cite{BOU13}.\href{http://link.aps.org/abstract/PRC/v87/e014606} {Copyright (2014) by the American Physical Society.}}
\end{figure}

The portion of the spectrum between 20 MeV and 150 MeV is linked to the end of the intranuclear cascade. Some spallation models add an intermediate step between the INC and the de-excitation, the pre-equilibrium, believing that the remnant nucleus from the cascade does not have all the characteristics of a compound nucleus. Figure \ref{Fig_neut-pre1} shows clearly below 100 MeV two groups of model. If we exclude the two codes JQMD (from the PHITS transport code) and CASCADEX, these two groups are characterized by having or not a pre-equilibrium phase. One way to highlight the usefulness or not of a pre-equilibrium model is to look at, for example, the neutron spectrum in the reaction p(63MeV) + Pb, where the incident energy is such that models with pre-equilibrium should give better results. On the upper part of Figure \ref{Fig_neut-pre2} are shown the results of the models with a pre-equilibrium step and on the lower part the models without pre-equilibrium. While the results, for three angles, of the models with pre-equilibrium are less scattered and give best results for the backward angle, the models without pre-equilibrium are generally better at the forward angle. The situation is more complicated for the transverse direction. The diversity of the results for models without pre-equilibrium is due to the special treatment of this low incident energy, outside the field of the spallation reactions. Besides, a model without pre-equilibrium, INCL4.5, gives interesting results if we ignore the spurious fall at $\sim$10 MeV. This defect in INCL4.5 was corrected in the following version INCL4.6 leading to good results, as shown in Fig.\ref{Fig_neut-pre-ab} . From all this it appears that the usefulness of a pre-equilibrium model is not obvious.

\begin{figure*}[htb]
\centering
\begin{minipage}{.9\textwidth}
\centering
\vspace*{-.3cm}
\resizebox{.9\textwidth}{!}{
\includegraphics{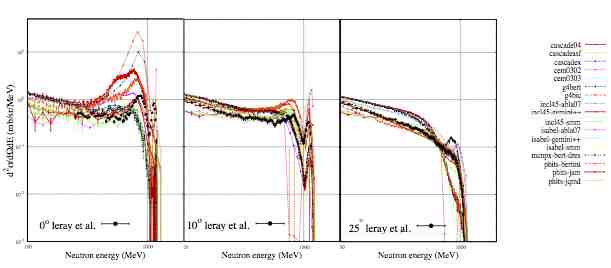}}
\caption{\label{Fig_neut-high-for}
Double differential cross-section production of neutrons at three forward angles (0{\degree} on the left, 10{\degree} in the middle and 25{\degree} on the right) from the reaction p (1200MeV) + Fe. This plot focuses on the high energy part (above 100 MeV). The seventeen models (listed in Tab. \ref{Tab_Part}) are plotted and the experimental data taken from \cite{LER02}.}
\end{minipage}
\begin{minipage}{.89\textwidth}
\centering
\resizebox{.48\textwidth}{!}{
\includegraphics{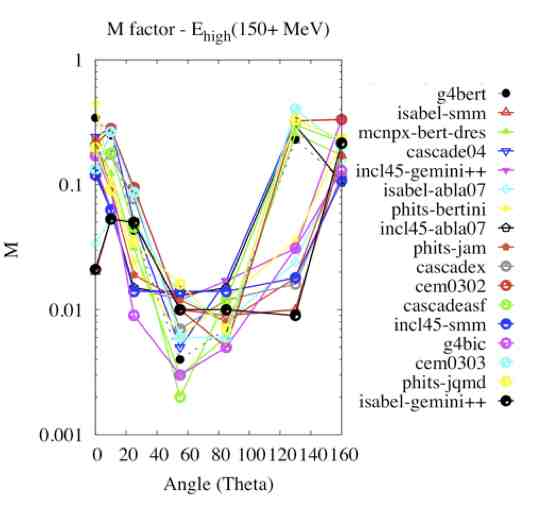}}
\caption{\label{Fig_neut-high-trans}
Angular distributions of the FoM $M$ related to the high-energy  ($\ge$ 150 MeV) neutrons produced in the reaction p~(800MeV) + Fe. The seventeen models (listed in Tab. \ref{Tab_Part}) are plotted and the experimental data taken into account come from \cite{LER02}.} 
\end{minipage}
\begin{minipage}{.9\textwidth}
\centering
\resizebox{.5\textwidth}{!}{
\includegraphics{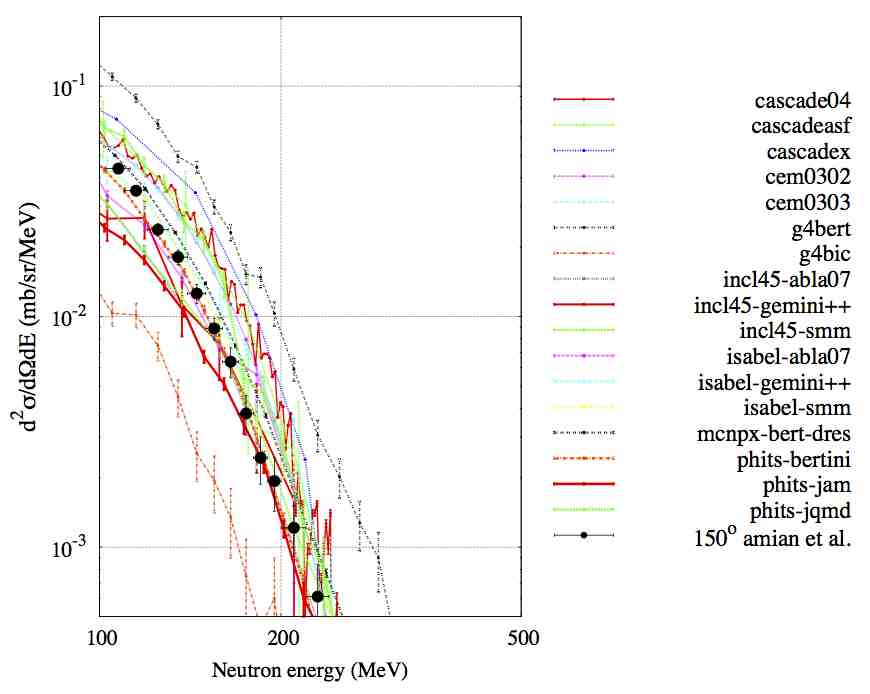}}
\caption{\label{Fig_neut-high-back}
Double differential cross-section production of high energy neutrons at 150{\degree} the reaction p(800MeV) + Fe. The seventeen models (listed in Tab. \ref{Tab_Part}) are plotted and the experimental data taken from \cite{AMI92}.}
\end{minipage}
\end{figure*}

The high-energy part of the neutron spectrum is very anisotropic, thus one must distinguish the forward angles ($\le$ 45\degree - Fig. \ref{Fig_neut-high-for}), the intermediate angles (45 $\le \theta \le$ 135\degree - Fig. \ref{Fig_neut-high-trans}) and the backward angles ($\ge$ 135\degree - Fig. \ref{Fig_neut-high-back}).

The forward angles are characterized by two peaks, which disappear gradually as the angle increases (Fig. \ref{Fig_neut-high-for}). The quasi-elastic peak is located at energies close to that of the incident projectile, it is the consequence of the direct interaction of the projectile with a neutron in the nucleus and thus is sensitive to the nucleon-nucleon cross-sections used, as well as to the spatial distribution of the nucleons in the nucleus. While the peak position is often correct, its height is a problem for many models. The quasi-inelastic peak is located slightly below the quasi-elastic peak and is significantly broader. This peak is due to the decay of the $\Delta$(1232) resonance and does not seem always easy to reproduce at all angles. At {25\degree} some models still exhibit a peak in contrary to the experimental data and at {0\degree} some differences of one order of magnitude can exist on its height between models and measurements. However it is worthwhile to note that at {0\degree} the simulations are difficult, because a compromise has to be found between the optimum aperture angle and the computational time.

The intermediate angles are the part of the high energy spectrum reproduced the best by all models. Angular distributions of the FoM $M$ (formula \ref{eq:forme-pos}, page \pageref{eq:forme-pos}) are shown in Fig. \ref{Fig_neut-high-trans} for the reaction p(800 MeV) + Fe and one clearly sees that the models reproduce much better the transverse angles ($M$ is close to 0) than the forward and backward angles. 

One example of spectra of neutrons emitted at high energies in the backward direction is given in Fig. \ref{Fig_neut-high-back}. If the shape of the distribution is fairly well reproduced, the models differ by the values of the cross-sections first, and, secondly, towards the "lower" energies some models deviate from the experimental data. This last point is probably due to the multiple scattering. Other data, not shown here, point out that these difficulties increase with the angle.

\begin{figure}[hbt]
\begin{center}
\resizebox{.43\textwidth}{!}{
\includegraphics{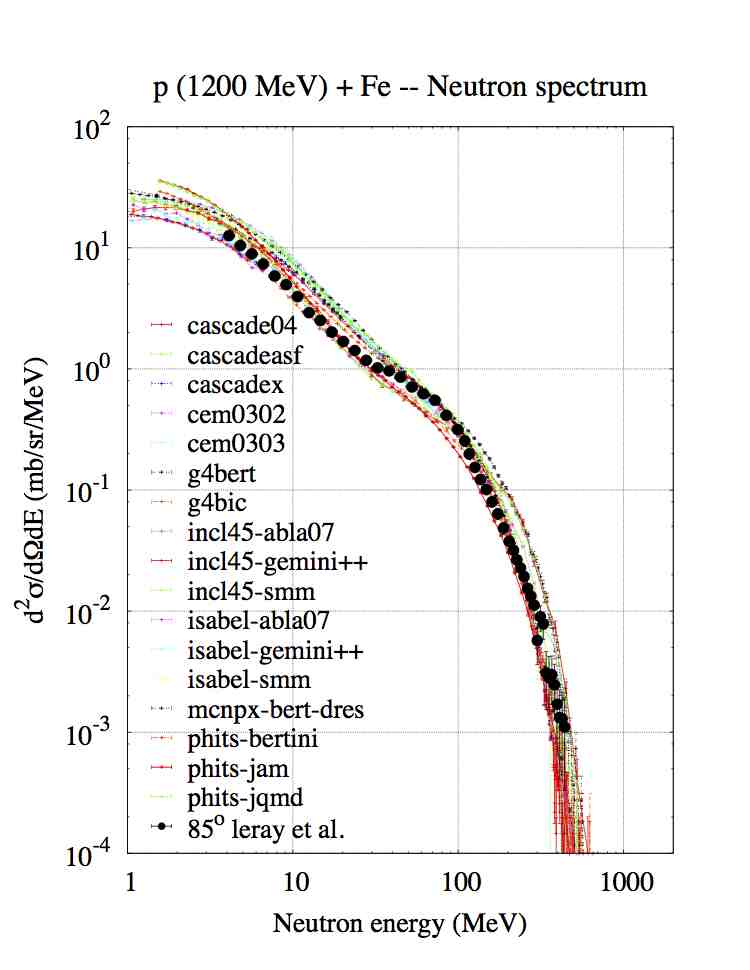}}
\resizebox{.43\textwidth}{!}{
\includegraphics{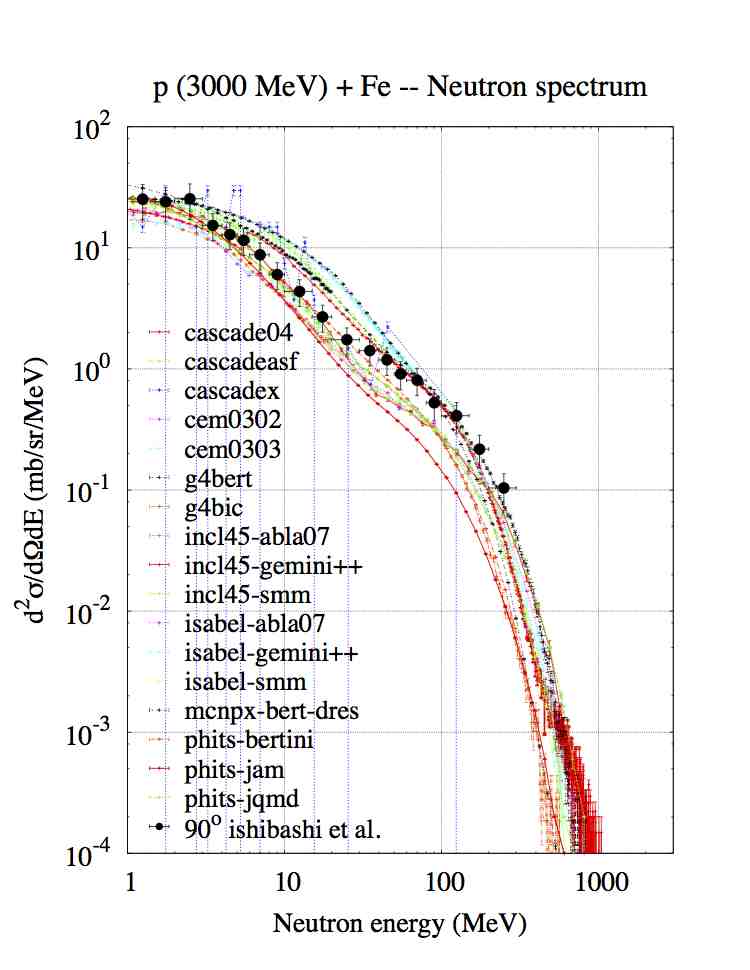}}
\end{center}
\caption{\label{Fig_neut-med-high}
Double differential cross-section production of neutrons at around 90{\degree} on an iron target. On the upper part with 1.2 GeV incident protons and on the lower part with 3.0 GeV incident protons. The seventeen models (listed in Tab. \ref{Tab_Part}) are plotted and the experimental data taken from \cite{LER02} (1.2 GeV) and \cite{ISH97} (3.0~GeV).}
\end{figure}

The neutron spectra can be used to address the high energy limit of the spallation models. This is illustrated by Fig. \ref{Fig_neut-med-high} where neutron spectra at the same angles (close to 90\degree) are plotted with the same target, but with two incident energies: 1.2 GeV and 3.0 GeV. This kinematics region of neutron production is, as previously stated, well reproduced by the models (see the left part), but when the incident energy increases and reaches the value of 3.0 GeV, the calculation results are scattered and less good. The reliability of the spallation models above 3.0 GeV depends strongly on the models. The point, to extend the models to higher energies, is to take carefully into account the needed channels in the intranuclear cascade (resonances?)  and in the de-excitation phase (multifragmentation?). This point is discussed in section \ref{spalhe}. 

\begin{figure}[hbt]
\begin{center}
\resizebox{.43\textwidth}{!}{
\includegraphics{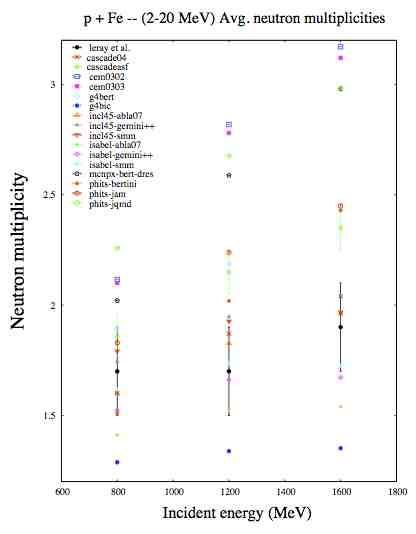}}
\resizebox{.43\textwidth}{!}{
\includegraphics{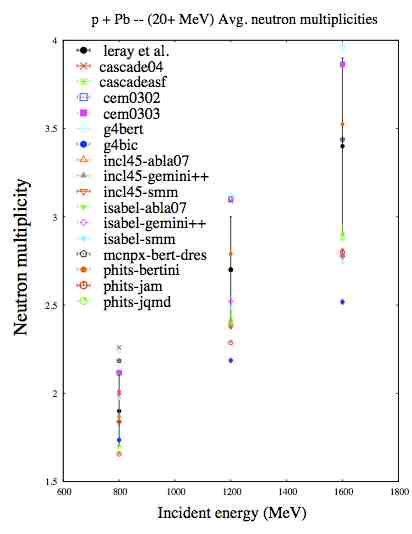}}
\end{center}
\caption{\label{Fig_neut-high-mult}
Average neutron multiplicities at three incident energies (800 MeV, 1.2 GeV and 1.6 GeV). On the upper part for low-energy neutrons (2 MeV $\le$ $\le$ 20 MeV) for the reaction p+Fe, and on the lower part for high-energy neutrons ($\ge$ 20 MeV) for the reaction p+Pb. Sixteen models (see Tab. \ref{Tab_Part}) are plotted and the experimental data taken from \cite{LER02}.}
\end{figure}

To conclude on the neutron production Figure \ref{Fig_neut-high-mult} shows six neutron multiplicities, three incident energies (800 MeV, 1.2 GeV and 1.6 GeV) for low-energy neutrons with an iron target and the same three incident energies for high-energy neutrons with a lead target. The same conclusions as those from the double differential spectra can be drawn. So, if in the case of iron some rare calculation results are within a factor of 2, whereas many models are within 30\% or better, in the case of lead nearly all points are in the error bars or very near. Actually the differences must probably come from the neutron energies and not from the target. The low-energy neutrons come from the de-excitation phase and are in competition with all other de-excitation channels. The multiplicities are most of the time overestimated, which could indicate the lack of other channels.

\begin{figure}[hbt]
\begin{center}
\resizebox{.45\textwidth}{!}{
\includegraphics{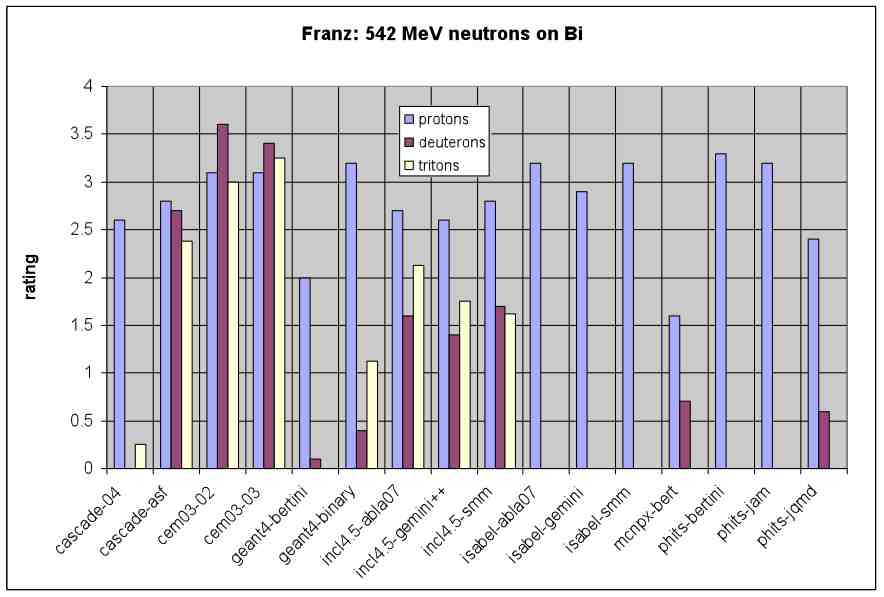}}
\resizebox{.45\textwidth}{!}{
\includegraphics{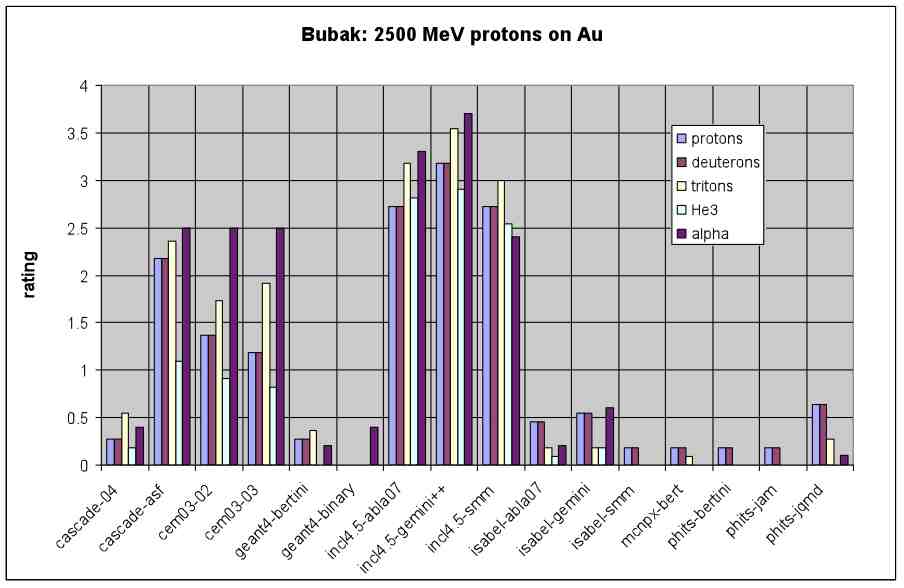}}
\end{center}
\caption{\label{Fig_lcp-note}
Rating results obtained using the method given in the upper part of Table \ref{tab:Tab_iaea-rating}, divided by the maximum number of points, for light charged particle double differential cross-sections. The upper part shows the reaction n(542 MeV) + Bi based on the experimental data from \cite{FRA90} and the lower part the reaction p(2500 MeV) + Au based on \cite{LET02,BUB07}.}
\end{figure}

The light charged particles considered during this benchmark were the proton, the deuteron, the triton, the $^3$He and the $\alpha$. The rating procedure, explained in the upper part of Tab. \ref{tab:Tab_iaea-rating}, was applied to each data set (no global rating). Fig. \ref{Fig_lcp-note} shows two examples, which allow the main conclusions to be drawn on the modeling of the light charged particle emission. While some models seem to be generally more efficient than others, the results depend very much on the reaction and on the given emitted particles. Thus, for the reaction n(542 MeV) + Bi  (upper part of Fig. \ref{Fig_lcp-note}) protons are reasonably reproduced by many models, which is not the case at higher energies, as seen with the reaction p(2500 MeV) + Au (lower part of Fig. \ref{Fig_lcp-note}). Although the emitted particle energy range is not exactly the same in both reactions, data at 542 MeV not covering the low-energy (evaporation) region, it is clear that the incident energy plays a role. The case of the composite particles (d, t, $^3$He and $\alpha$) clearly divides the models. There are those that emit the charged particles throughout the spallation reaction and those which do not have a suitable mechanism for high energy emission (intranuclear cascade). Every developer agreed on the need for a coalescence model in the INC to be able to emit cluster particles.

\clearpage

\medskip 
\clearpage

\begin{figure}[hbt]
\begin{center}
\resizebox{.43\textwidth}{!}{
\includegraphics{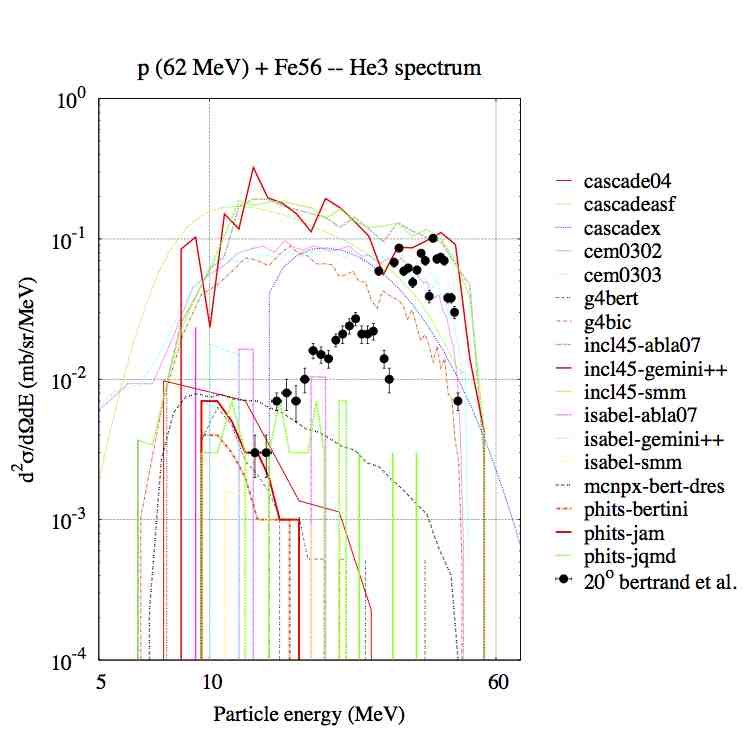}}
\resizebox{.43\textwidth}{!}{
\includegraphics{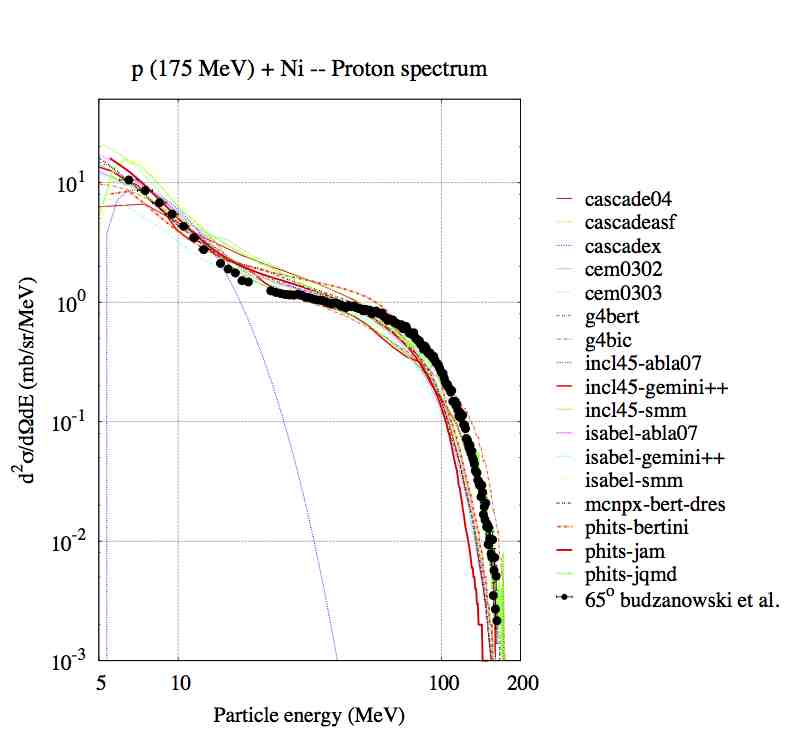}}
\end{center}
\caption{\label{Fig_lcp-casextrem}
Worst and best results on light charged particle emission obtained by the models within the Benchmark of Spallation Models. On the upper part the double differential cross-section production of $^3$He at 20{\degree} in the reaction p(62 MeV) + Fe (experimental data from \cite{BER73}) and on the lower part the double differential cross-section production of protons at 65{\degree} in the reaction p(175 MeV) + Ni (experimental data from \cite{BUD09}).}
\end{figure}
\begin{figure}[hbt]
\begin{center}
\resizebox{.38\textwidth}{!}{
\includegraphics[trim= 1.7cm 2.7cm 1cm 1cm, clip]{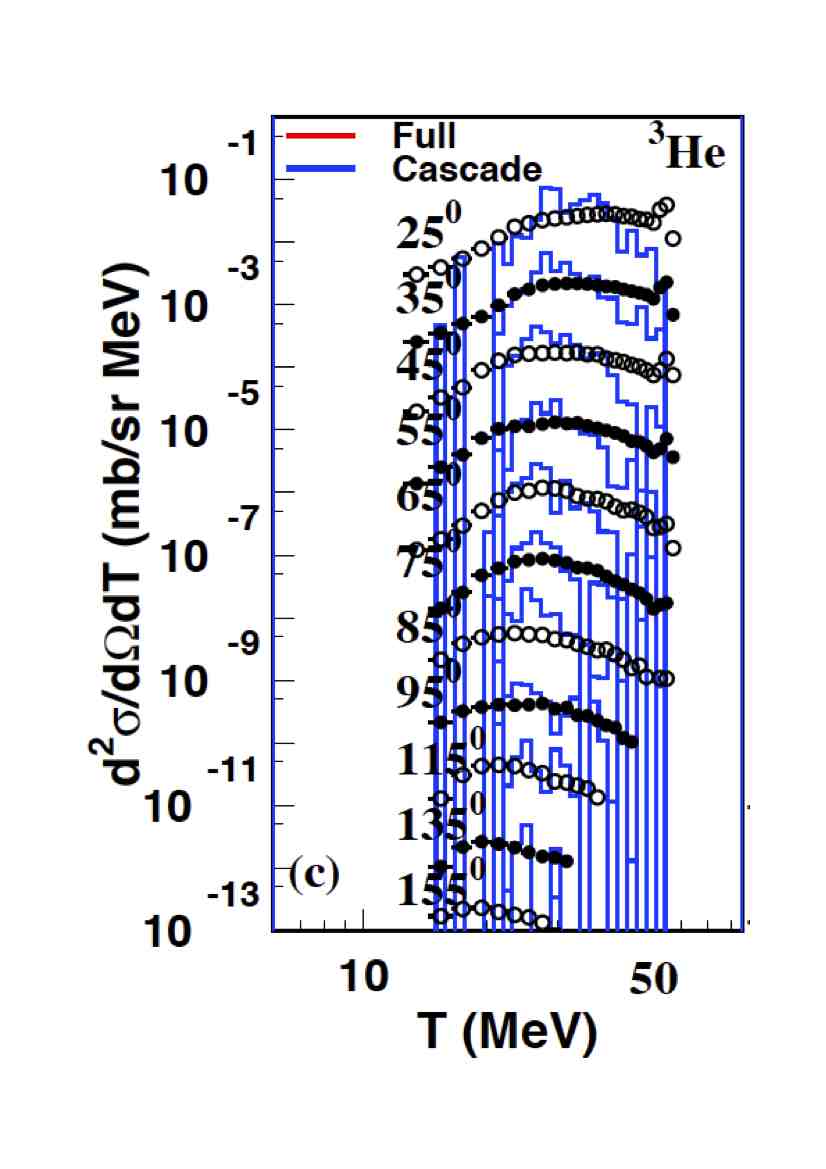}}
\end{center}
\caption{\label{Fig_ab-he3-fe62}
Double differential cross-sections of $^3$He production in reaction p + $^{208}$Pb colissions at 63 MeV (experimental data from \cite{GUE05}). Figure from \cite{BOU13}.\href{http://link.aps.org/abstract/PRC/v87/e014606} {Copyright (2014) by the American Physical Society.}}
\end{figure}

Figure~\ref{Fig_lcp-casextrem} shows two extreme cases encountered during this benchmark and illustrates the dispersion of the calculation results and the capability of the models to reproduce the experimental data. While, for the proton emission, all models are able to give good results in some cases as in the reaction p(175 MeV) + Ni, the situation is totally different for the $^3$He emission for low-energy reactions as the p(62 MeV) + Fe case, where no model could fit the experimental data. This latter case combines the difficulties of a composite particle and an incident energy below the usual assumptions for the spallation reactions. Nevertheless, after this benchmark some models tried to improve these kinds of results and some good results were obtained (see Fig. \ref{Fig_ab-he3-fe62}).

\begin{figure}[hbt]
\begin{center}
\resizebox{.47\textwidth}{!}{
\includegraphics{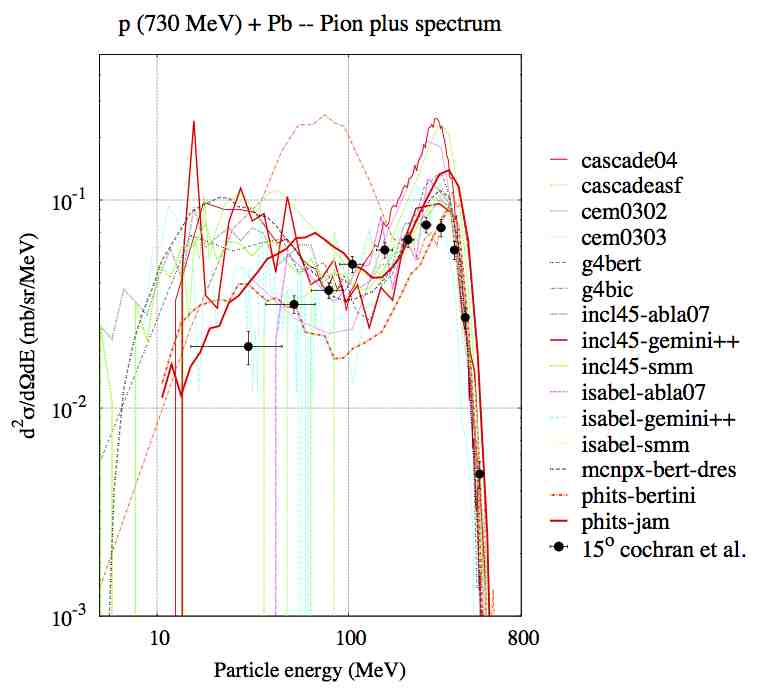}}
\resizebox{.47\textwidth}{!}{
\includegraphics[trim= 0cm 0cm 0cm -1.3cm, clip]{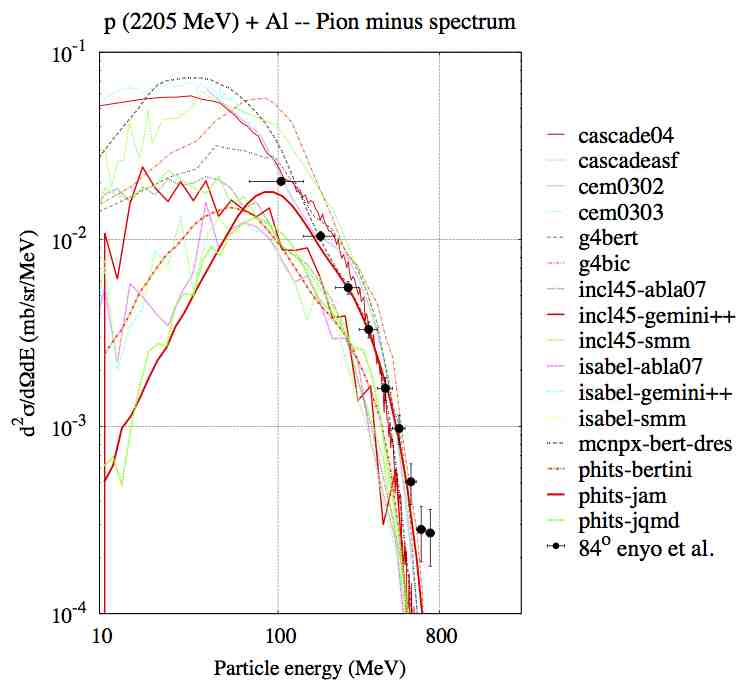}}
\end{center}
\caption{\label{Fig_pion}
Double differential cross-section production of pions. On the upper part $\pi^+$ at 15{\degree} from p(730 MeV) + Pb (exp. data taken from \cite{COC72}) and on the lower part $\pi^-$ at 84{\degree} from p(2205 MeV) + Al (exp. data taken from \cite{ENY85}).}
\end{figure}

Pions could be included in the light charged particles, but they proceed from very different mechanisms of production and their production cross-sections are lower. Their comprehensive analysis has unfortunately not been undertaken, but, taking Figure \ref{Fig_pion} for example, we can draw the following conclusions: while for high-energy pions models give fairly comparable results, more or less good, the structure at the forward angles is not always well reproduced, and furthermore, strong differences between the models appear when the pion energy decreases.
\medskip

Regarding residue production two types of observables are available: isotopic (mass and charge) distributions and excitation functions (isotope production according projectile energy).  R. Michel used the scoring system described in the lower part of Tab. \ref{tab:Tab_iaea-rating} only for the distributions, and lumped the residual nuclides into several mass domains, because the productions in each domain were possibly related to specific mechanisms:
\begin{list}{$\bullet$}{}
\item near-target products,
\item spallation products with masses exceeding half the target mass,
\item light products with masses (much) smaller than half the target mass,
\item fission products (for lead and uranium), and
\item light complex nuclei (A $\le$10).
\end{list}

\begin{figure}[hbt]
\begin{center}
\resizebox{.5\textwidth}{!}{
\includegraphics{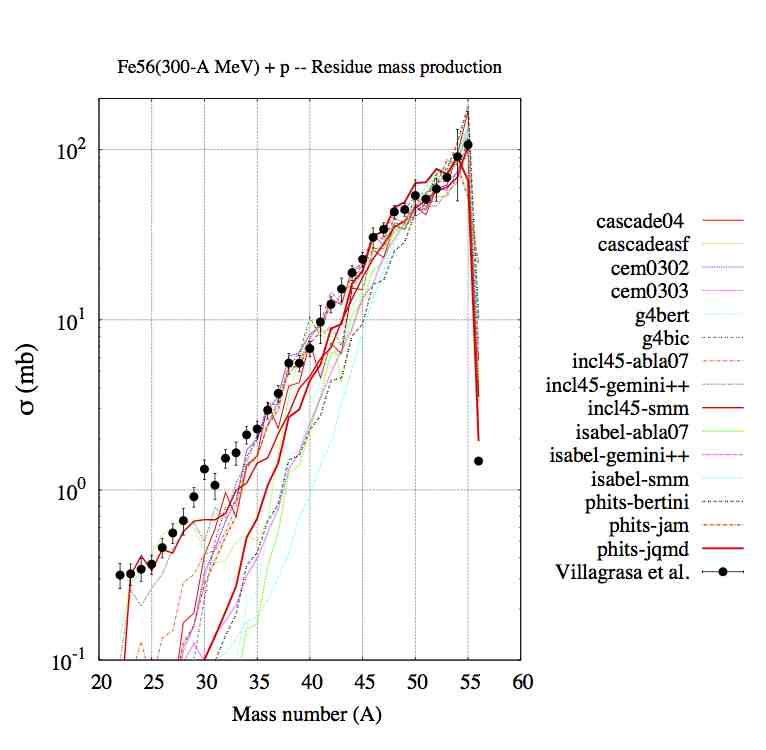}}
\end{center}
\caption{\label{Fig_res-mass-fe300}
Mass distributions of the nuclide production given by sixteen models for the reaction $^{56}$Fe(300-AMeV)+p (exp. data taken from \cite{VIL07}).}
\end{figure}

\begin{figure}[hbt]
\begin{center}
\resizebox{.5\textwidth}{!}{
\includegraphics{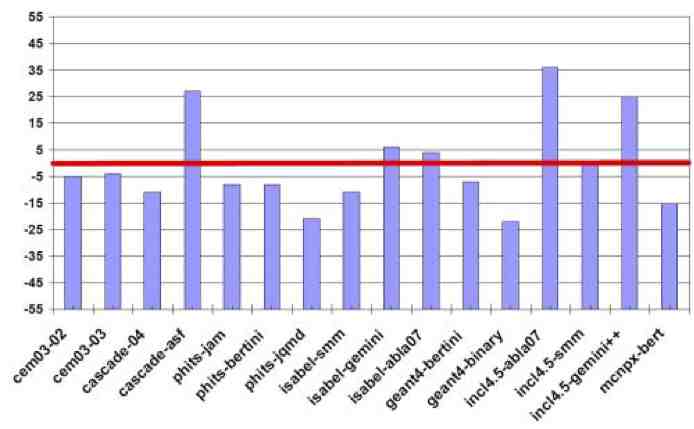}}
\end{center}
\caption{\label{Fig_res-note-AZ}
Rating results obtained using the method given in the lower part of Table \ref{tab:Tab_iaea-rating} for the mass and charge distributions and based on the experimental data taken from \cite{VIL07,NAP04,AUD06,ENQ01,TAI03,BER03,RIC06,BER06}).}
\end{figure}

Five reactions have been chosen concerning the distributions: $^{56}$Fe(300-A.MeV; 1000-A.MeV)+p, $^{208}$Pb(500-A.MeV; 1000-A.MeV)+p and $^{238}$U(1000-A.MeV)+p. The experiments were done at GSI in inverse kinematics. Almost all models give rather good results regarding the near-target products, while differences appear and become more and more significant when the residual nucleus become lighter. This is illustrated with Fig. \ref{Fig_res-mass-fe300}. A study of each model regarding the mass and charge distributions and using the different domains listed previously yielded the results shown in Fig. \ref{Fig_res-note-AZ}. An absolutely perfect model should get 50 points and the worst result should be scored -50.
A first conclusion on residue production is that no model reproduces the whole data. The main problems are:  
\begin{list}{$\bullet$}{}
\item describing the increase of cross-sections for light complex nuclides (A $\le$ 10) with decreasing product masses,
\item the fission channel, both for lead and uranium,
\item the competition between fission and spallation in case of uranium, and
\item spallation products for heavy target elements.
\end{list}
It should be noted, however, that three models provide much better results than the rest, even if three others are at the frontier between moderately good and moderately bad, and that some models have sizable problems. The role of the de-excitation modeling is very important in the production of residues, as shown by different combinations of intranuclear cascade models with de-excitation models.

\begin{figure}[hbt]
\begin{center}
\resizebox{.5\textwidth}{!}{
\includegraphics{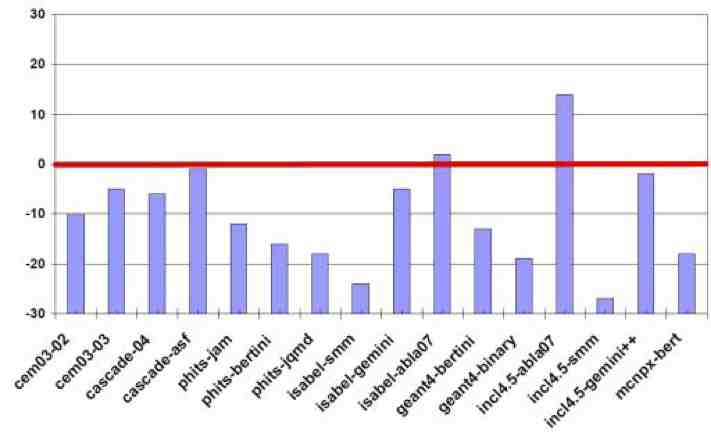}}
\end{center}
\caption{\label{Fig_res-note-isot}
Rating results obtained using the method given in the lower part of Table \ref{tab:Tab_iaea-rating} for the isotopic distributions and based on the experimental data taken from \cite{VIL07,NAP04,AUD06,ENQ01,TAI03,BER03,RIC06,BER06}).}
\end{figure}

The isotopic distributions are more difficult to reproduce than the distributions in mass and in charge, since these latter are the result of the sum on the charges or masses respectively, which may lead to some compensations between overestimations and underestimates. Fig.~\ref{Fig_res-note-isot} shows the scores obtained by the models. An absolutely perfect model should get 28 points and the worst result should be scored -28. This figure clearly shows the difficulty of all models to simulate the isotopic distributions. Three models are close to the red line (between moderately good and moderately bad) and only one is much better than the others. Moreover, some codes are much worse than in the case of the mass and charge distributions, which proves the effect of compensation and the limits of the integrated observables.

The main problems encountered on isotopic distributions simulation include:
\begin{list}{$\bullet$}{}
\item the heavy target elements regarding the fission channel in general,
\item the spallation products for the heavy target elements, and, in particular,
\item the competition between spallation and fission in the case of the target element uranium.
\end{list}

\begin{figure}[hbt]
\begin{center}
\resizebox{.35\textwidth}{!}{
\includegraphics{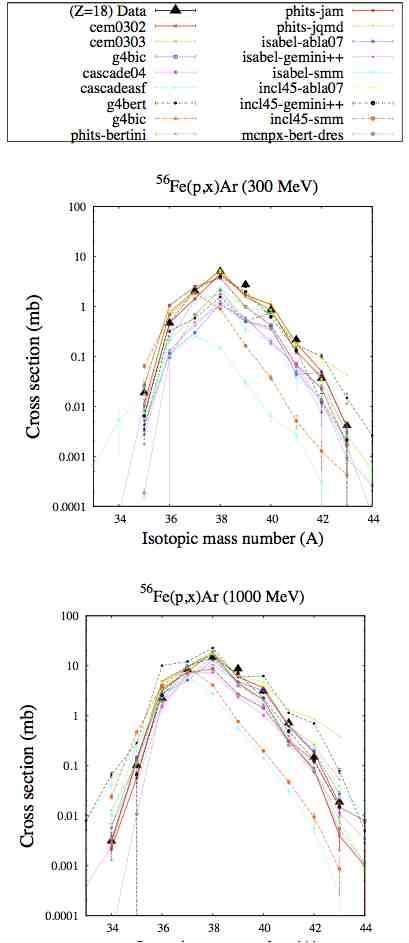}}
\end{center}
\caption{\label{Fig_res-isot}
Mass distributions of the argon production given by sixteen models for the reactions $^{56}$Fe(300-AMeV)+p (on the upper part) and  $^{56}$Fe(1000-AMeV)+p (on the lower part). The experimental data respectively taken from \cite{VIL07} and \cite{VIL07,NAP04}.}
\end{figure}

A last point can be mentioned, which is the importance of the incident energy in the capability of models to reproduce isotope productions. Fig. \ref{Fig_res-isot} gives an example with the production of argon isotopes in the reaction p+Fe at two incident energies: 300 MeV and 1000 MeV. 

The excitation functions are the right observable for judging the quality of models depending on the incident energy. They are the production cross-sections of isotopes versus the incident energy, thus the price to pay is they are still more difficult to modeled than the isotopic distributions.

For the analysis of excitation functions the method had to take into account the large number of figures and very different behaviors of the models. Thus R. Michel focused on specific isotopes again to characterize some particular mechanisms. Some isotopes are produced by both targets, i.e. Fe and Pb:
\begin{list}{$\bullet$}{}
\item $^4$He , $^7$Be, $^{10}$Be as light complex product nuclides (A$\le$10)
\item $^{22}$Na, $^{24}$Na, $^{46}$Sc, $^{48}$V as light product nuclides,
\item $^{54}$Mn and $^{56}$Co as near-target products in case of the iron target and as light products in case of the lead target.
\end{list}
Some others are related only  to the Pb target:
\begin{list}{$\bullet$}{}
\item $^{75}$Se as a light product nuclide,
\item $^{88}$Zr and $^{95}$Zr as fission products to demonstrate the differences between neutron-poor and neutron-rich fission products,
\item $^{127}$Xe and $^{128}$Xe to demonstrate the problem occurring in the transition from fission to spallation,
\item $^{149}$Gd, $^{175}$Hf, and $^{188}$Pt as spallation products, and 
\item $^{200}$Pb and $^{204}$Bi as near-target products.
\end{list}

\begin{figure}[hbt]
\begin{center}
\resizebox{.42\textwidth}{!}{
\includegraphics{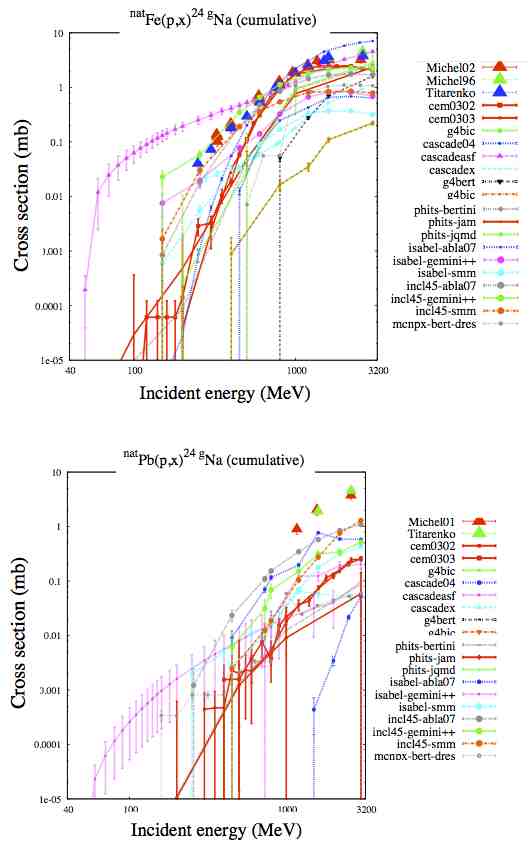}}
\end{center}
\caption{\label{Fig_res-na24}
 $^{24}$Na cumulative production versus the incident proton energy with an iron target (upper part) and a lead target (lower part). The experimental data respectively taken from \cite{MIC95,MIC02,TIT08} and \cite{GLO01,TIT06}.}
\end{figure}

The results, concerning very light nuclei (A$\le$10), have been strongly improved since the previous NEA benchmark \cite{MIC97}, even if all the models can do even better and if some have still serious deficiencies. Fig.~\ref{Fig_res-na24} displays an example of a light nucleus ($^{24}$Na). The two main features are the large dispersion of results, pointing out basic differences between the models in this domain, and the dependence with the target, signing different production mechanisms. Regarding the fission products, as previously mentioned, there is the group of models reproducing reasonably well the experimental data and the others. It may be added that the neutron-rich isotopes are less well reproduced than the neutron-poor, probably because the neutron-rich isotopes start to be produced at a low incident energy (below 100 MeV), not well handled by most of the models. Moreover, some nuclei can be produced by different mechanisms according to the incident energy (e.g. the Xe nuclei can be obtained after a long evaporation process or as fission products); some models include those right mechanisms of production but this is a domain where improvements are to be made by many models. The nuclei originating from the evaporation process divide the models into two categories: those that give very accurate results, and the others that have sometimes almost one order of magnitude across the range, especially when the nucleus is light. Finally, the nuclei near the mass of the target are generally well reproduced unless one considers nuclei with a higher charge than the target where there is still progress to do.
\medskip

From this inter-comparison some models or combinations of models seem on average better than the others, but none is yet capable of being better than everyone else on the whole set of observables. This largely explains why so many {\it different} models still exist. It is these differences that may allow us to better identify, better understand the mechanisms, formalisms, ingredients that must be taken into account for the modeling of spallation reactions. The pre-equilibrium phase and the high-energy composite light charged particles emission are two examples. It was found that the utility of an intermediate phase between the cascade and the de-excitation does not seem absolutely necessary, in the first case, but there was a consensus on the use of a coalescence model within the cascade to reproduce correctly the spectra over the entire energy range in the second case. Other topics were also studied during this benchmark, as the advantage of the Hauser-Feshbach formalism compared to the Wei\ss kopf formalism to describe the evaporation channel and the promise of QMD models as more sophisticated models for the fast phase. No clear difference appeared when using one formalism or another for the evaporation and unfortunately the benchmarked JQMD model suffers from too long a computation time to be competitive in the case of incident nucleons and the emission of composite particles was not as good as expected. Besides, regarding the emission of composite particles, developers of PHITS have built recently a new INC code including a coalescence model (INC-ELF (Emission of Light Fragments)) \cite{SAW12}.

The choice of the experimental data set is important, because it sets the space where the models are surveyed and some mechanisms require specific observables to be carefully validated. The multifragmentation process does not seem necessary in light of experimental data used here, since the production of intermediate mass nuclei apparently can be reproduced more or less correctly with a mechanism of evaporation or binary breakup. However, its usefulness may be more evident at higher incident energies. Some other examples of useful data can be mentioned. The excitation functions obtained with incident neutrons would be very interesting to compare carefully against those obtained with protons, because most of the time people assume the same result for all produced isotopes. Such data are very scarce, but some efforts have been done recently in this sense \cite{YAS13,LEY11}. The study of double differential production of nuclei heavier than $\alpha$ (Lithium, Beryllium, etc.) would help to go deeper into the study of the production of light nuclei (A$\le$10). Some data were already available during this benchmark \cite{BUB07}, but as additional data, therefore no calculation results were sent. Finally the study of spallation reactions with beams made of composite particles would not only be interesting but also necessary for the models implemented in transport codes, as proved by a study \cite{DAV13} on the production of nuclei with a charge two units higher than the target nucleus.

\subsection{Thick targets}
\label{thickbench}

The spallation reactions take place most of the time in thick targets where secondary reactions must be accounted for. Therefore, the users have to validate the spallation models within a transport code to get an integrated result. Three types of benchmark related to thick target can be mentioned: the international exercises opened to everyone like the NEA benchmark in the nineties \cite{BLA93}, the validations done by the users themselves on their own projects, and the benchmarks performed by the developers of transport codes.

Implementation of the spallation models in a transport code and long computation times are two constraints that have to be added to the ones mentioned previously about the set up of a benchmark dealing with only the elementary process. This probably explains the rare attempts of international benchmarks. Nevertheless we can cite two exercises.

The first one already mentioned, {\it Thick target benchmark for Lead and Tungsten} \cite{FIL95}, was conducted in the 1990s and supervised by the NEA. The primary goal was to start with a non-fissionable target, then to continue with a fissionable target later. Two cylindrical targets were chosen, one made of tungsten and another made of lead. Experimental data from LANSCE (LANL) \cite{ULM95} were foreseen to be used to validate the models, but they were published in 1995 just after the end of this benchmark, then only an inter-comparison of the models was done. Twelve laboratories were involved and were supposed to provide calculation results for neutron spectra, neutron leakages and distributions in mass and charge of the spallation products in several places in the target. On average the laboratories provided more or less three quarters of the information requested. Most of the time the codes used by the participants were based on the HETC code for the domain of the spallation reactions, but other codes were used as GEANT, FLUKA and NMTC/JAERI, which will become PHITS. References on these versions can be found in \cite{FIL95}. The results were compiled as tables of numerical values and graphs to compare the models, but unfortunately no comments were made. However it was clear that, even for neutron productions, significant differences existed at that time between the codes. Recommendations and physics analysis of the results are difficult with thick targets, because many effects occur: in addition to the projectile interaction with the target nucleus, the production and the interaction of the secondary particles play a role and the slow down of all these particles must be taken into account. Nevertheless these benchmarks help the users to choose the right transport codes, or at least to know the reliability of their codes in a given domain, and can also point out the difficulty for the spallation models to simulate particular particle emission or residue production governed by unusual reactions in term of projectile and/or energy.

Since the accelerator-driven systems are an option to reduce the amount of nuclear waste through transmutation, the need arises to have good computational tools to design these facilities using a spallation target. It is in this context that a study of spallation models, associated with the transport code in which they are implemented, was undertaken both as a task of the European project EUROTRANS/NUDATRA and as an IAEA project (Coordinated Research Project on Analytical and Experimental Benchmark Analyses of Accelerator Driven Systems).

\begin{figure}[hbt]
\begin{center}
\resizebox{.5\textwidth}{!}{
\includegraphics{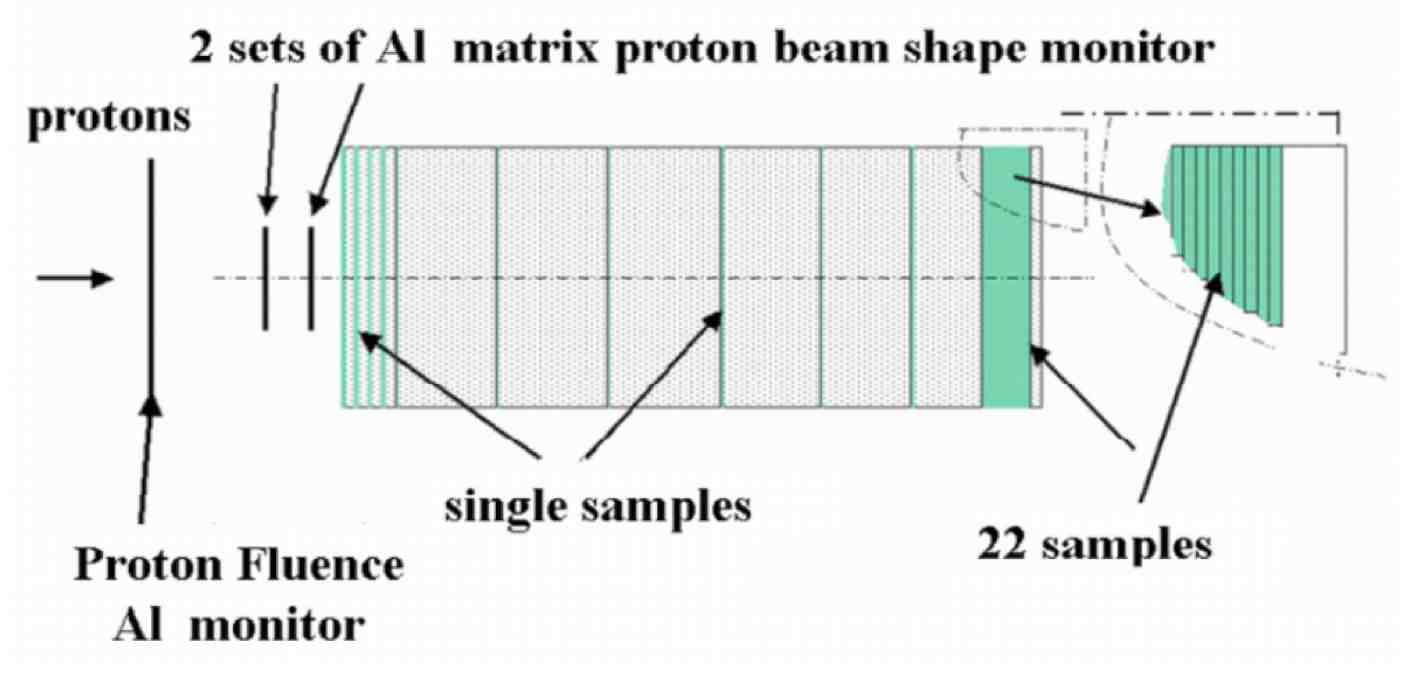}}
\end{center}
\caption{\label{Fig_manip-jerzy}
Schematic view of the lead target used in the experiment at JINR. Figure is from \cite{POH07} (Courtesy of W. Pohorecki).}
\end{figure}

One of the materials most often mentioned as a good candidate for a spallation target is lead. Experimental measurements with which to compare the results of calculation being scarce (for particles, but even more true for residual nuclei), an experiment was developed for the axial distributions of radionuclide activity in a lead cylinder bombarded by a proton beam of 660 MeV at JINR (Dubna - Russia) \cite{JAN11}. Fig. \ref{Fig_manip-jerzy} shows the thirty-two locations where measurements of activities were made. The lead cylinder had a length of 30.8 cm and a diameter of 8 cm. The isotopes whose activities were measured are: $^{46}$Sc, $^{59}$Fe, $^{60}$Co, $^{65}$Zn, $^{75}$Se, $^{83}$Rb, $^{85}$Sr, $^{88}$Y, $^{88}$Zr, $^{95}$Nb, $^{95}$Zr, $^{102m}$Rh, $^{102}$Rh, $^{110m}$Ag, $^{121m}$Te, $^{121}$Te, $^{139}$Ce, $^{172}$Hf, $^{172}$Lu, $^{173}$Lu, $^{175}$Hf, $^{183}$Re, $^{185}$Os, $^{194}$Au, $^{194m2}$Ir, $^{195}$Au, $^{203}$Hg et $^{207}$Bi. Calculations of heat released were also requested, but no measurement has been made, so no comparison calculation/experiment was performed.

The benchmark was first introduced to the ND2007 conference \cite{POH07}. Sixteen laboratories had indicated their interest in this exercise, but unfortunately only four have actually participated (organizers included, namely the University of Krakow). The other three laboratories were: IPEN (Brazil), CIEMAT (Spain) and CEA (France). The computation time can be one possible reason for the low number of participants, because, for example, the time required by INCL4.2-Abla was forty days on one processor to get satisfactory statistical uncertainties. In addition, in this case, another task had to be added, that is calculation of activities taking into account the time of irradiation and cooling before measurement. Publication of the results was done in 2011 in a report \cite{JAN11}.
Four models or model combinations available in MCNPX2.5.0 \cite{MCN05}, the default option (Bertini-Dresner) and those known to give the best results (INCL4.2-Abla, Isabel-Abla and CEM2k) were tested, as well as CEM03 in MCNPX2.6 \cite{MCN08} and the FLUKA code \cite{FAS03,FER05}. Actually, different institutes have sometimes used the same models, but the use of different code handling the decay of  the radionuclides during the times of irradiation and cooling and the possible use of different options in the transport code, probably explains the differences in the results. Nevertheless a careful study of the consistency of these results was performed and led to exclude one set of calculations.

Regarding the production of radionuclides in the whole target, all models were most of the time within a factor of two compared to the experimental data and in some cases even better. The best models were Isabel-Abla and FLUKA and the worst Bertini-Dresner. However, an interesting point of the experiment was the axial distributions of those activities. Only four models were used to compute the activities all along the target and could be compared to the experimental data: Bertini-Dresner, CEM2k, INCL4.2-Abla and Isabel-Abla. While the figures showing distributions were studied for the analysis, two deviation factors were also used to help the analysis,
\begin{eqnarray}\label{eq:forme-pos2}
          &D =&   \frac{1}{N}\sum_{i=1}^{N}\left|1-\frac{C_i}{E_i} \right|, \nonumber \\
          &H =&   \sqrt{ \frac{1}{N}\sum_{i=1}^{N}\left( \frac{E_i-C_i}{\sigma_{E,i}+\sigma_{C,i}} \right)^2 }, \nonumber
 \end{eqnarray}
with $C_i$ ($E_i$)  the calculated (measured) activity and $\sigma_{C,i}$ ($\sigma_{E,i}$) the associated uncertainty. These factors for different isotopes and models (see e.g. Tab.~\ref{Tab_benchjerzy-H} for the $H$ factor) helped achieve the same overall conclusions as for the whole target, i.e. that Isabel-Abla was the best model followed by INCL4.2-Abla, CEM2k and Bertini-Dresner. 
\begin{table}
\center
\begin{tabular}{c}
\resizebox{.4\textwidth}{!}{\includegraphics{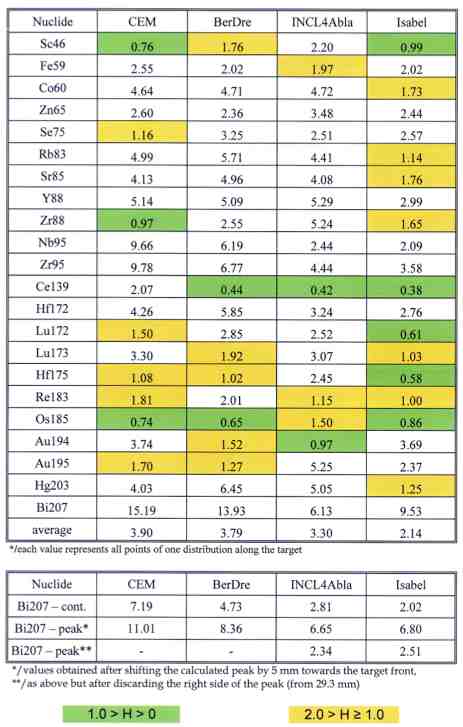}} \\
\end{tabular} 
\caption{\label{Tab_benchjerzy-H}
H-factor values for the models Bertini-Dresner, INCL4.2-Abla, Isabel-Abla and CEM2k available in MCNPX2.5.0 for several isotopes. Figure taken from \cite{JAN11} (Courtesy of J. Janczyszyn).}
\end{table}
However, the same figure shows more details. For example, CEM2k is the best for $^{88}$Zr, while it is Bertini-Dresner for $^{185}$Os. Isabel-Abla in rare cases can give the worst results (e.g. $^{194}$Au) and INCL4-Abla is the model with the lowest fluctuations in the prediction of isotope production. All models could still be improved and some of them will be (INCL4.2-Abla and CEM2k).

This benchmark on a thick lead target bombarded by a 660 MeV proton beam was the opportunity, first, to know the reliability of the models on the isotope production, thus on the radioactivity prediction, and, second, to point out the different features of the models (overall quality, variation according to the isotopes) and so to stress that the reliability makes sense only in given domains. Finally this kind of exercise, as with any benchmark, should of course be redone again whenever significant improvements are made in the models. If one can regret that not so many transport codes were involved in this benchmark, one must mention that some works are or have been performed elsewhere. 

These works concern mainly neutron emission. One can cite the works \cite{ADA05,MAJ06,MAJ07,KRA10} performed partly by the Nuclear Physics Institute of the ASCR, (Czech Republic) and the JINR (Russia) to characterize the neutron field around a spallation target in an ADS. They irradiated several thick targets and doing so they obtained experimental data also needed for benchmarking the transport codes used in designing such facilities. For example the PhD thesis of M. Majerle \cite{MAJ09} described two experiments, one made of a lead cylinder irradiated with 660 MeV protons and another one more complex where a uranium blanket surrounds a spallation lead target. The experimental data obtained with the first set up provides a good base to validate the transport codes, but the second one was also used. This was done by comparing several spallation models in MCNPX.2.5.0 and FLUKA to those data. All models gave predictions within the systematical uncertainties, and the best models seemed to be FLUKA and INCL4.2-Abla. Another, much recent, benchmark \cite{OH11} on neutron production has been performed using older data with proton beams of 113 MeV \cite{MEI89} and 256 MeV \cite{MEI90}  and several targets. Here again MCNPX (here MCNPX2.6.0 \cite{MCN08}) and FLUKA (version 2008.3b) were tested, as well as PHITS. The default options of the models were used in MCNPX and, for PHITS, the Bertini and the JQMD models were benchmarked. The main results were that FLUKA and PHITS give better results than MCNPX if a model is used instead of the LA150 library, and at backward angles (150\degree) the JQMD models give better results than Bertini. This latter conclusion is confirmed by a study \cite{IWA10} published just before, dealing with a benchmark on thin target of models in PHITS on proton-induced (140 MeV) neutron production at backward angles. The data from \cite{MEI90}, in  particular, are also used by the GEANT4 collaboration to benchmark their codes on thick targets with proton or deuteron projectiles, and always below 256 MeV. As said in section \ref{benchtrans}, the results are put on their website, but with no analysis. Finally other experimental data exist on angular distributions of neutrons from thick targets like those obtained at Saturne \cite{MEN98,VAR99}  and KEK \cite{MEI99}. A study using both \cite{DAV03} showed once again that neutron emission was much easier to simulate than isotope production, if we compare with the results of the NUDATRA/IAEA benchmark previously discussed.

\subsection{Conclusions on validations}
\label{concbench}

For the last twenty years numerous benchmarks have been done to validate the spallation models either to assess their reliability on a specific project or to know if they include in the right way the needed physics mechanisms. One outcome is the great improvements of some models and now the best ones are probably all implemented in one or several transport codes. However, the estimates of the uncertainties of the calculation results obtained for a complex target are still a difficulty to solve. While the statistical uncertainties are quite easy to obtain, the main problem lies in the intrinsic reliability of the models. This point is addressed in section \ref{tooless} by means of an estimate of the radionuclide production in the spallation target of the future ESS facility. Finally, if some details can still be studied as for example the light charged particle spectra in the all energy range or the intermediate mass residue production, the capabilities of the models can also be extended to lower and/or higher energies with the need to treat correctly the new open channels. These extensions are discussed in section \ref{front}. 


\section{Spallation as a tool}
\label{tool}


Modeling and validation are the two ways to understand the reaction mechanisms; they deal with the knowledge of spallation reactions. However these reactions can also be considered from another angle: a necessary tool in a particular study. In this latter case the outcomes of the reaction are used and two possibilities exist to get them: experimental data or computational results. In some rare cases the experimental data are used, but the broad energy range covered by the spallation reactions, as well as the numerous types of targets, projectiles and emitted particles involved, make the simulations indispensable.

\begin{table*}[hbt]
\begin{minipage}[c]{17cm}
\center
\begin{tabular}{ccccc} \hline\hline
\vspace{-0.25 cm}    
  & & & & \\ 
  & & & projectile & \\ \cline{3-5}\vspace{-0.25 cm}
  & & & & \\ 
  Library & Energy range                                              & neutron                     & proton                      & others \\  \cline{3-5}\vspace{-0.25 cm}
  & & & & \\
              &  (MeV)                                                         &                                  & Z\_target range        &   \\  
              &                                                                     &                                  & (number of nuclides)&  \\  \hline
  & & & & \\ 
  TENDL-2013\cite{KON12} & $\rightarrow$ 200              & 1 $\rightarrow$ 110  & 3 $\rightarrow$ 110   & 3 $\rightarrow$ 110 (d, t, $^3$He, $\alpha$)                 \\
                                       &                                           & (2630)  & (2625)   &  (2625, 2625, 2624, 2624)                 \\ \hline
  & & & & \\ 
  HEAD-2009\cite{KOR10}  & 150 $\rightarrow$ 1000 & 1 $\rightarrow$ 84   & 1 $\rightarrow$ 84     &                   \\
                                            &                                       & (682)                       & (682)                          &                    \\ \hline
  & & & &  \\ 
  JENDL/HE-2007\cite{WAT11}  & $\rightarrow$ 3000 & 1 $\rightarrow$ 95  & 1 $\rightarrow$ 95     &                    \\    
                                                   &                                & (107)                       & (107)                          &                    \\ \vspace{-0.25 cm} 
  & & & &\\  \hline\hline
\end{tabular} 
\caption{\label{tab:Tab_lib} Libraries available for the spallation reactions. Inspired from Tables 1 and 2 of Korovin in \cite{KOR10}.}
\end{minipage}
\end{table*}

Spallation codes provide various elementary cross-sec\-tions: reaction cross-section, emission cross-section of a specific particle ($\sim$ multiplicity, when total, or differential in energy and/or angle), residual nucleus production cross-section. This information can be used to build up databases. The first subsection discusses this point. The spallation codes, once implemented in a particle transport code, can also be used as event generator. When a particle hits a macroscopic target, transport codes are used to simulate creation/absorption/emission of all particles and nuclei, and thus a spallation code is invoked when a spallation reaction occurs, in order to treat the reaction and to deliver the results, making the transport code able to process the next interactions. This is the subject of the second subsection.

\subsection{Cross-section generator}
\label{toolxsgen}

Spallation codes can generate databases, especially cross-sections, which will be used afterwards in studies where spallation reactions take place. These databases are more or less complete. Those containing all necessary information can be used in transport codes, in place of the spallation models themselves, and the others aim at specific studies. The former are described in the first subsection while for the latter the case of cosmogenic nuclides will be discussed. Other examples, not described in this paper, could be mentioned, such as the database built by Konobeyev and Fischer on the gas production in spallation targets (278 target nuclei from Z=3 to Z=83) \cite{KON14}, where several spallation codes have been used and compared to available experimental data to provide the best production estimate of p, d, t, $^3$He and $\alpha$ with proton beams from 62 MeV to 1200 MeV.

\subsubsection{Libraries}
\label{toollibraries}


In transport codes the two ways to treat the reactions are the reaction models or the databases (libraries). For a long time, below 20 MeV the libraries were used for incident neutrons and the models for the higher energies and the other particles. These libraries have been extended and now, for some target nuclei, evaluated data exist for neutron and proton up to 150 MeV. Above 150 MeV, the realm of spallation reactions, the broad spectra of energies, of emitted particle types and of residual nuclides make the evaluation much more difficult, since it is not conceivable to launch all required experiments to map the domain. This explains why the transport codes call a spallation model instead of using a library. Nevertheless some libraries exist as shown in Table \ref{tab:Tab_lib}. This Table lists the three databases developed to be used in transport codes or in another way. The energy ranges covered are different. Two of them are in the spallation regime, HEAD-2009 and JENDL/HE-2007, while TENDL-2013 goes up 200 MeV only. We mention it first because the low-energy limit of the spallation reactions is not so clear (see section \ref{cascade}) and second because it is the most complete one in terms of projectile type and target nucleus. The correlation between the energy range and these two other parameters is obvious. The other two libraries deal with broader energy ranges, but fewer target and projectile types. A few words are given hereafter about HEAD-2009 and JENDL/HE-2007.

The HEAD-2009 library, motivated by the transmutation of nuclear wastes, has been developed to provide the nuclear data required to design the dedicated facilities. Since the nuclear data libraries for a nucleon with an energy below 150 MeV already exist, the idea was to focus on the high energy part, i.e. above 150 MeV. The strategy rests on the comparison of calculated results obtained by several spallation codes with experimental data taken from the EXFOR library  \cite{EXFOR}. The codes or the combinations of codes are those implemented in MCNPX2.6.0 \cite{MCN08} (Bertini-Dresner, Bertini-Abla, Isabel- Dresner, Isabel-Abla, INCL4.2-Dresner, INCL4.2-Abla and CEM03.01) and the CASCADE code in CASCADE/INPE \cite{BAR99}. 
The deviation factor used to define the best model, in a given target mass number range, is the product of the two factors $F$ and $H$ described in section  \ref{benchiaea} page \pageref{eq:forme-pos}. As one rejection criterion, the authors decided to remove an experimental datum {\it if at least one of the calculated cross-sections was equal to zero}. This rule is debatable, because the potentialities of a code will not be used if they are not shared by all the other codes. It could also be added that the idea of splitting the target mass number into nine sets should be applied to the type of emitted particles as well, for example, but the work would rapidly become tedious. This last point highlights the difficulty of building such libraries and the requirement for necessary criteria. The final result is a library, available on request, of data files written in the ENDF-6 format \cite{ENDF6}, the most common format used by the transport codes. Inevitably some of the computational models have been improved since then, and such a work should be done regularly to take into account these improvements, as it has been already mentioned by the authors \cite{KOR10}.

\begin{figure}[hbt]
\begin{center}
\resizebox{.5\textwidth}{!}{
\includegraphics{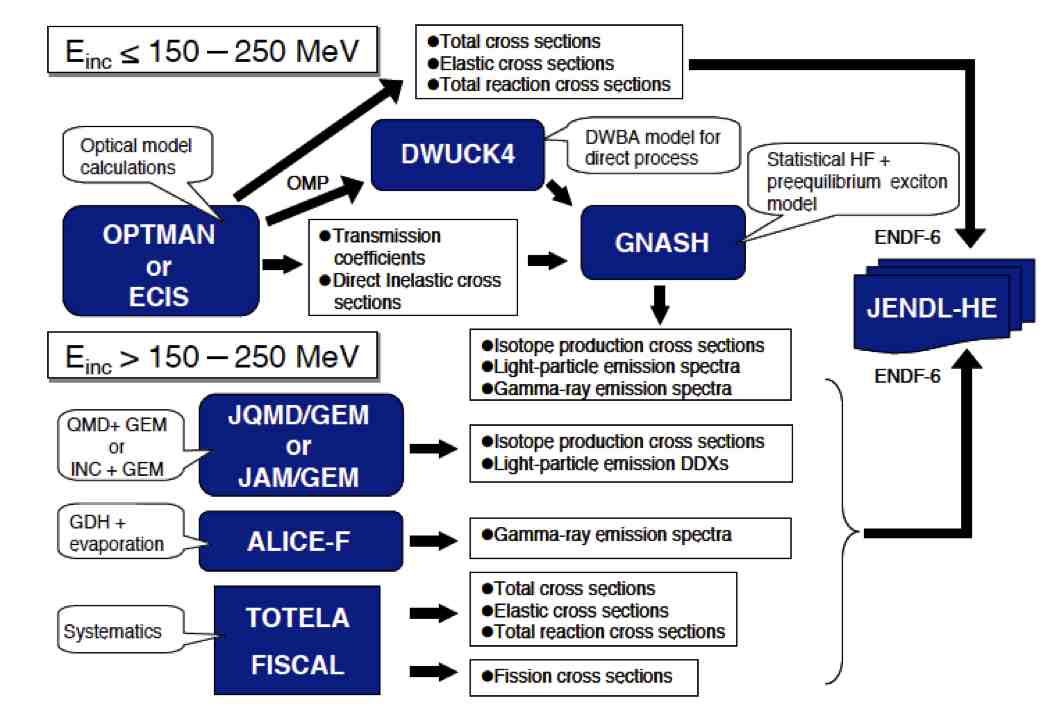}}
\end{center}
\caption{\label{Fig_JENDLHE}
The codes used to build the JENDL/HE library. Figure drawn from  \cite{WAT04}. \href{http://extras.springer.com/2005/978-0-7354-0254-6/cdr_pdfs/indexed/stage4_copyr/326_1.pdf}{(Courtesy of Y. Watanabe).}}
\end{figure}

The JENDL/HE-2007 library was also motivated by the extension of the low-energy libraries (below $\sim$150 MeV) up to 3 GeV. A first version, JENDL/HE-2004 \cite{WAT04} was released in 2004 and contained 66 nuclides from hydrogen to mercury. An updated version published in 2011 and named JENDL/HE-2007 \cite{WAT11} contains now 107 nuclides up to americium. Building up such a library on a broad energy range requests several dedicated codes to generate the data and Fig. \ref{Fig_JENDLHE} illustrates schematically the way used by JENDL/HE developers. Above 150 MeV - 250 MeV the combinations of spallation codes, included at that time in the transport code PHITS, JQMD\cite{NII95}/GEM\cite{FUR00} and JAM\cite{NAR01}/GEM are used, except when enough measurements are available. In this case a fitting procedure is applied. Experimental data are also used for tuning the parameters of the models.  As in HEAD-2009 only neutron and proton are considered as projectiles and the cross-sections are tabulated in the ENDF-6 format. In their paper \cite{WAT11} the developers of the library show some comparisons of the content of JENDL/HE-2007 with experimental data, but also compare the neutron flux produced from a thick target obtained with this library and some others (ENDF/B-VII.0\cite{CHA06}, TENDL-2009\cite{KON09}), but unfortunately at rather low energy (113 MeV). Nevertheless, other benchmarks of the JENDL/HE-2007 have been carried out, and one can mention the work of Takada {\it et al.} \cite{TAK09}, where the validations rest on thick targets made of heavy elements and bombarded with protons with energies up to 2.83 GeV.  Among the conclusions of this paper, two of them can be applied to all libraries. The first one is the difficulty inherent in the fact of using a specific model for each given energy range, that leads to the appearance of a jump in some cross-sections as it is sometimes the case in JENDL/HE-2007 at 250 MeV (already known in \cite{WAT11}), but the second one is the proof that a library speeds up the calculation compared to a reaction model. 

Building a nuclear data library in the spallation domain is a huge task. The different types of projectile, the broad energy spectrum and all possible nuclides as target explain the status of the three main libraries, if we include TENDL. TENDL-2013 goes only up to 200 MeV, but it is the only one that provides calculation results with composite particles up to $\alpha$ as projectile and for more than 2000 target nuclei. JENDL/HE-2007 is the only one that delivers cross-sections with projectile energies up to 3 GeV, but in return, at the time being, only 107 target nuclei have been considered. In-between there is HEAD-2009, from 150 MeV to 1 GeV with 682 targets. It is worth saying that they are all less than ten years old, more or less, and so they should include more and more data in the next few years. The advantage of such databases is both to speed up the calculation, avoiding the use of models, and sometimes, in rare cases, to get more reliable results than with models thanks to cross-sections drawn from experimental data. The other side of the coin is the necessary use of the spallation models to get most of the ingredients, and, as long as the models are improved, the libraries have to be regularly updated, which makes the work even more laborious. Moreover the correlations disappear when using a library.


\subsubsection{Dedicated databases for cosmogenic nuclides}
\label{toolcosmogeny}

\begin{figure}[hbt]
\begin{center}
\resizebox{.4\textwidth}{!}{
\includegraphics{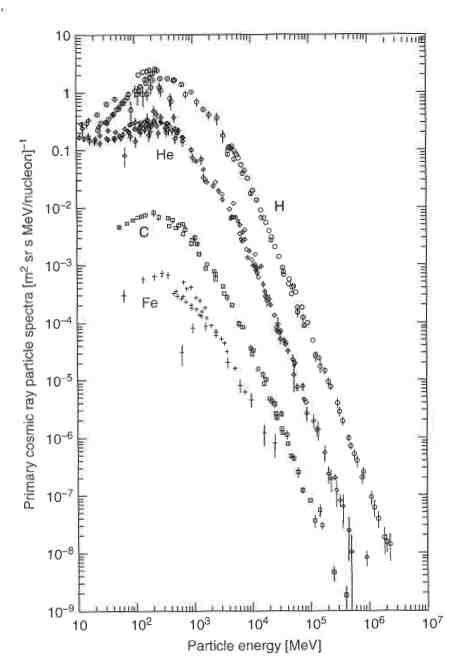}}
\end{center}
\caption{\label{Fig_GCR-spectra}
The energy spectra of the cosmic rays measured at Earth. Figure taken from \cite{FIL09} and originated from \cite{SIM83}.}
\end{figure}

The Galactic Cosmic Ray (GCR) consists of protons ($\sim$87\%), $\alpha$ ($\sim$12\%) and other ions ($\sim$1\%) with a broad energy spectrum, which is peaked around 1 GeV (Fig. \ref{Fig_GCR-spectra}). 
The interaction with different types of bodies in the space leads to the studies of, for example, meteorites, micrometeorites (but here that is the Solar Cosmic Ray that plays a more important role with a lower energetic spectrum \cite{TRA13}), or planetary atmospheres. The case of meteorites has been, and is still, extensively studied. The ability to simulate the amount of nuclei created in a meteorite due to its irradiation by the GCR can determine the size of the meteorite, its time of exposure to radiation, its possible residence time on earth, even the GCR flux. Details on this type of study can be found e.g. in reference \cite{AMM09}. The formula for the cosmogenic nucleus production rate $P$ is
\begin{eqnarray}\label{eq:ammon}
          P_j(R,d,M) = \sum\limits_{i=1}^N c_i \frac{N_A}{A_i} \sum\limits_k \int\limits_0^\infty \sigma _{i,j,k}(E)  J_k(E,R,d,M) dE, \nonumber
\end{eqnarray}
with $j$ the given nucleus, $R$ and  $d$ the radius and the depth of the meteorite at which the rate is determined, $M$ the solar modulation parameter\footnote{Energy loss by heliocentric distance of a particle entering the solar system.}, $i$ a target element, $c_i$ its concentration and $A_i$ its mass number, $N_A$ the Avogadro's constant, $k$ the type of particle inducing the reaction (primary and secondary particles), $\sigma _{i,j,k}(E)$ the production cross-section of the isotope $j$ by interaction of the particle $k$ with the target $i$ and with an incident energy $E$, and finally $ J_k(E,R,d,M)$ the density flux of the particle $k$. Comparisons of measured and calculated production rates from samples enable information about the studied meteorite to be drawn. The needed ingredients, production cross-sections and particle flux in the meteorite, are provided by experiments or calculations. Previously some calculations using spallation codes through transport codes were already completed. One can mention the studies by Reedy and Masarik \cite{REE94} and Leya and Masarik \cite{LEY09} for meteorites using LAHET and Masarik and Beer \cite{MAS99} for the terrestrial atmosphere using GEANT3. However, these studies used the codes to calculate the particle flux and not the production cross-sections. The latter were either experimental data or were determined using different codes. Therefore, according to the availability and quality of the experimental data, codes are used and some authors give a quality factor to the results. For example in \cite{LEY09} the factors go from A, when numerous high quality data are available with protons on thin targets and thick targets, these latter ones allowing to draw the neutron-induced cross-sections, up to E when codes are the last solution. It is worth noting that this rating is just to stress the point that the results rest on experiments or modeling, with no consideration given to the reliability of the models.

Excitation functions (or production cross-sections vs projectile energy) were calculated and compared to experimental data mainly to benchmark the codes, but also to complete the databases needed for the previous studies about cosmogenic nuclides.  Due to the required accuracy on the cross-sections used in such works, the experimental data were preferred to computational results, even until recently, as can be read in \cite{AMM09}. However, as seen with the IAEA {\it Benchmark of Spallation Models} in section \ref{benchiaea}, the spallation models have been greatly improved and moreover the conclusions showed the way to refine and also extend them. 

Three types of projectiles must be taken into account in such meteorites' studies: protons, neutrons and $\alpha$. Protons, because they are the main part of the cosmic ray and some secondaries can induce other reactions. Neutrons, as the most numerous secondary particles and with enough energy to generate cosmogenic nuclides. The $\alpha$ are known to be stopped easily and should not penetrate  the meteorites so deeply, but they are a significant part of the cosmic ray and can be responsible of the production of particular nuclides. Regarding calculation results, but also experimental data, the situation with these projectiles has evolved in recent years.

\begin{figure}[hbt]
\begin{center}
\resizebox{.38\textwidth}{!}{
\includegraphics{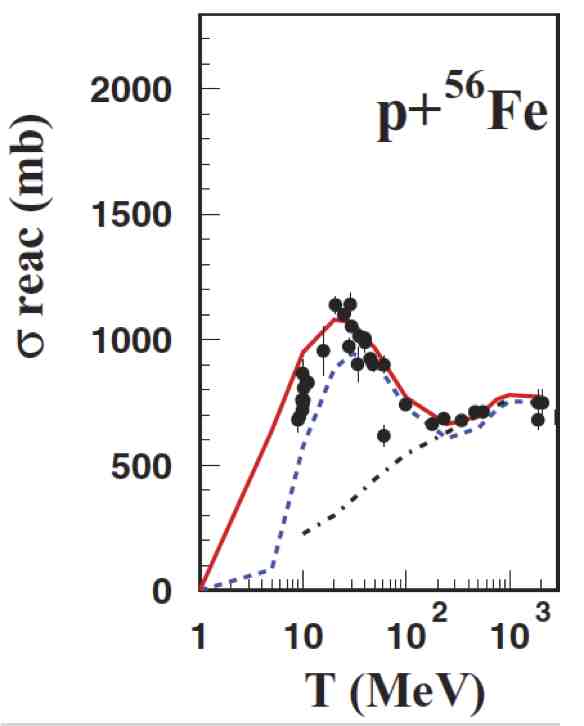}}
\end{center}
\caption{\label{Fig_p-reac}
Reaction cross-sections for the reaction p+$^{56}$Fe. Calculated results are in solid red line for INCL4.6, dashed blue line for INCL4.5 and dashed-dotted line for INCL4.2, and data are from \cite{CAR96,PRA97,BAR93}.  Figure taken from \cite{BOU13}. \href{http://link.aps.org/abstract/PRC/v87/e014606} {Copyright (2014) by the American Physical Society.}}
\end{figure}

\begin{figure}[hbt]
\begin{center}
\resizebox{.4\textwidth}{!}{
\includegraphics{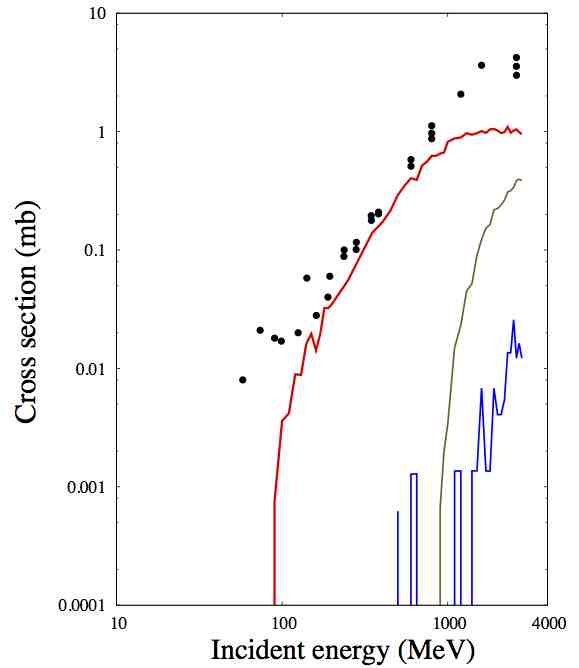}}
\end{center}
\caption{\label{Fig_fenat-be10}
Excitation function of the reaction p + $^{nat}$Fe $\rightarrow$ $^{10}$Be + X. Calculated results are in red for INCL4.5-Abla07, in blue for INCL4.2-Abla and in green for Bertini-Dresner, and data are from \cite{MIC02}. Figure drawn from \cite{DAV11c}. \href{http://sait.oat.ts.astro.it/MSAIt820411/PDF/2011MmSAI..82..909D.pdf} {Copyright {\copyright} SAIt 2011.}}
\end{figure}

\begin{figure}[hbt]
\begin{center}
\resizebox{.4\textwidth}{!}{
\includegraphics{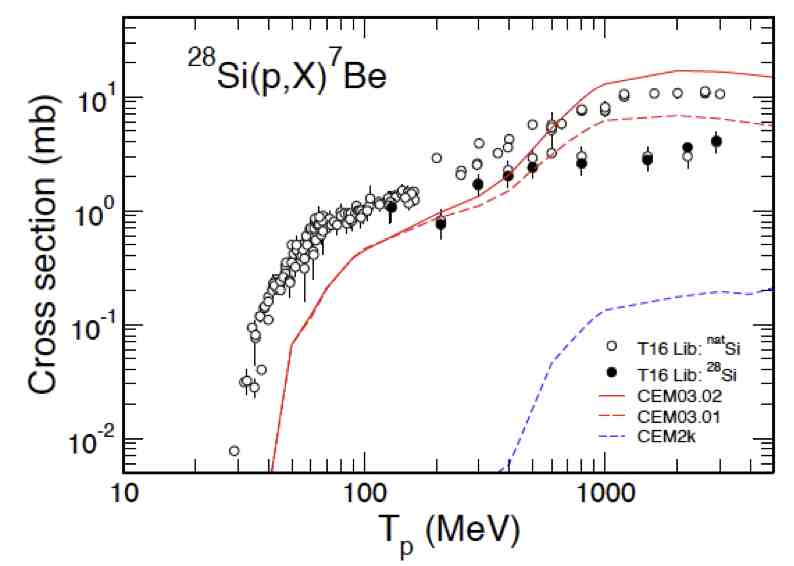}}
\end{center}
\caption{\label{Fig_cem-be7}
Excitation function of the reaction p + $^{28}$Si $\rightarrow$ $^{7}$Be + X. Calculated results are in solid red line for CEM03.02 \cite{MAS07}, in dashed red line for CEM03.01 \cite{MAS05} and in dashed blue line for CEM2k \cite{MAS00b}, and data are from \cite{MAS98}. Figure taken from \cite{MAS08}. (Courtesy of S. Mashnik). }
\end{figure}

As stated by R. C. Reedy in \cite{REE13} the cross-sections available for proton-induced reactions allow good agreements to be obtained with the studied samples for six main cosmogenic nuclides ($^3$He, $^{10}$Be, $^{14}$C, $^{21}$Ne, $^{26}$Al, $^{36}$Cl). Nevertheless the models have been continuously developed and we say one word below about improvements achieved recently by the INCL4 and Abla07 codes. First, the reaction cross-sections calculated were very often reliable only above say 100 MeV, not surprisingly knowing the hypotheses of the modeling (see section \ref{modeling}, page \pageref{modeling}). The efforts made in INCL4.6 \cite{BOU13} to carefuly treat the first interaction between the projectile particle and the target nucleus resulted in getting much better reaction cross-sections. An example is given in Fig. \ref{Fig_p-reac}, showing the improvements from INCL4.2 to INCL4.6.
One of the main shortcomings of the spallation codes was the production of Intermediate Mass Fragments (A $\le$ 20). Improvements have been made on both sides, that is by using a coalescence model in the intranuclear cascade and by taking into account light nuclides emission via evaporation and/or multifragmentation. Fig. \ref{Fig_fenat-be10} shows the case of $^{10}$Be production from an iron target, and, since this nuclide is principally produced during the de-excitation phase, it measures especially the great improvement made by Abla07 in this area. Similar improvements with the various versions of the CEM code were reported first in \cite{MAS08} as illustrated with Fig. \ref{Fig_cem-be7}, and more recently in \cite{MAS14} where a detailed study has been performed on the fragmentation of light nuclides, pointing out the role of the coalescence model and of the Fermi breakup. 

\begin{figure}[hbt]
\begin{center}
\resizebox{.4\textwidth}{!}{
\includegraphics{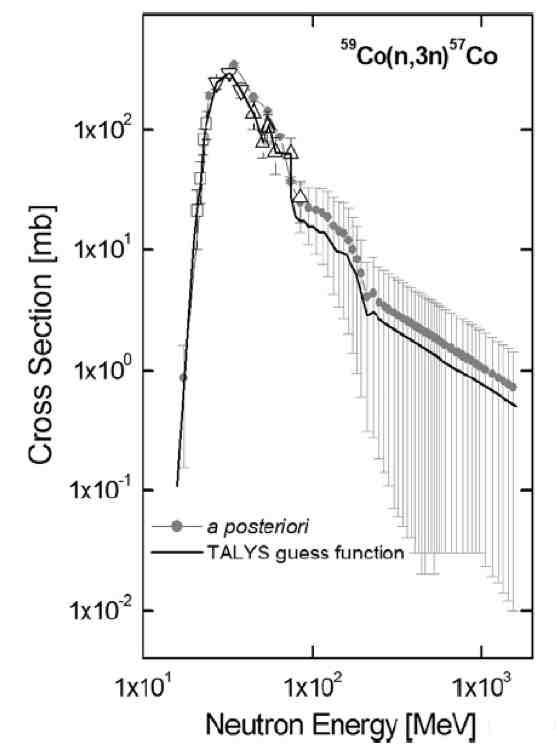}}
\end{center}
\caption{\label{Fig_fex-n-leya}
Neutron-induced excitation function of $^{57}$Co obtained with a $^{59}$Co target. The gray symbols are the results obtained with the guess function (TALYS calculations) plotted as a solid black line. Experimental data from other experiments \cite{KIM99,VEE77,UNO96} are plotted as open symbols. Figure taken from  \cite{LEY11} (Courtesy of I. Leya). \href{http://www.sciencedirect.com/science/article/pii/S0168583X11006550} {Copyright {\copyright} 2014 Elsevier B.V}.}
\end{figure}
\begin{figure}[hbt]
\begin{center}
\resizebox{.4\textwidth}{!}{
\includegraphics{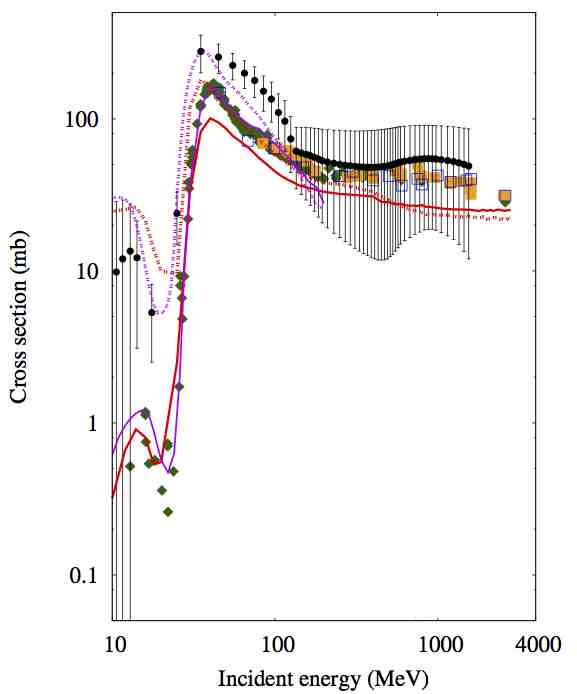}}
\end{center}
\caption{\label{Fig_fex-fenat-n}
Neutron- and proton-induced excitation functions of $^{54}$Mn obtained with a $^{nat}$Fe target. Dotted lines are for neutron calculations and solid lines for proton. Red color is dedicated to INCL4.5-Abla07 and purple color to the TENDL-2010\cite{KON10} library obtained with the TALYS code. Data taken from \cite{LEY11} (neutron - black points) and \cite{MIC02,TIT08} (proton - colored points). Figure drawn from  \cite{DAV11c}. \href{http://sait.oat.ts.astro.it/MSAIt820411/PDF/2011MmSAI..82..909D.pdf} {Copyright {\copyright} SAIt 2011.}}
\end{figure}
Despite the great improvements in the proton channel, the only, even if sizeable, benefits for studies like cosmogenic nuclides are still the cases where no experimental data exist. The situation concerning the neutron is very different. On the one hand, the contribution of neutrons to the production of some nuclides can reach 95\%, and, on the other hand, our knowledge of the neutron-induced production cross-sections is not as good as in the case of protons. The reason lies in the difficulty to get mono-energetic neutron beams, especially when the energy increases. Just below 200 MeV some experimental data exist (e.g. \cite{KIM98,KIM99,SIS07}), but above this the data are very scarce. One can mention the recent data obtained with 287 MeV and 370 MeV neutron beams at the Research Center for Nuclear Physics (RCNP - Osaka, Japan) \cite{YAS11,SEK11}, some being dedicated to cosmogenic nuclides \cite{NIN11}. An alternative to the neutron beams, proposed by Leya and Michel \cite{LEY11}, is the use of thick targets irradiated by a proton beam. Assuming firstly that only the primary and secondary protons and secondary neutrons are responsible for the nuclide production, secondly that the proton production cross-sections are well known, thirdly that the calculated particle spectra are reliable, and using a guess function when not enough data has been measured in the whole energy range, i.e. up to 1.6 GeV, they obtained neutron-induced production cross-sections by subtracting the proton-induced production rates from the measured data. Fig. \ref{Fig_fex-n-leya} shows an example of the results obtained with a cobalt target. Two regions stand out. Below 100-200 MeV, where the TALYS code, used to give the guess function, is reliable, the results are good, matching the experimental data from other experiments using neutron beams and with reasonable uncertainties. Above 200 MeV, i.e. outside the domain of validity of TALYS, the results have huge error bars. Therefore, this method, which gives good results where the guess function is supposed to be reliable, should probably use a reliable spallation model to get much better results above 200 MeV, and at least to reduce the uncertainties. Fig. \ref{Fig_fex-fenat-n} is another example of excitation functions using the data obtained in \cite{LEY11} for neutrons, but in addition with a comparison to the same nuclide produced with protons. One can notice again the increase of uncertainties above 200 MeV, but several other remarks can be made. Below 200 MeV the difference between TALYS and INCL4.5-Abla07 is not substantial, especially at very low energy, while above only the spallation model is able to give results quite close to the experimental data. Moreover, above 200 MeV the neutron and proton excitation functions are rather similar, whereas this is not true below this value. The assumption that proton-induced production cross-sections can act as a substitute of neutron-induced production cross-section is not always true. This was already mentioned e.g. in \cite{REE13}. To conclude about neutrons, the spallation models can help in the production of excitation functions with a neutron projectile in two ways: either by the direct calculations or by playing the role of a guess function (in the whole energy range or above a given energy) in the method developed by Leya and Michel. 

\begin{figure}[hbt]
\begin{center}
\resizebox{.4\textwidth}{!}{
\includegraphics{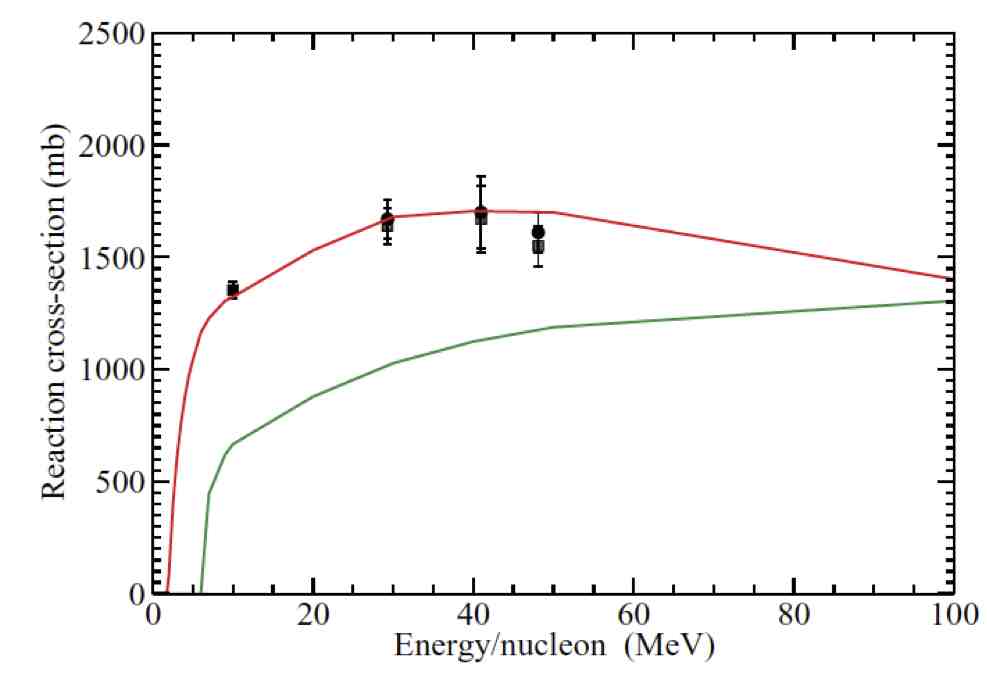}}
\end{center}
\caption{\label{Fig_ab-alpha}
Total reaction cross-section as function of the incident kinetic energy per nucleon for $\alpha$ on Ni target nucleus. The predictions of the INCL4.2 (INCL4.6) model are given by the green (red) curves. Data are taken from \cite{ING00,IGO63}. Figure taken from \cite{BOU13}. \href{http://link.aps.org/abstract/PRC/v87/e014606} {Copyright (2014) by the American Physical Society.}}
\end{figure}

\begin{figure}[hbt]
\begin{center}
\resizebox{.4\textwidth}{!}{
\includegraphics{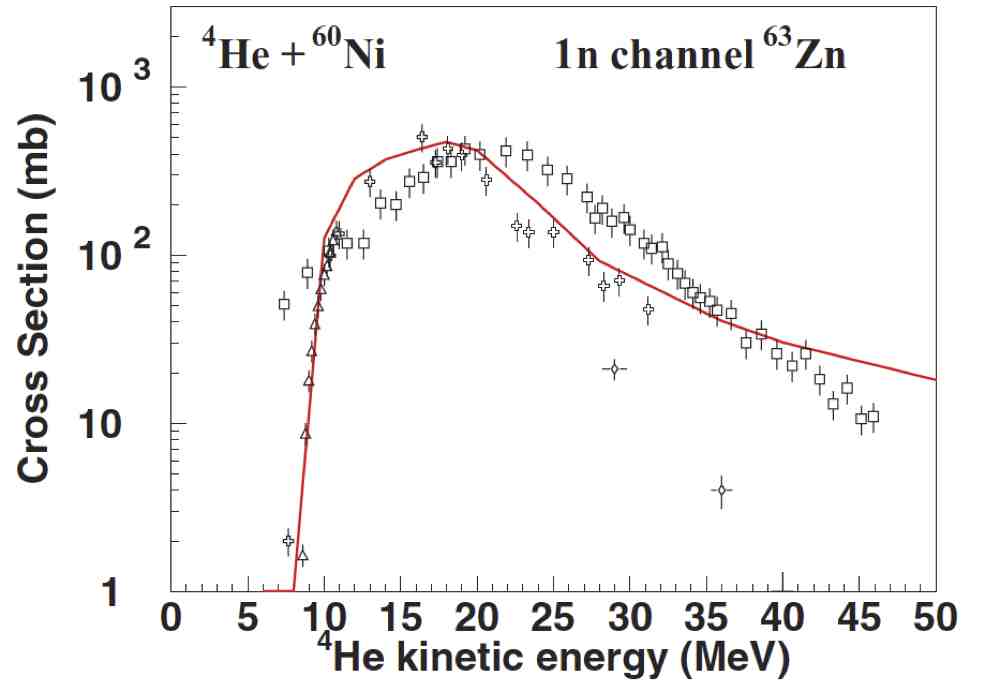}}
\end{center}
\caption{\label{Fig_ab-alpha-ni}
$\alpha$-induced excitation function of $^{63}$Zn obtained with a $^{60}$Ni target. The solid red line represents the INCL4.6-Abla07 results. Data taken from \cite{STE64,LEV91,YAD08,TAN60}. Figure drawn from \cite{BOU13}. \href{http://link.aps.org/abstract/PRC/v87/e014606} {Copyright (2014) by the American Physical Society.}}
\end{figure}

The last particle of interest is the $\alpha$. It represents only 12\% of the cosmic ray and can be rapidly slowed down, but can be responsible for specific nuclides. Moreover the experimental data are so scarce that the interaction of an $\alpha$ with a target nucleus is either neglected or treated with the assumption that the $\alpha$ is actually two neutrons and two protons containing 25\% each of the $\alpha$ energy. Doing so the production rate can be calculated with protons only (assuming neutrons and protons are similar) and corrected with a factor 1.55\footnote{$\tau_{tot}$, $\tau_p$($\sigma_{p}$) and $\tau_\alpha$($\sigma_{\alpha}$) respectively the total production rate, the production rates (cross-sections) due to protons and to $\alpha$.  \\
$\frac{\tau_{tot}}{\tau_p} = \frac{(\tau_{p}  + \tau_\alpha)}{\tau_{p}}$ = $\frac{0.87*\sigma_{p} + 0.12*\sigma_{\alpha}}{\sigma_{p}}$ = $\frac{0.87*\sigma_{p} + 0.12*4*\sigma_{p}}{\sigma_{p}}$ =1.55 } \cite{LEY00}. This assumption is reasonable for nuclides produced by high- to medium-energy $\alpha$, but is not valid with low-energy $\alpha$ \cite{AMM09}. In spallation modeling the interactions with composite particles like $\alpha$ have not been studied as carefully as the ones with nucleons. For example in Fig. \ref{Fig_ab-alpha} the reaction cross-sections calculated by INCL4.2 underestimated the experimental data. As with neutrons the data available are seldom.  Only recently an effort has been made and, for instance, the version INCL4.6 is now able to give good results (red line in  Fig. \ref{Fig_ab-alpha}). Some excitation function results are also good as shown in Fig. \ref{Fig_ab-alpha-ni}. In the case of $\alpha$-induced reactions the spallation models could surely be used nowadays to build databases dedicated to cosmogenic nuclides. 

The last point must be stressed. Two types of nuclide production cross-sections exist: independent or cumulative. The independent cross-section is the direct production, i.e. the result of the interaction, while the cumulative cross-section takes into account also the decay of the progenitors. These are the cumulative cross-sections that are used in such databases and most of the time also measured. Regarding measurements two definitions of the cumulative cross-sections can be read in the literature. For Yu. Titarenko \cite{TIT02} {\it cumulative} means simply the sum of the cross-sections, daughters and mother, while for R. Michel \cite{GLO01} the definition is drawn from the formula of the activity and takes into account the ratio of the half-lives and leads to $\sigma^{cum}_D = \sigma^{ind}_D+\sigma^{cum}_M$ $\frac{\lambda_M}{\lambda_M-\lambda_D}$, where $D$($M$) means daughter(mother) and $ind$($cum$) independent(cumulative). Therefore, as mentioned in \cite{TIT02} where the definition of Michel is named {\it supracumulative}, the definition of the cross-sections must be given unambiguously to avoid in some, fortunately rare, cases erroneous calculations and comparisons. 

Finally, thanks to recent improvements, the spallation modeling should probably be a useful and essential tool to build databases regarding cosmogenic nuclides, and/or to be of use in the neutron case as a guess function.



\subsection{Event generator}
\label{tooleventgen}

In this section the spallation codes are directly used in a particle transport code as event generator. The simulations replace, most of the time, experiments infeasible because they are too complicated, too expensive or too long to be carried out.
Three studies are discussed to point out which results can be obtained with a transport code using a spallation model as event generator. The way to proceed is very similar in the three studies, but each one has been chosen to underline a particular point. The first deals with the designing of a facility with the Eurisol Design Study, the second treats the prediction of calculations regarding the Megapie project, a spallation target which was operated four months in 2006, and the third one tackles the question of the uncertainties via the feasibility works of the European Spallation Source target.

\subsubsection{Designing - Eurisol-DS}
\label{tooleurisol}



\begin{figure}[hbt]
\begin{center}
\resizebox{.5\textwidth}{!}{
\includegraphics{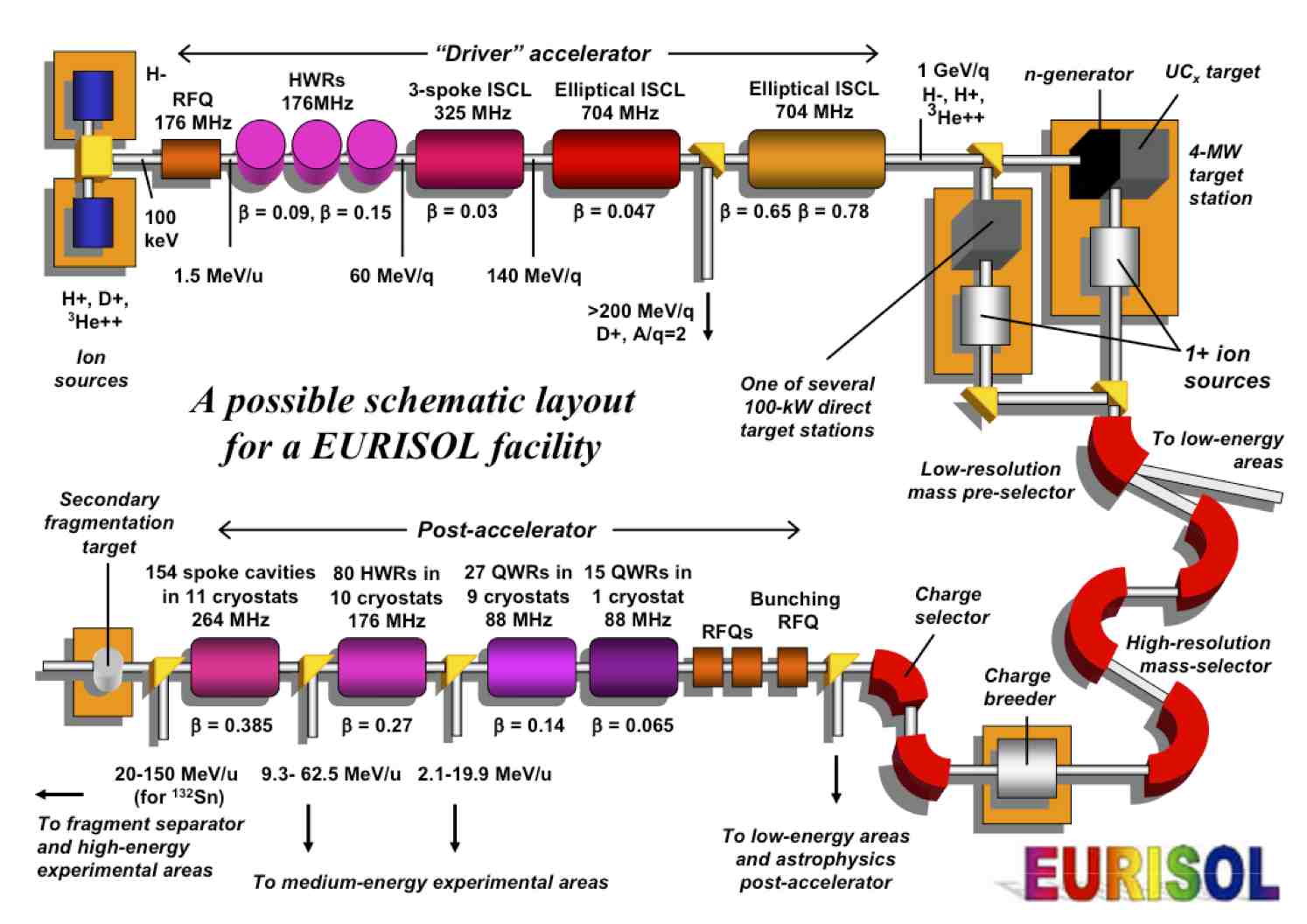}}
\end{center}
\caption{\label{Fig_eurisol-scheme}
A possible schematic view of the EURISOL facility. Figure taken from  \cite{EUR09}.}
\end{figure}
\begin{table*}
\begin{minipage}[c]{17cm}
\center
\begin{tabular}{lccccc} \hline
Target material  & Al$_2$O$_3$    & SiC                 & Pb (molten) & Ta                  & UC$_3$ \\ \hline
$\rho$               & 2.0                     & 3.2                 & 11.4              & 12.5              & 2.418    \\ \hline
L                       & 50-75-100-125   & 32-48-64-80  & 9-18-27-36   & 18-16-24-32  & 40-60-80-100    \\ \hline
R                       & \multicolumn{5}{c}{9.0-12.7-18.0-25.5}    \\ \hline
$\sigma$           & \multicolumn{5}{c}{R/3}    \\ \hline
Beam particles  & \multicolumn{5}{c}{protons}    \\ \hline
E                       & \multicolumn{5}{c}{0.5-1.0-1.5-2.0}    \\ \hline
P                       & \multicolumn{5}{c}{100}    \\ \hline
\end{tabular} 
\caption{\label{tab:Tab_eurisol-direct} Parameters used in the in-target isotope production yield calculations. Target density $\rho$ given in g/cm$^{3}$, length L in cm, radius R in mm, FWHM $\sigma$ in mm, incident particle energy E in GeV, beam power P in kW. Table drawn from \cite{CHA10}. \href{http://link.springer.com/article/10.1140/epja/i2010-10989-7} {With kind permission of The European Physical Journal (EPJ).}}
\end{minipage}
\end{table*}
The EURISOL (EURopean Isotope Separation On-Line) project is a European project  aiming at providing exotic radioactive ion beams. This facility, based on the ISOL method, should be the next step after the similar facilities such as SPIRAL2 at GANIL and HIE-ISOLDE at CERN, but with intensities several orders of magnitude higher. A complementary project under construction in Germany, FAIR (Facility for Antiproton and Ion Research), uses the method of in-flight separation. While FAIR uses heavy ions as primary beam, EURISOL will use particles like the proton. The exotic ions will be produced either directly in a target by spallation reactions, or as fission products in an actinide target surrounding a mercury converter providing the needed spallation neutrons to induce fission. Fig. \ref{Fig_eurisol-scheme} gives a schematic view of what could be the EURISOL facility, with the proton accelerator on the top left, the target stations, source of radioactive ions, on the top right, the beam preparation on the bottom right, and finally the post-accelerator and the experimental areas on the bottom left.

The spallation reactions take place principally in the two types of target stations already mentioned: direct targets (three in parallel) and fission target. The direct targets will provide the proton-rich ions and the fission target the neutron-rich ions. The complementarity of the two options has been checked in \cite{ENE09b}. The two types of study regarding such targets are the in-target production yields and the safety issues. Regarding the in-target production yields, the ions produced in the fission target are induced by low-energy neutrons, whose energy spectra are generally well reproduced by the codes, and their yields are well known. The production yield calculations in the direct targets according to the free design parameters were, on the other hand, more challenging. Concerning safety, the 4 MW delivered in the mercury converter (the neutron source for the fission target station) will lead to a much higher radioactivity than in the direct targets (100 kW). Thus the safety studies were focussed on this converter. The references of the other studies can be found in \cite{EUR09}, especially in sections {\it Calculated Yields of Exotic Ions} and {\it Radiation Safety}.

Assessments of the in-target production yields of radioactive isotopes in the direct targets are detailed in \cite{CHA10} and we give here just the methodology, the main results and the conclusions which can be drawn for the spallation modeling. The study was based on a preliminary work\cite{STO05} dedicated to the feasibility of such targets, addressing the engineering questions and providing nominal values for the most relevant parameters. The goal was twofold: first supplying an estimate of the best isotope production yields for the eleven elements of interest, seven given by NUPECC \cite{COR03} (Be, Ar, Ni, Ga, Kr, Sn and Fr) and four more suggested by the EURISOL collaboration \cite{PAG07} (Li, Ne, Mg and Hg), and second for which target configurations. Since the amount of calculation increases dramatically considering the materials, the dimensions of the targets and the projectile energies, a compromise had to be found and a set of configuration parameters was chosen. They are reported in Table \ref{tab:Tab_eurisol-direct}.

The choice of different materials rests on the aim to optimize the range in charge and mass of the produced nuclei. The targets were defined as cylinders, but four lengths and four radii for each material were calculated, in order to further optimize the performances, i.e. to find out a balance between a big target where numerous nuclei are produced, but where the extraction efficiency is poor and the opposite case of a small target, with less nuclei, but a better release efficiency. The study was restricted to the proton beam and a constant power of 100 kW, but with four energies and therefore four associated intensities. Finally the transport code MCNPX2.5.0 was chosen, but only two spallation codes were used: INCL4.2-Abla and CEM2k. This choice was dictated by several constraints and goals: i) the possibility to compare different simulations, ii) the reliability of the models \cite{DAV07a,DAV07b}, and iii) the computation time, because each codes was run for 320 configurations.

\begin{figure}[hbt]
\begin{center}
\resizebox{.34\textwidth}{!}{
\includegraphics[trim=0cm 0.4cm 0cm 0.8cm ,clip]{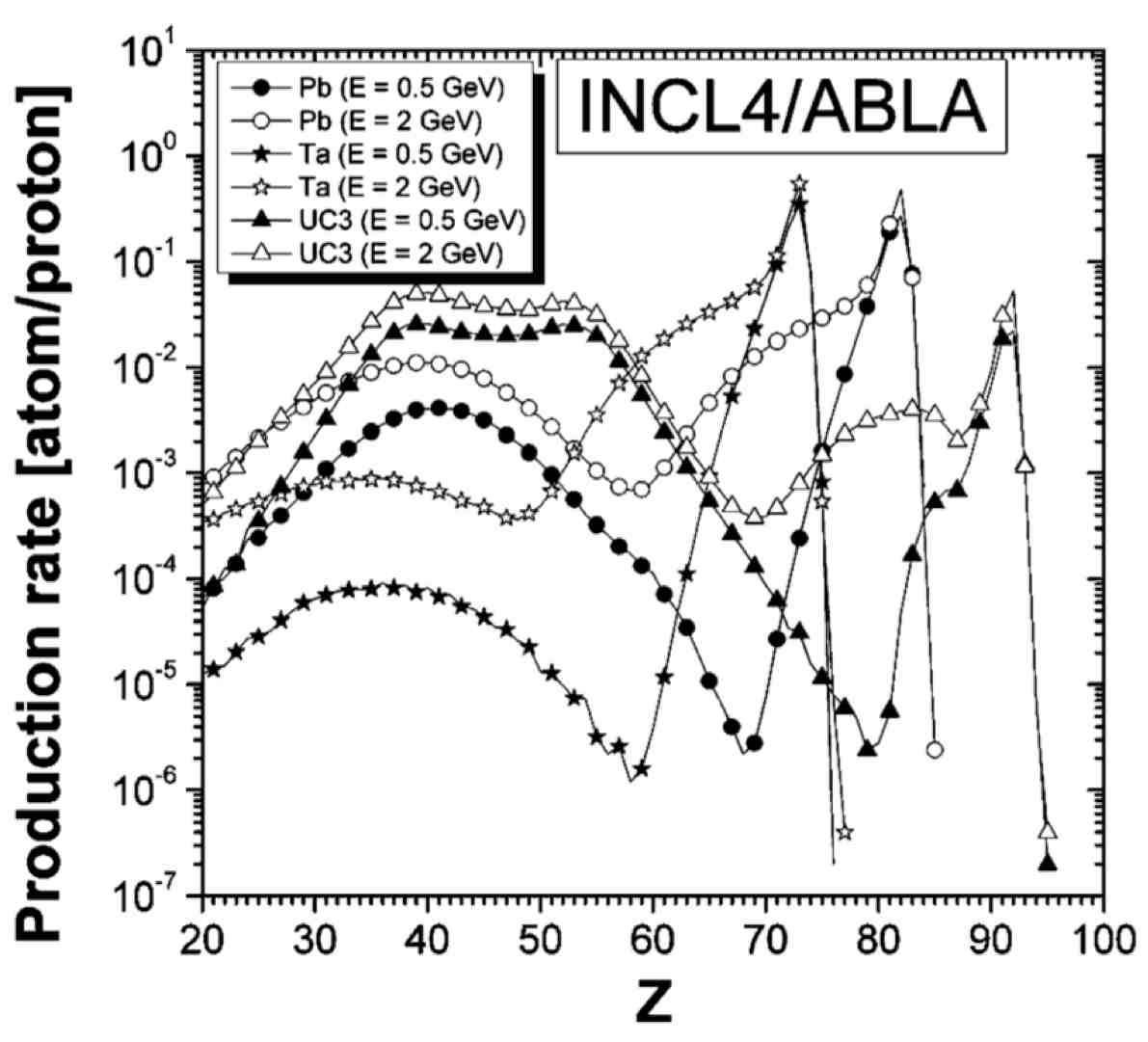}}
\end{center}
\caption{\label{Fig_eurisol-direct1}
Charge distributions of the production rates per incident proton (atom/proton) for targets with R = 18 mm and a mass $\sim$ 2 kg. Calculations done with INCL4-Abla in MCNPX2.5.0. Figure taken from  \cite{CHA10}. \href{http://link.springer.com/article/10.1140/epja/i2010-10989-7} {With kind permission of The European Physical Journal (EPJ).}}
\end{figure}

\begin{figure}[hbt]
\begin{center}
\resizebox{.37\textwidth}{!}{
\includegraphics[trim=0.5cm 4cm 0cm 4cm ,clip]{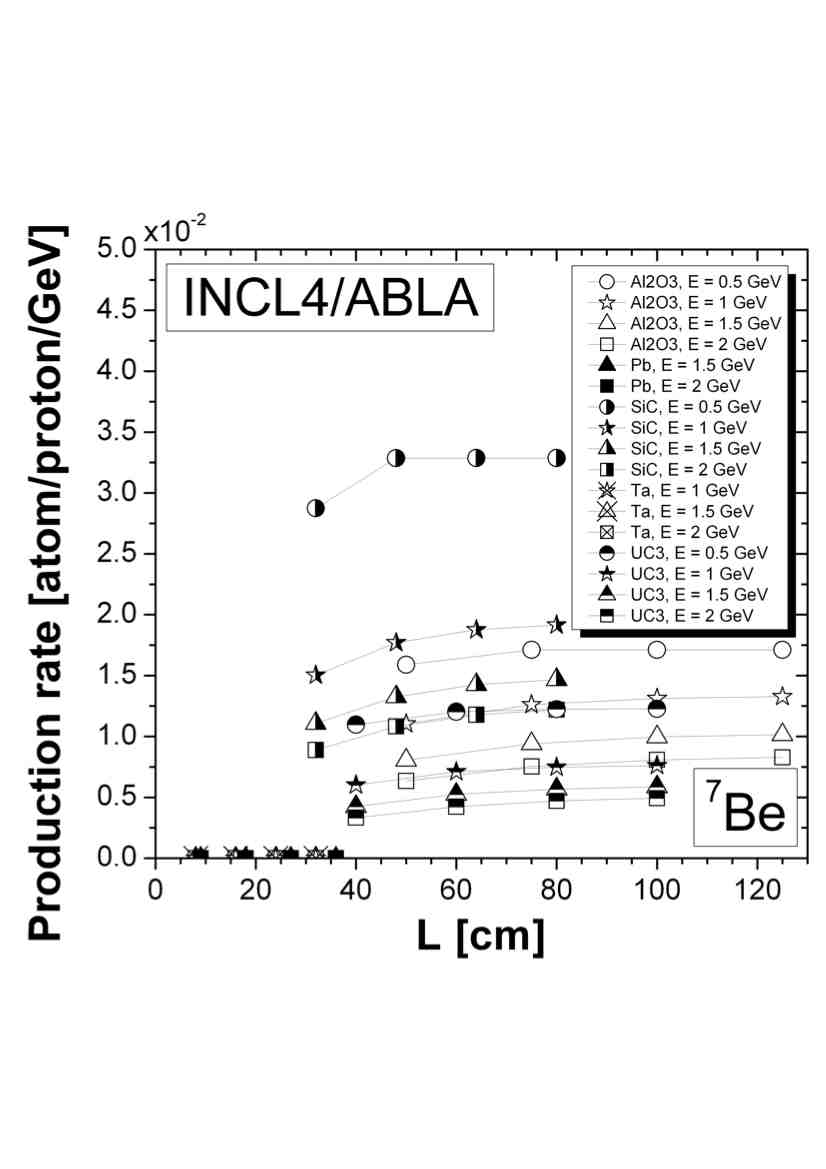}}
\end{center}
\caption{\label{Fig_eurisol-direct3}
$^{7}$Be production rate per incident proton per GeV function of the target length L (R = 18 cm). Calculations done with INCL4-Abla in MCNPX2.5.0. Figure taken from  \cite{CHA10}. \href{http://link.springer.com/article/10.1140/epja/i2010-10989-7} {With kind permission of The European Physical Journal (EPJ).}}
\end{figure}

The strategy to determine the optimal configurations consisted of studying the criteria one after the other. The first was the material of the target according to a given isotope, the second the incident proton energy and the third the dimensions, length and radius of the target. The type of graphs used for selecting the materials was the distributions in charge or in mass. Fig.~\ref{Fig_eurisol-direct1} gives an example. 
\begin{table}
\center
\begin{tabular}{c}
\resizebox{.45\textwidth}{!}{\includegraphics{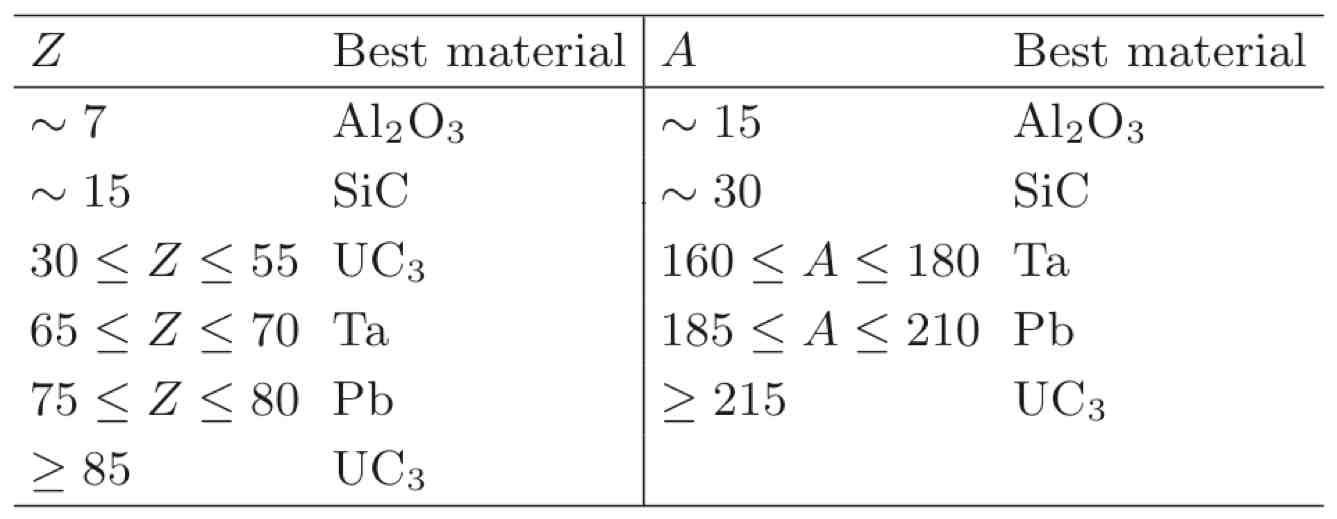}} \\
\end{tabular} 
\caption{\label{Fig_eurisol-direct2}
Optimal target materials for the in-target production of radioactive nuclei from calculations with INCL4-Abla and CEM2k in MCNPX2.5.0. Table drawn from  \cite{CHA08}.}
\end{table}
The findings of this study are summarized in Tab.~\ref{Fig_eurisol-direct2}. Sometimes several materials are good candidates for  a charge or a mass, in this case the A and Z were not reported. Within the framework of this project it was said that increasing the intensity was more interesting than increasing the energy, for reasons of cost. This led to study of the yields per incident energy unit. Fig.~\ref{Fig_eurisol-direct3} shows the example of $^7$Be. 
\begin{figure}[hbt]
\begin{center}
\resizebox{.4\textwidth}{!}{
\includegraphics{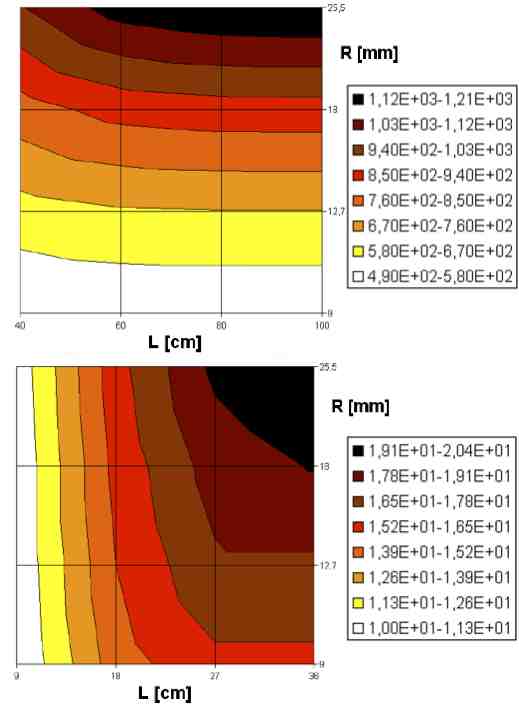}}
\end{center}
\caption{\label{Fig_eurisol-direct4}
$^{92}$Kr (top) and $^{180}$Hg (bottom) production rates per incident proton [$10^{-6}.s^{-1}$] as function of target radius R and length L, with respectively a UC$_3$ target and a proton energy of 0.5 GeV and a Pb target and a proton energy of 1 GeV. Calculations done with CEM2k in MCNPX2.5.0. Figure drawn from \cite{CHA08}.}
\end{figure}
While in general the higher the energy is, the better are the yields, the beams of 0.5 GeV are the most efficient option when the yields per unit of energy are considered. Finally, the choice of  the target dimensions was made from graphs displaying production rates depending both on the length and on the radius of the target. Two examples are given in  Fig.~\ref{Fig_eurisol-direct4}. In one case the gain is mainly obtained by  increasing the radius ($^{92}$Kr), while in the other it is obtained by increasing the length ($^{180}$Hg). One possible explanation for this difference is the production mode. The $^{92}$Kr is produced by fission in UC$_3$, with the beam, but also with the secondary particles distributed throughout the volume, whereas the $^{180}$Hg from a Pb target requires high incident energy and therefore it is produced preferentially along the axis of the primary beam.


Conclusions of this study are summarized in Tab.~\ref{Fig_eurisol-direct5}, which gives for each isotope of interest  the maximum yield and the best materials for an 1 GeV beam and a target with a radius of 18 mm. Some notes were added because, sometimes the statistic was too low to draw strong conclusions, sometimes it was the quality of the model for the isotope and the target considered, and sometimes the experimental data did not allow the validity of the models to be judged. The purpose of these remarks is to point out the sensitive cases for the future  studies regarding the target design of the EURISOL facility. Obviously, once produced in the target the isotopes must be extracted and the in-target production yields have to be corrected for the release and ionisation efficiencies. Their assessments need specific studies and a summary is given in \cite{EUR09} (in "Additional material" - Task 11). These factors are rarely higher than around ten percent, most of the time a few percent, and for some isotopes even lower. 

\begin{table*}
\begin{minipage}[c]{17.5cm}
\hspace{-0.5 cm}
\center
\scriptsize
\begin{tabular}{lllllll} \hline\vspace{-0.15cm}
 & & & & & & \\ 
 \hspace{-0.3 cm} {Isotope} & \hspace{-0.3 cm} Optimal target(s) &\multicolumn{2}{c}{\hspace{-0.3 cm} Max. yield [atom/s]}                                  & \multicolumn{2}{c}{MCNP statistical error [\%]} &  \hspace{-0.2 cm} {Additional remarks}\\\cline{3-4}\cline{5-6}\vspace{-0.15cm}
 & & & & & & \\               &\hspace{-0.3 cm}         for E = 1GeV      & \hspace{-0.3 cm} INCL4/ABLA                          & 	\hspace{-0.3 cm} CEM2k                                    & INCL4/ABLA	& CEM2k & \\\hline\vspace{-0.15cm}
 & & & & & & \\ 	
 \hspace{-0.3 cm} $^{9}$Li	                     & \hspace{-0.3 cm} Al$_2$O$_3$; SiC     & \hspace{-0.3 cm} 2.5 10$^{11}$; 1.1 10$^{11}$   & \hspace{-0.3 cm} 4.5 10$^{11}$; 1.3 10$^{12}$    & 3; 4	        & 2; 1	                          & \hspace{-0.4 cm} Remark (1) \\
 \hspace{-0.3 cm} $^{11}$Li	                     & \hspace{-0.3 cm} Al$_2$O$_3$           & \hspace{-0.3 cm} 1.8 10$^{9}	 $                          & \hspace{-0.3 cm} 3.6 10$^{10}$	                       & 30	                & 6	                                         &\hspace{-0.4 cm}  Remark (1) \\
 \hspace{-0.3 cm} $^{7}$Be	                     & \hspace{-0.3 cm} SiC	                         & \hspace{-0.3 cm} 1.2 10$^{13}$	                    & \hspace{-0.3 cm} 1.2 10$^{13}$	                       & $<$ 1	        & $<$ 1	                                 & \hspace{-0.4 cm} Remark (1) \\
 \hspace{-0.3 cm} $^{11}$Be	                    & \hspace{-0.3 cm} Al$_2$O$_3$	         & \hspace{-0.3 cm} 1.3 10$^{11}$	                   & \hspace{-0.3 cm} 4.4 10$^{11}$	                      & 3	                & 2	                                         & \hspace{-0.4 cm} Remark (1) \\
 \hspace{-0.3 cm} $^{12}$Be	                   & \hspace{-0.3 cm} Al$_2$O$_3$	         & \hspace{-0.3 cm} 3 10$^{10}	$                           & \hspace{-0.3 cm} 2.3 10$^{11}	$                      & 7	                & 3	                                         & \hspace{-0.4 cm} Remark (1) \\
 \hspace{-0.3 cm} $^{17}$Ne	                   & \hspace{-0.3 cm} SiC	                        & \hspace{-0.3 cm} 1.3 10$^{10}$	                  & \hspace{-0.3 cm} 3.7 10$^{9}$	                      & 30	               & 20	                                         & \hspace{-0.4 cm} Remark (2), (3) \\
 \hspace{-0.3 cm} $^{18}$Ne	                   & \hspace{-0.3 cm} SiC	                        & \hspace{-0.3 cm} 1.9 10$^{12}$	                  & \hspace{-0.3 cm} 8.6 10$^{10}$	                      & $<$ 1	               & 5	                                  & \hspace{-0.4 cm} Remark (2), (3) \\
 \hspace{-0.3 cm} $^{25}$Ne	                   & \hspace{-0.3 cm} Al$_2$O$_3$; Pb      & \hspace{-0.3 cm} 1.2 10$^{10}$; 3.7 10$^{8}	$  & \hspace{-0.3 cm} 2.4 10$^{9}$; 3.9 10$^{9}$	       & 13; 100	       & 25; 30	                          & \hspace{-0.4 cm} Remark (3) \\
 \hspace{-0.3 cm} $^{20}$Mg	                   & \hspace{-0.3 cm} SiC	                        & \hspace{-0.3 cm} 2.3 10$^{11}$	                  & \hspace{-0.3 cm} 2.4 10$^{10}$	                      & 3	               & 8	                                         & \hspace{-0.4 cm} Remark (3) \\
 \hspace{-0.3 cm} $^{30}$Mg	                   & \hspace{-0.3 cm} Pb ; UC$_3$	        & \hspace{-0.3 cm} 2.5 10$^{8}$; 2.5 10$^{8}$	  & \hspace{-0.3 cm} 5.1 10$^{9}$; 3.1 10$^{9}	$       & 100; 100     & 20; 20	                                 & \hspace{-0.4 cm} CEM2k: overestimates, factor $\geq$ 3 \cite{DAV07a,DAV07b} \\
 \hspace{-0.3 cm} $^{46}$Ar	                   & \hspace{-0.3 cm} Pb ; UC$_3$	        & \hspace{-0.3 cm} 1.2 10$^{8}$; 8.7 10$^{8}$	  & \hspace{-0.3 cm} 2.7 10$^{10}$; 3.1 10$^{10} $     & 100; 40	       & 10; 8	                         & \hspace{-0.4 cm} CEM2k: overestimates, factor $\geq$ 3 \cite{DAV07a,DAV07b} \\
 \hspace{-0.3 cm} $^{56}$Ni	                   & \hspace{-0.3 cm} Pb ; UC$_3$	        & \hspace{-0.3 cm} 1.2 10$^{8}$; no event	          & \hspace{-0.3 cm} 2.8 10$^{10}$; 2 10$^{10}$	       & 100; no event& 10; 9	                             & \hspace{-0.4 cm} CEM2k: overestimates, factor $\approx$ 100 \cite{DAV07a,DAV07b} \\
 \hspace{-0.3 cm} $^{72}$Ni	                   & \hspace{-0.3 cm} UC$_3$	               & \hspace{-0.3 cm} 1.1 10$^{10}	$                          & \hspace{-0.3 cm} 1.1 10$^{11}$	                      & 15	                & 4	                                         &\hspace{-0.4 cm}  CEM2k: overestimates, factor $\geq$ 5 \cite{DAV07a,DAV07b} \\
 \hspace{-0.3 cm} $^{63}$Ga	                   &\hspace{-0.3 cm}  Pb ; UC$_3$	        & \hspace{-0.3 cm} 2.5 10$^{8}$; no event	          & \hspace{-0.3 cm} 2.1 10$^{10}$; 2.4 10$^{10} $     & 100; no event& 10; 9	                             &\hspace{-0.4 cm}  CEM2k: overestimates, factor $\approx$ 100 \cite{DAV07a,DAV07b} \\
 \hspace{-0.3 cm} $^{81}$Ga	                   & \hspace{-0.3 cm} UC$_3$	               & \hspace{-0.3 cm} 2.1 10$^{10}	$                          & \hspace{-0.3 cm} 1.8 10$^{11}$	                      & 10	                & 3	                                         & \hspace{-0.4 cm} CEM2k: overestimates, factor $\geq$ 5 \cite{DAV07a,DAV07b} \\
 \hspace{-0.3 cm} $^{74}$Kr	                   & \hspace{-0.3 cm} Pb ; UC$_3$	       & \hspace{-0.3 cm} 1.2 10$^{8}$; no event	          & \hspace{-0.3 cm} 1.9 10$^{10}$; 1.5 10$^{10}  $    & 100; no event& 9; 10	                                 & \hspace{-0.4 cm} CEM2k: overestimates, factor $\approx$ 100 \cite{DAV07a,DAV07b} \\
 \hspace{-0.3 cm} $^{92}$Kr	                   & \hspace{-0.3 cm} UC$_3$	              & \hspace{-0.3 cm} 1.1 10$^{12}$	                          & \hspace{-0.3 cm} 8 10$^{11}	    $                          & 2	                & 2	                                  & \\
 \hspace{-0.3 cm} $^{107}$Sn	          & \hspace{-0.3 cm} Pb ; UC$_3$	      & \hspace{-0.3 cm} 2.5 10$^{8}$; 1.2 10$^{8}$	  & \hspace{-0.3 cm} 1.1 10$^{10}$; 1.6 10$^{10} $     & 60; 100	        & 15; 10	                          & \hspace{-0.4 cm} CEM2k: overestimates, factor $\approx$ 10 \cite{DAV07a,DAV07b} \\
 \hspace{-0.3 cm} $^{132}$Sn	          & \hspace{-0.3 cm} UC$_3$	              & \hspace{-0.3 cm} 2.3 10$^{11}	 $                         & \hspace{-0.3 cm} 3.3 10$^{11}	$                      & 4	                & 3                                           & 	 \\
 \hspace{-0.3 cm} $^{180}$Hg	          &\hspace{-0.3 cm}  Pb	                      & \hspace{-0.3 cm} 1.5 10$^{10}	 $                         & \hspace{-0.3 cm} 1.2 10$^{10}	$                      & 10	                & 12                                  & 	 \\
 \hspace{-0.3 cm} $^{206}$Hg	          & \hspace{-0.3 cm} Pb	                      & \hspace{-0.3 cm} 5.7 10$^{11}	$                         & \hspace{-0.3 cm} 1.9 10$^{10}	$                      & 2	                & 10	                                       & \hspace{-0.4 cm} INCL4/ABLA: overestimates, factor $\geq$ 5 \cite{BOU02} \\
 \hspace{-0.3 cm} $^{205}$Fr	                 & \hspace{-0.3 cm} UC$_3$	              & \hspace{-0.3 cm} 1.7 10$^{9}$	                          & \hspace{-0.3 cm} 9.6 10$^{10}	$                      & 30	                & 4	                       &  \hspace{-0.4 cm} INCL4/ABLA: underestimates \cite{DAV07a,DAV07b};  \\
                  &      &             &               & 	                &                       &  \hspace{-0.4 cm} CEM2k: overestimates \cite{DAV07a,DAV07b} \\\hline
\end{tabular} 
\caption{\label{Fig_eurisol-direct5}
In-target maximal nuclei yields (atom/s) expected within EURISOL single-stage targets for 1 GeV protons. Proton beam energy = 1 GeV, beam power = 100kW, spallation models = INCL4-Abla and CEM2k, target radii = 18 mm, low energetic secondary neutron flux contribution taken into account, CINDER'90 \cite{WIL98} evolution time = 1 ms. Note that E = 1 GeV is not always the best energy for maximizing the nuclei yields. Remark (1): Nuclei obtained after a long evaporation process are most of the time overestimated by CEM2k and underestimated by INCL4-Abla (see the mass distribution for iron in \cite{DAV07a,DAV07b}). Remark (2): For neutron-poor nuclei the yield can fall down of more than one order of magnitude for two isotopes with only 1 mass difference [8] (here $^{17}$Ne and $^{18}$Ne). Remark (3): A specific benchmark should be done for light targets (needs of data) to disentangle between both models. Table drawn from \cite{CHA10}. \href{http://link.springer.com/article/10.1140/epja/i2010-10989-7} {With kind permission of The European Physical Journal (EPJ).}}
\end{minipage}
\end{table*}

Such a study needs a good knowledge of the spallation models to be able to choose between the two codes when the results are different, but in return is a good exercise to understand what kinds of mechanism are important for the modeling and what should be improved. Two examples of the analysis performed in \cite{CHA10} can be reported. The first one concerns the $^{25}$Ne production. This isotope can be produced either from a UC$_3$ target via the fission mechanism or from an Al$_2$O$_3$ target via the interaction of the proton projectile on the $^{27}$Al nucleus with emission of three protons and one $\pi^+$. The fission mechanism was better reproduced by INCL4-Abla than CEM2k \cite{DAV07a}, so the results of INCL4-Abla for the first option were more reliable. The second channel needs a good handling of $\pi^+$ emission, thus, since INCL4-Abla was known to overestimate this production, its predictions must be considered carefully. The second example deals also with pion production, since it is the $^{231}$Fr production from a UC$_3$ target. If the interaction is done with a $^{235}$U nucleus, five protons and one $\pi^+$ must be emitted. If the target nucleus is a $^{238}$U, the main mechanisms are emission of six protons and two neutrons, or five protons, three neutrons and one $\pi^+$. More than one $\pi^+$ can be emitted, but the probabilities fall down dramatically. CEM2k was not able to produce such an isotope and the results of INCL4-Abla were questionable, at least for the same reason as previously mentioned on the pion production. Since then both models have been improved concerning the pion emission \cite{MAS08,BOU13,FIL08,LER11}, but also they include new mechanisms as cluster emissions, which could be another way to release the needed mass and charge in the $^{238}$U case. To conclude on this study, at least three aspects of the uncertainties have been mentioned: the statistical errors reported in Tab.~\ref{Fig_eurisol-direct5}, the estimates with two spallation models as a cross-check and the use of experimental data to benchmark the codes \cite{DAV07a,DAV07b}. This topic is discussed in-depth in section \ref{tooless}.

The safety issues were important in the EURISOL Design Study project, especially regarding the mercury converter. Numerous aspects had nothing to do with spallation reactions and all details can be read in the various references cited in \cite{EUR09}. Here two results related to the spallation modeling are discussed. The first one deals with the radioactivity level in the mercury converter \cite{RAP10} and the second one with the air activation \cite{ENE09c}.
\begin{figure}[hbt]
\begin{center}
\resizebox{.45\textwidth}{!}{
\includegraphics{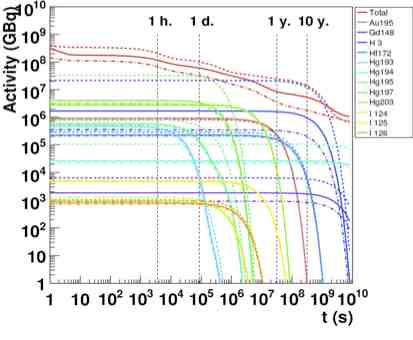}}
\end{center}
\caption{\label{Fig_eurisol-fission1}
Activity estimates (total and from major contributors) in the mercury target, function of cooling time after 40 years of irradiation with 1 GeV protons and a beam intensity of 2.28 mA. Calculations were performed with three models of spallation available in MCNPX2.5.0 (Isabel-Abla (solid line), CEM2k (dashed line) and INCL4.2-Abla (dashed-dotted line)) and the activation code CINDER'90. Figure taken from \cite{RAP10}. \href{https://www.oecd-nea.org/science/egsaatif/satif8.pdf} {Copyright {\copyright} OECD 2010.}}
\end{figure}
 Fig. \ref{Fig_eurisol-fission1} gives the total activity of the mercury target at the shut-down after an irradiation time of 40 years with a 1 GeV proton beam delivering 4 MW (7 months per year), and also during the cooling. Estimates  were made by the use of three spallation models available in the transport code MCNPX2.5.0, Isabel-Abla, CEM2k and INCL4.2-Abla, each followed by the activation code CINDER'90 \cite{WIL98}. The first result was the very high activity generated. It  is comparable to that of a 20 MW research reactor \cite{MOO03,MOO08} and with the cooling time may become greater due to specific contributors, which play an important role. All models confirm this high radioactivity, although some differences exist, which vary in time and can reach almost one order of magnitude. Actually the differences are due to specific nuclei that are not produced with the same rate according to the model used.
\begin{table*}
\center
\begin{tabular}{|c|c|c|c|c|c|c|} \hline
 & \multicolumn{3}{c|}{1 year after irradiation} & \multicolumn{3}{c|}{10 years after irradiation} \\ 
\cline{2-7}     & ISABEL-ABLA & CEM2k & INCL4-ABLA & ISABEL-ABLA & CEM2k & INCL4-ABLA \\ \hline
Total activity  & 8.4 10$^6$ & 2.4 10$^7$ & 3.5 10$^6$ & 5.2 10$^6$ & 1.4 10$^7$ & 1.7 10$^6$ \\ \hline
$^{195}$Au   & 2.4 10$^5$ & 2.1 10$^5$ & 2.2 10$^5$ & 1.8              & 1.6              & 1.7              \\ \hline
$^{148}$Gd   & 1.8 10$^3$ & 6.4 10$^3$ & 9.0 10$^2$ & 1.7 10$^3$ & 5.9 10$^3$ & 8.3 10$^2$ \\ \hline
$^{3}$H         & 1.6 10$^6$ & 2.0 10$^7$ & 3.3 10$^5$ & 9.6 10$^5$ & 1.2 10$^7$ & 2.0 10$^5$ \\ \hline
$^{172}$Hf   & 1.6 10$^5$ & 2.0 10$^5$ & 1.5 10$^5$ & 5.6 10$^3$ & 7.1 10$^3$ & 5.2 10$^3$ \\ \hline
$^{194}$Hg  & 2.6 10$^4$ & 1.0 10$^5$ & 2.2 10$^4$ & 2.6 10$^4$ & 1.0 10$^5$ & 2.2 10$^4$ \\ \hline
\end{tabular} 
\caption{\label{Fig_eurisol-fission2}
Contributions to the activity (GBq) of the mercury target of some important isotopes. Estimates from three models available in MCNPX2.5.0: Isabel-Abla, CEM2k and INCL4.2-Abla. Table taken from \cite{RAP10}. \href{https://www.oecd-nea.org/science/egsaatif/satif8.pdf} {Copyright {\copyright} OECD 2010.}}
\end{table*}
Tab.~\ref{Fig_eurisol-fission2} gives the contributions of some radionuclides that play an important role in the radiation protection of the mercury target. Two of them, $^{148}$Gd and $^{3}$H, deserve attention. The first one is not a great contributor to the total radioactivity, but it is an $\alpha$ emitter. The three calculation results reported in Tab.~\ref{Fig_eurisol-fission2} show large differences. From a Hg target nucleus the $^{148}$Gd isotope is obtained after a long evaporation process and, for example, INCL4.2-Abla was known to fail to reproduce such isotopes. 
The other isotope is the tritium. This particle is emitted during either the intranuclear cascade or the de-excitation phase. Isabel, INCL4.2 and Abla were not able to emit this particle and the production came only from the low-energy neutron activation, handling with CINDER'90. Actually tritium in some modeling can also be produced during the pre-equilibrium process, this intermediate phase included in some models. An example is the yield calculated by CEM2k, larger and probably more realistic. This particular point was studied later and is discussed in the next section. With these two examples it is clear that, even with a spallation model that has been considered reliable or at least better than others \cite{DAV07a,DAV07b}, in specific domains its use must be avoided. In return such a study points out some weak points, which must be fixed to improve and/or extend the model.

A second topic related to spallation and safety issues is the air activation between the target and the shielding. In \cite{ENE09c} authors explain the two ways used to estimate air activation, both based on the same fact: the low density prevents from using the usual and direct Monte-Carlo method for the residual nucleus production, which requires too long computation times. The production yields $Y$ of the radionuclides are then calculated with the following equation
\begin{eqnarray}
Y_i = \sum_{j,k} N_j \int \sigma_{ijk}\Phi_k(E)dE \nonumber ,
\end{eqnarray}
where $N_j$ is the density of element $j$, $\sigma_{ijk}$ the production cross-section of the isotope $i$ from the interaction of the particle $k$ with the target nucleus $j$ and $\Phi_k(E)$ the flux of the particle $k$ with an energy $E$. The difference between the two ways lies mainly in the cross-sections. The first method used the transport code MCNPX to get the flux and the cross-sections are drawn from evaluated databases \cite{NNDC}. Unfortunately these databases provide only neutron-induced production cross-sections. The second method is based on the PHITS transport code \cite{IWA02} that treats directly the above equation, using its calculated spectra and the needed excitation functions estimated in previous studies \cite{FUR01}. This way PHITS includes the contributions of the proton-induced reactions. The  difference between the two methods appeared clearly when calculating the radiation activity: PHITS was a factor 3 higher than the other approach. Moreover, one of the main contributors, the radioisotope $^{11}$C, which was missing in the method using MCNPX, was known to be underestimated by a factor close to 3 by PHITS around 1 GeV. This result shows the strong underestimate of the first method. Simulations are not only required to save time and to study various configurations of a facility where experiments would be difficult and expensive, but they also enable the developers to improve  the capabilities of the reaction models by pointing out the deficiencies. In addition the confidence level increases when the reliability of the reaction models is known.

Safety issues and cosmogenic nuclides, discussed in section \ref{toolcosmogeny}, meet here with the needs of excitation functions and the benefit of spallation codes.

\subsubsection{Predicting and validating - Megapie}
\label{toolmegapie}

 \begin{figure}[hbt]
\begin{center}
\resizebox{.3\textwidth}{!}{
\includegraphics{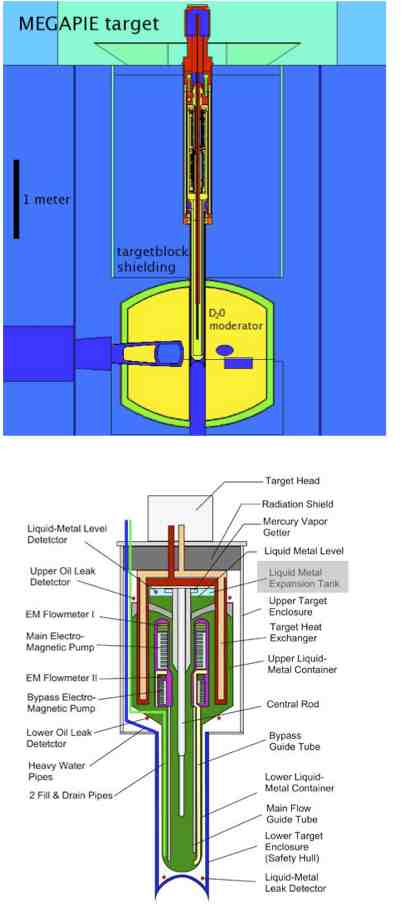}}
\end{center}
\caption{\label{Fig_megapie-design}
Schematic views of the MEGAPIE spallation target. Upper part, geometry used in the transport code 
MCNPX2.5.0 with the target and the other elements of the system such as the heavy water and, lower part, 
main components of the target. Figures drawn from \cite{ZAN08}. \href{http://www.iaea.org/inis/collection/NCLCollectionStore/_Public/40/109/40109493.pdf}{(Courtesy of L. Zanini).}}
\end{figure}
MEGAPIE (MEGAwatt PIlot Experiment) is a prototype of liquid lead-bismuth (LBE for Lead Bismuth Eutectic) spallation target. This international project was launched in 2001 \cite{BAU01}. It aimed at designing, building and operating a liquid spallation target of about 1 MW power at PSI (Paul Scherrer Institute - Switzerland) in the SINQ spallation neutron source facility. Fig. \ref{Fig_megapie-design} shows two views of the target. Irradiation lasted four months from August to December 2006 with 575 MeV protons and an average intensity of 0.947 mA.

Neutronic performance, delayed neutrons, gas production and target activation are four topics that were studied and described in the report "Neutronic and Nuclear Post-Test Analysis of MEGAPIE" \cite{ZAN08} as well as in the feedback experience reported in \cite{ZAN11}. Regarding neutronic performance, measurements and calculations were satisfying, except a discrepancy between calculations and measurements in the central rod, where calculations were 2 or 3 times higher than the measurements. A careful study concluded that the reason lay in the geometry description, where the details are important, and not in the spallation reaction modeling \cite{SEN10}. Concerning the delayed neutrons only measurements were performed and they revealed a safety issue in such a liquid metal target, where the delayed neutron contributors in the target loop can generate a non-negligible neutron flux in ancillary components. The analysis showed that the contributors were light mass fragments (e.g. $^{17}$N) and fission products ($^{87}$Br and $^{88}$Br) originating from spallation reactions. These kinds of nuclei are known to be difficult to compute, with a not well defined mechanism (extended evaporation or very asymmetric fission) for the first ones, and because they are very neutron-rich isotopes for the other two. It would then be interesting to compare these experimental data with calculation results of some recent spallation models, to see if the modeling has been improved in these areas since previous calculations \cite{RID07}. The two other topics, target activation and volatiles, are discussed with more details below.

\begin{figure}[hbt]
\begin{center}
\resizebox{.5\textwidth}{!}{
\includegraphics[trim=0.5cm 1cm 0cm 2cm, clip]{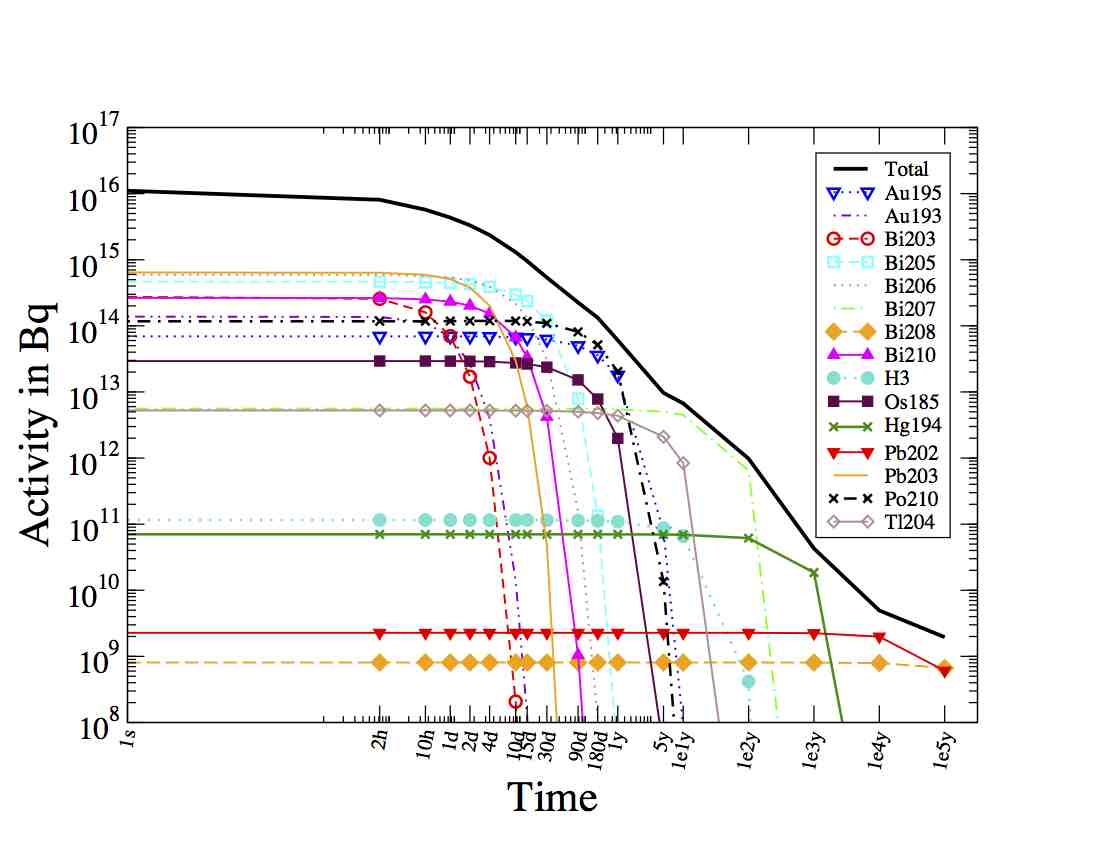}}
\resizebox{.5\textwidth}{!}{
\includegraphics[trim=0.5cm 1cm 0cm 2cm, clip]{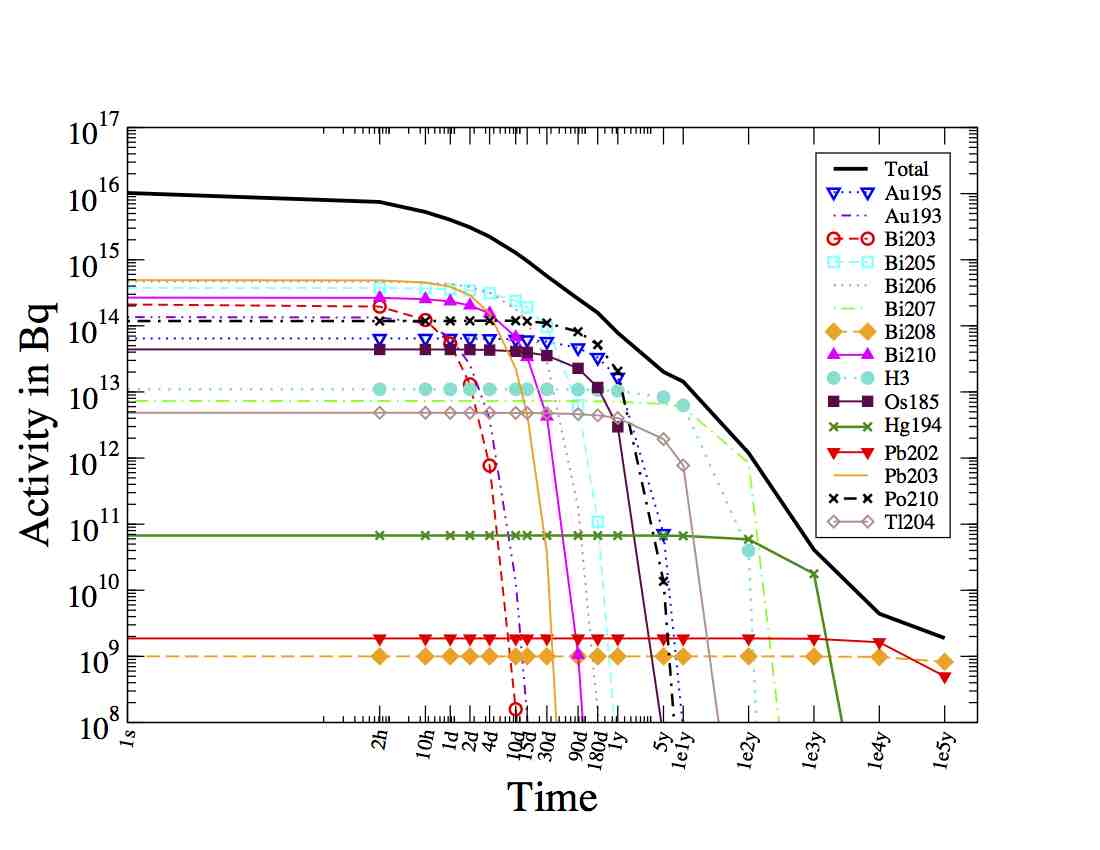}}
\end{center}
\caption{\label{Fig_megapie-contrib}
Main contributors to the activity of the LBE MEGAPIE target, as a function of cooling time after 123 days of irradiation with 575 MeV protons at 0.947 mA using INCL4.2-Abla (top panel) and Bertini-Dresner (bottom panel) models. Figures taken from \cite{DAV08}.}
\end{figure}
Target activation includes activation of the spallation target itself, of its container and of the window between the spallation target and the accelerator. Calculations were made with several codes and the result of the comparison  was that differences in activities were low for each component. For example Fig. \ref{Fig_megapie-contrib} displays the main contributors in the case of the liquid lead-bismuth spallation target, given by INCL4.2-Abla (top) and Bertini-Dresner (bottom). The total activity is due to many contributors and the most important ones, especially after the shut-down, are nuclei with charges and masses close to those of lead and bismuth. This explains the similar results, since, in general, the codes produce these elements with rather similar rates. However a notable difference appears after around ten years of cooling and is due to tritium. This difference, already encountered in the EURISOL study, was explained by the non-emission in INCL4.2 and Abla of tritium. While the activity of the container, surrounded by a heavy water tank, comes essentially from the low-energy neutrons, washing out the differences between the spallation models, one again sees the difference due to tritium in the window.

As already mentioned light nucleus emission has been included in more recent versions of INCL4 and Abla: in INCL4.5 (and next versions) via a coalescence model \cite{BOU13} and in Abla07 principally thanks to an improved evaporation modeling in the case of hydrogen and helium (section Abla07 in \cite{FIL08}). These new versions, implemented in a beta version of MCNPX2.7 \cite{MCN11}, were used to estimate activation of the window as a function of the cooling time, as it was done with INCL4.2-Abla. The results were as expected closer to the other models for a cooling time around ten years. However, a careful study attempting to validate the obtained results using elementary data, as tritium production cross-sections, showed that first the new calculation were reliable, but also that the activation due to tritium was partly due to the $^3$He isotope \cite{DAV11d,DAV12}. The reason comes from the combination of the four months of irradiation, resulting in a non-negligible production of $^3$He via spallation, and the rather high flux of low-energy neutrons in the window area. Since the tritium production cross-section via low-energy neutron on $^3$He is 1000 barns and the neutron flux around ten times higher than the proton flux, the contribution of $^3$He to the tritium production is estimated between 20\% and 30\% (The full calculation with MCNPX gives  29,7\%, and a rough estimate knowing fluxes, nucleus densities and production cross-sections gives 20.5\%). In addition to the validation of the new results, this study pointed that, in case of a high neutron flux and a long irradiation time, a good estimate of $^3$He production was required to get a reliable tritium production. In a more general way, the truism saying that a model must be as complete as possible is here proved, and the way to improve the best spallation models is the extension of their capabilities.

\begin{figure}[hbt]
\begin{center}
\resizebox{.5\textwidth}{!}{
\includegraphics{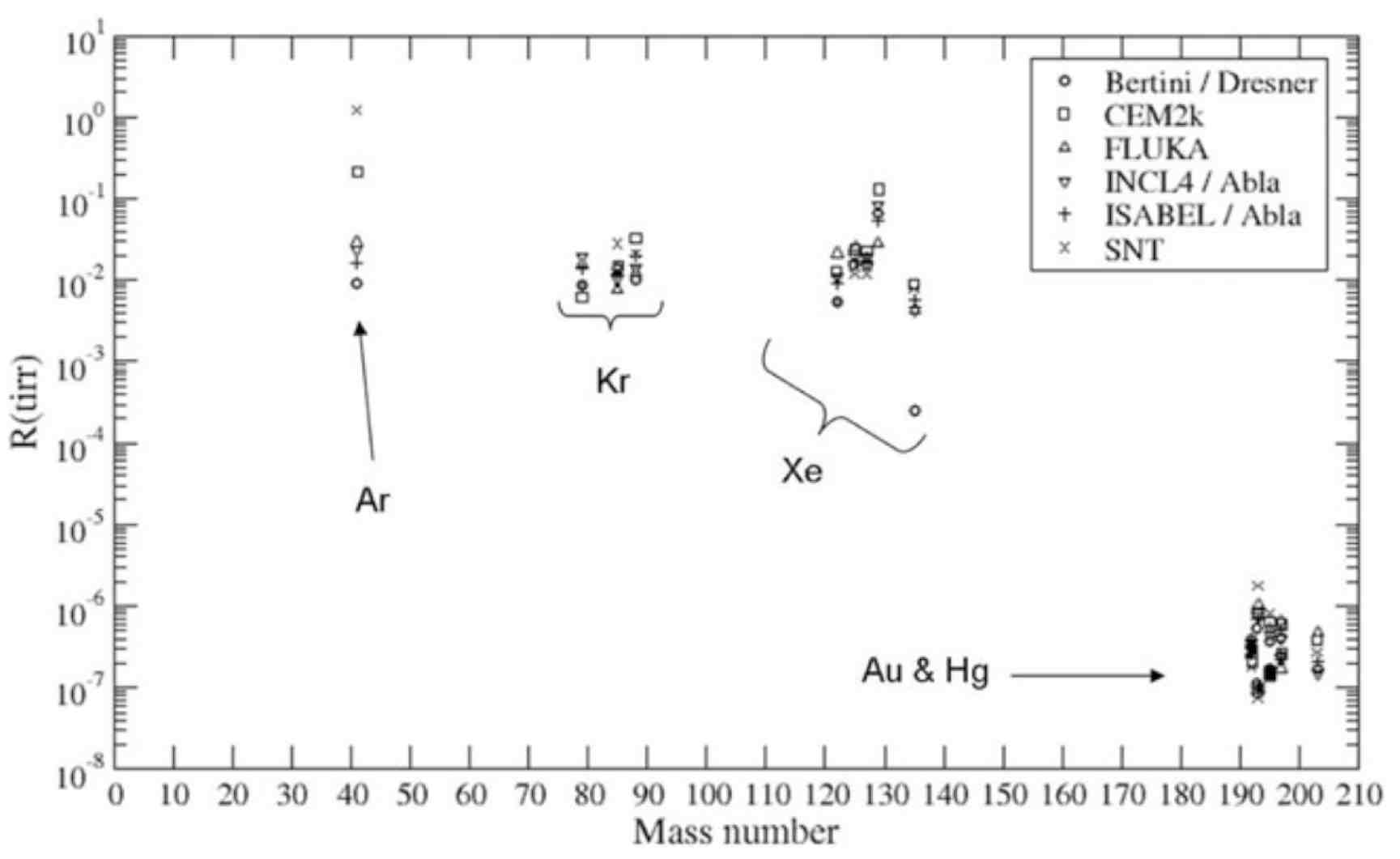}}
\end{center}
\caption{\label{Fig_megapie-volatile}
Ratio R between experimental activity in the expansion volume and simulated in-target production as a function of isotope mass number. This ratio can be seen as a release fraction. It is not meant as a comparison between measurements and calculations, given that measurements are performed outside the target, while the simulation is done in-target. Figure drawn from \cite{THI11} (Courtesy of N. Thiolli\`ere). \href{http://epubs.ans.org/?a=12522}{Copyright 2014 by the American Nuclear Society, La Grange Park, Illinois.}}
\end{figure}
The volatile elements do not have an important weight in the total activities, but they remain a risk during operation in case of release, and therefore their quantities must be known. Measurements at MEGAPIE were performed and they consisted of the recovery of released gas and measurement of isotope activities at the cover gas system (CGS) at the top of the target \cite{THI11}. Calculations of the in-target productions were done with several models (FLUKA \cite{FLU07a,FLU07b}, SNT \cite{KOR92} and MCNPX2.5.0 \cite{MCN05} associated to CINDER'90 \cite{WIL98}), but unfortunately the direct comparison with the measured activities was not possible, because the release fractions were not known. However, Fig.~\ref{Fig_megapie-volatile} shows the ratios obtained between activities calculated in the target and measurements at the CGS. Those ratios could be interpreted as the release fractions, since the noble gas isotopes have roughly the same ratio, while the mercury family have a much smaller ratio, which is what one would expect. Moreover the models all give the same results within a factor of 3, except for Ar. This latter case reflects the fact, already mentioned, that the models differed in their results concerning intermediate mass fragments. In \cite{THI11} a delicate point is stressed: the importance of metastable nuclei on activity due to volatiles. The production rate of metastable nuclei is rarely calculated by spallation codes themselves. However, there is a possibility in MCNPX to get the production of metastable nuclei (\cite{MCN05}, page C-11), but this estimate, based on tabulated fractions, which depend on the excitation energies, should be studied more specifically. Other codes consider even a crude ratio of 0.5. Nevertheless, as mentioned previously in section \ref{evaporation} (page \pageref{evaporation}), recent developments on this topic have been published \cite{PHI13}.

\begin{figure}[hbt]
\begin{center}
\resizebox{.4\textwidth}{!}{
\includegraphics{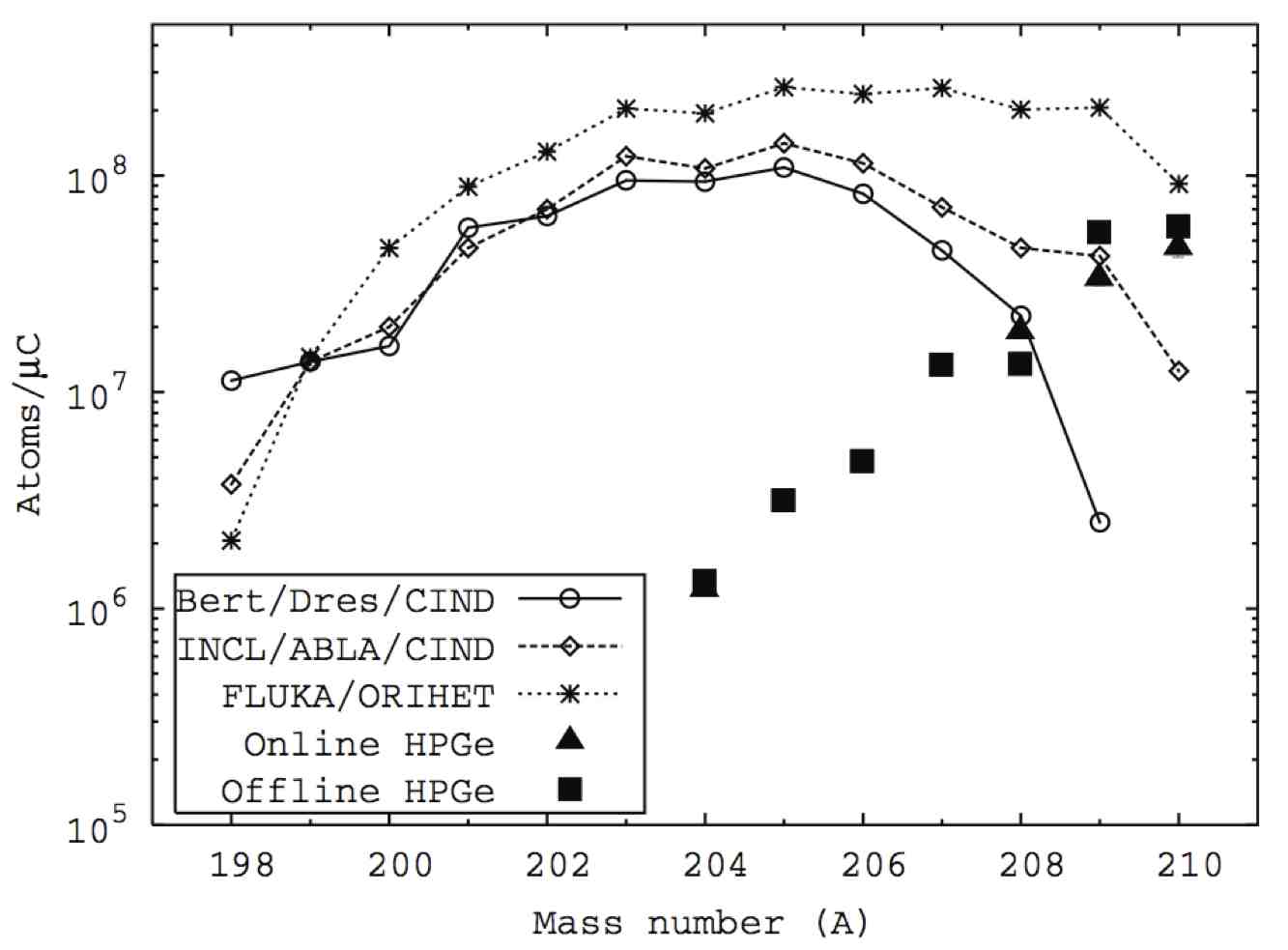}}
\end{center}
\caption{\label{Fig_astate2007}
Release rate of astatine isotopes from the ISOLDE LBE target at 1.4 GeV measured by \cite{TAL08} and compared with the results of different transport codes. Figure taken from \cite{TAL08}. \href{http://nd2007.edpsciences.org/articles/ndata/abs/2007/01/ndata07762/ndata07762.html}{(Courtesy of L. Zanini).}}
\end{figure}

Specific studies, related to MEGAPIE, on the release of volatile species in a liquid lead-bismuth spallation target were carried out at ISOLDE (CERN): the IS419 experiment (2004-2005). The target was a cylinder (20 cm long and 1 cm radius) bombarded with protons of 1.0 and 1.4 GeV. A first publication \cite{TAL08} showed the production rates vs the mass number measured for noble gases, iodine, mercury, and astatine, as well as the comparisons with calculation results. While the models used reproduce more or less the shapes for noble gases, mercury and iodine (this last one being only partially released, in-target and release rates differ by several orders of magnitude), the case of astatine is different and interesting. Although astatine is produced with a low rate and its isotopes have short half-lives, some of them decay into polonium with highly radiotoxic isotopes, and so the production of astatine becomes a safety issue, especially because of the higher volatility of astatine compared to polonium at the MEGAPIE temperatures. Unfortunately no models were able to match the experimental data, especially the shape of the mass distribution, as shown in Fig.~\ref{Fig_astate2007}. 
The difficulty to simulate astatine production from a lead-bismuth target comes first of all from the two production channels. The first one is related to the emission of one and only one $\pi^-$ in the interaction between the proton beam and a bismuth nucleus, $^{209}$Bi(p, $\pi^-$xn)$^{210-x}$At, and the second one brings into play the secondary helium nuclei, mostly the $\alpha$, via  $^{209}$Bi($^3$He,xn)$^{212-x}$At and $^{209}$Bi($\alpha$,xn)$^{213-x}$At, and with an incident energy below around 100 MeV \cite{RAM59,LAM85a,LAM85b,BAR74,DEC74,KEL49,PAT99,RIZ90,STI74,SIN94,HER05,RAT92}. In \cite{DAV13} new calculations are reported with the version INCL4.6. The new version gives good results for the elementary reactions responsible for the astatine production, but the mass distributions measured at ISOLDE in the liquid lead-bismuth target can be reproduced only if a diffusion time is applied. In this study this characteristic time is the time of the release of half the produced quantity. Only an estimate of the order of magnitude of the correction is looked for in \cite{DAV13}, thus a simple model is used and no specific effects are taken into account (e.g. geometrical effect, effect of depletion of the target). The diffusion time can be seen as an additional decay process in the Bateman equation. Let $\Phi$ be the flux of incoming beam intercepting $N_0$ target nuclei, $\sigma$ the average production cross section of a given isotope, and $T_{1/2}$ and $T_d$ respectively the decay and diffusion times. The number of nuclei at a given time inside the target is called $N_{in}(t)$ 
\begin{equation}
\frac {dN_{in}} {dt}=\Phi \sigma N_0- \frac{\ln 2}{T_{1/2}} N_{in} - \frac{\ln 2}{T_d} N_{in}, \nonumber
\label{in}
\end{equation}
and $N_{out}(t)$ is the number of nuclei that escaped from the target up to time t
\begin{equation}
\frac {dN_{out}} {dt}= \frac{\ln 2}{T_d} N_{in}. \nonumber
\label{out}
\end{equation}
Considering that the measurements integrate a short duration compared to $T_{1/2}$ and $T_d$, the measured yields per incident proton, $Y_{out}$,  is then
\begin{equation}
Y_{out}\approx \frac{\ln 2 \: \sigma N_0}{\lambda T_d} [1-{e^{-\lambda t_{irr}}}], \nonumber
\label {Yout}
\end{equation}
with $\lambda = \ln2 \:  (\frac{1}{T_d} +\frac{1}{T_{1/2}})$ and $t_{irr}$ the irradiation time. When the diffusion time is small compared to the decay time, $\lambda{T_d}$ becomes equal to $\ln2$, this is the case of instantaneous release. Whereas, when the decay time is small compared to the diffusion time, $\lambda{T_d}$ becomes equal to $\frac{T_d}{T_{1/2}}\ln2$ and the time to escape must be taken into account. More details are given in \cite{DAV13}.

\begin{figure}[hbt]
\begin{center}
\resizebox{.38\textwidth}{!}{
\includegraphics[trim=2.1cm 2.5cm 2.1cm 2.5cm, clip]{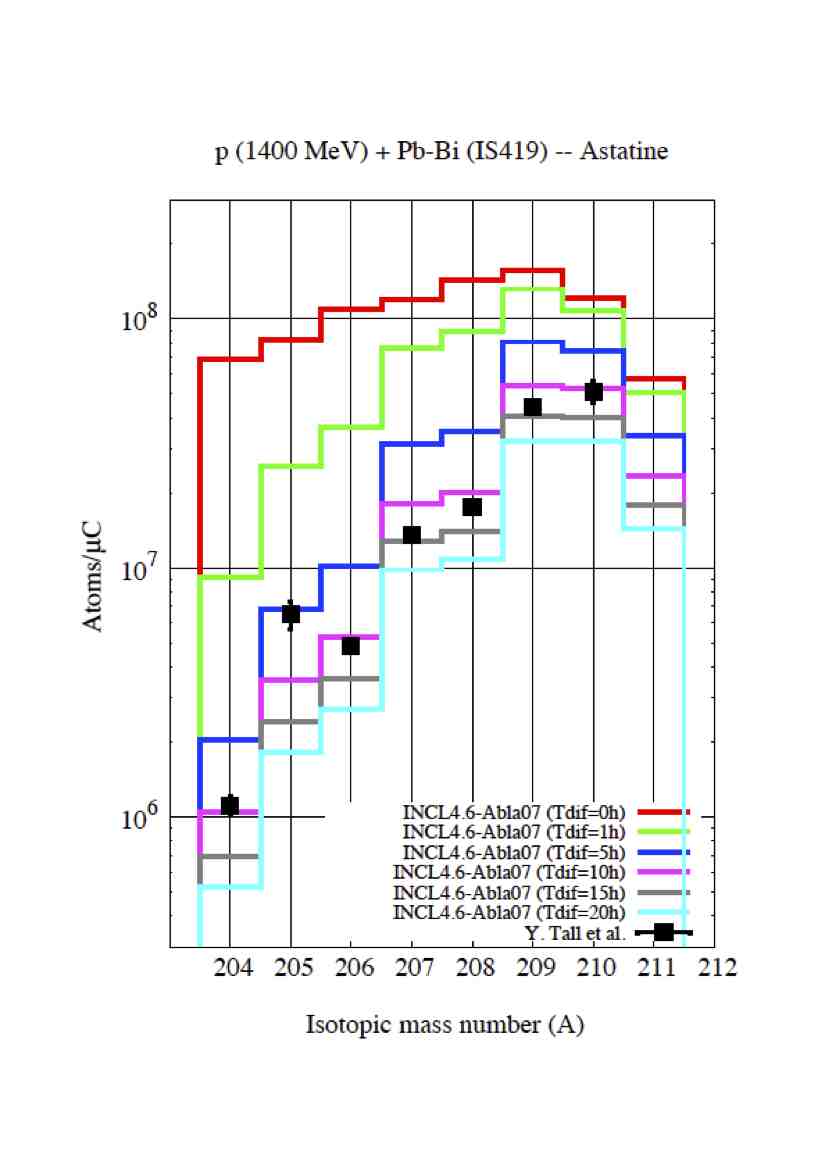}}
\end{center}
\caption{\label{Fig_astatenew-rate}
Astatine release rates measured by the ISOLDE collaboration \cite{ZAN14} at 1.4 GeV compared to MCNPX simulations, using INCL4.6-ABLA07. The solid red line is the yield given by MCNPX, the other lines are corrected to take into account the diffusion and decay times  (as explained in the text). Figure drawn from \cite{DAV13}. \href{http://link.springer.com/article/10.1140/epja/i2013-13029-4} {With kind permission of The European Physical Journal (EPJ).}}
\end{figure}
Fig. \ref{Fig_astatenew-rate} displays the results obtained with a proton beam of 1.4 GeV for various diffusion times and this seems to indicate that a time of around 10 h is required. The same value was found with the 1.0 GeV beam and a similar approach allows to match much better the experimental data for mercury when using a diffusion time of 10 min, which is what has been measured \cite{ZAN14}. The two main conclusions on this study are maybe that, first, the extension towards low incident energies, i.e. below $\sim$200 MeV, can be well performed for the nucleons and also for the composite particles and, secondly, a reliable spallation model can be used, not only to predict the common cross-sections, but also for {\it exotic} studies like the assessment of the release time for an element in a liquid target. Moreover, coming back to Fig. \ref{Fig_astate2007}, it would be interesting to know the results of the other models, and especially FLUKA, when applying this diffusion time. Finally it is worth noting that the experimental points in \cite{TAL08} and on Fig. \ref{Fig_astate2007} were preliminary and so differ a little from those of Fig. \ref{Fig_astatenew-rate} published in \cite{ZAN14}.

\subsubsection{Feasibility and uncertainties - ESS}
\label{tooless}

\begin{figure}[hbt]
\begin{center}
\resizebox{.4\textwidth}{!}{
\includegraphics{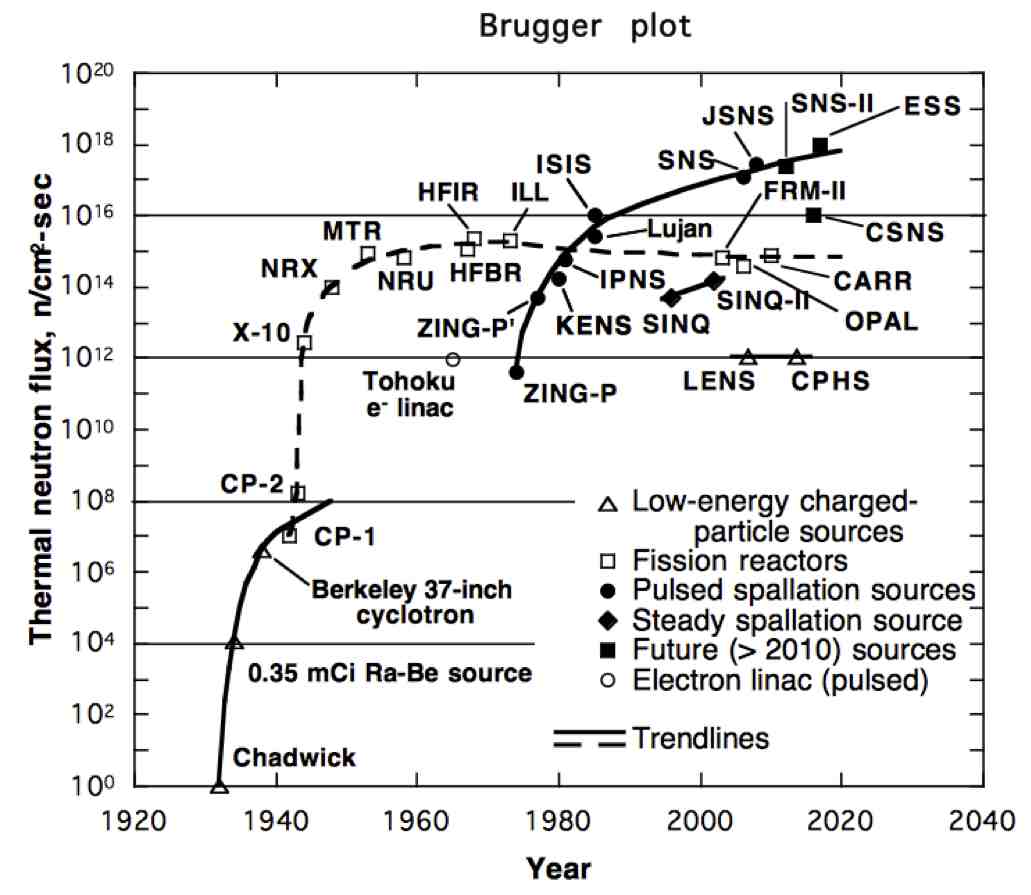}}
\end{center}
\caption{\label{Fig_brugger-plot}
Flux available at the neutron source facilities from the first ones up to ESS. Figure taken from \cite{CAR13}.}
\end{figure}

The ESS (European Spallation source) is a European project involving seventeen countries. The ESS (European Spallation source) is a European project involving seventeen countries. The facility will be built in Lund (Sweden and Denmark are the host nations) and should be operational around 2020.. ESS will join then the last two recent spallation sources SNS \cite{SNS} in USA and JSNS \cite{JSNS} in Japan, and will become the brightest neutron source (Fig.~\ref{Fig_brugger-plot}).
A step before building the facility was the careful study of the Technical Design \cite{ESSTDR}, which aimed at demonstrating the feasibility of the project. Regarding the spallation domain, one of the topics was the assessment of radionuclides produced in the target, linked to the safety issues. The goal was twofold, first the estimate with the most recent computational tools to decide on the feasibility, and second an attempt to characterize the reliability of the calculation results. The details of this study are reported in \cite{LEP14}, and we give below only the main points of interest.

\begin{figure}[hbt]
\begin{center}
\resizebox{.5\textwidth}{!}{
\includegraphics{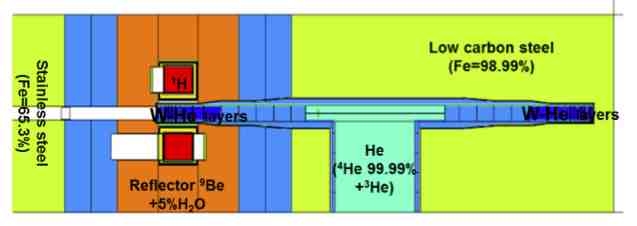}}
\end{center}
\caption{\label{Fig_ess-target}
Drawing of a MCNPX geometry of the ESS tungsten target. The 2.5 GeV proton beam comes from the left and the target is a wheel made of 11 tungsten layers of various thicknesses, surrounded by 2 mm of helium for cooling. Figure from \cite{LEP14}.}
\end{figure}
\begin{figure*}[hbt]
\begin{minipage}{17cm}
\begin{center}
\resizebox{.75\textwidth}{!}{
\includegraphics{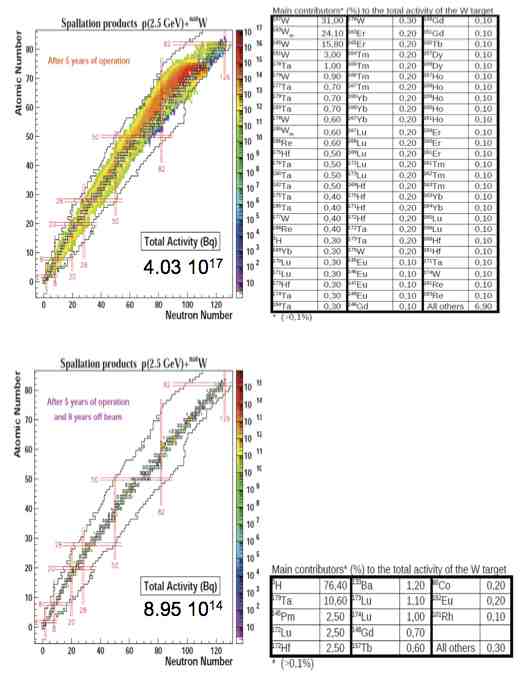}}
\end{center}
\caption{\label{Fig_ess-activities}
Radiation activities (Bq) of the ESS tungsten target bombarded with a 2.5 GeV proton beam delivering 5 MW on a chart of nuclides (on the left) and tables of the main contributors (on the right). Calculations were done with INCL4.6-Abla07 in MCNPX and CINDER'90. Two scenarios: on the top at the shut-down after five years of operation and on the bottom for five years of operation and after nine years of cooling. Figure drawn from \cite{LEP14}.}
\end{minipage}
\end{figure*}

Radiation activities in the target (schematic view in Fig.~\ref{Fig_ess-target}) were computed with two spallation models in the transport code MCNPX2.7 \cite{MCN11}, CEM03 and INCL4.6-Abla07 (private implementation for this latter), and with the activation code CINDER'90 \cite{WIL98}. Fig. \ref{Fig_ess-activities} shows the activity in the target of each isotope as well as the main contributors to the total radiation activity. Five years of irradiation are considered and two cooling times are displayed on this figure: at the shut-down (no cooling time) on the top and after nine years of cooling on the bottom. Obviously the same conclusions as the ones for the mercury converter of EURISOL (section \ref{tooleurisol}) and for the MEGAPIE target (section \ref{toolmegapie}) can be drawn here. The level of radiation is very high (4.03 $10^{17}$ Bq at the shut-down), the main contributors change with the cooling time (near-target radionuclides at the shut-down and  a particular role for the tritium later) and also the non-negligible role of some isomer states, which is not so well known. Therefore radiation activity, volume of gases and production of other hazardous nuclei, in case of accident, produced in the target can be computed and their amounts can be used to define the solutions regarding safety. However, one more piece of information is needed to get the {\it best} solutions: what is the confidence level related to these values?

The second part was then the estimate of the uncertainties. Several methods exist to assess the uncertainties related to a Monte-Carlo calculations. Three of them are the comparison of two different models, the use of experimental data and a sensitivity analysis. The latter, which needs to run numerous calculations where the main ingredients of the model vary within their own uncertainties around their nominal value, is time consuming and focuses on the precision of the model, not on its accuracy, and so only the two first ones have been used in this study. Actually, the comparison of two codes is the very first criterion of whether the results are reliable or not. The comparison between CEM03 and INCL4.6-Abla07 has been done for the total activity in the tungsten target at different cooling times (Fig. 83 in \cite{LEP14}). From the shut-down to a period of up to nine years of cooling the models agree within a factor of 2, and with only 14\% of difference at the end of irradiation. The variation comes from the most important contributors, which change during the cooling. Another comparison regarding tritium in the target also showed a factor about 2, a result that was expected from Fig. 4 of \cite{LER10}. The main effort in this study was indeed dedicated to the second method, based on experimental data. The activity at a given cooling time is due to various numbers of main contributors as shown in Fig.~\ref{Fig_ess-activities}. Thus the uncertainty estimate of the activity means determining the production yields uncertainties of the main contributors. Those latter uncertainties come from the uncertainties of the projectile fluxes and of the elementary production cross-sections. The projectiles being principally nucleons, the uncertainties of the calculated fluxes are assumed to be negligible compared to the uncertainties related to the isotope production cross-sections (see e.g. section \ref{benchiaea}), and the efforts were focussed on those latter ones. 

The availability of experimental data according to the energy domains, concerning the isotope production cross-sections, forces splitting the case of neutrons into three energy parts: below 25 MeV, between 25 MeV and 200 MeV and above 200 MeV. The first domain concerns the low-energy neutrons, handled by the activation code CINDER'90, which uses a library of evaluated cross-sections. Uncertainties related to this library are assumed to be negligible. The intermediate energies are the most tricky domain, because very few experimental data exist (see e.g. section \ref{toolcosmogeny}, page \pageref{toolcosmogeny}). For the last domain, above 200 MeV, the production from a neutron is supposed to be the same as the one from a proton, which allow to use the experimental data obtained for proton-induced reactions (even if it is not always the case as stated in section \ref{toolcosmogeny}). Moreover, if there is more data with protons, in some cases the target nucleus is not exactly the one in question, but close in mass and charge. Obviously this substitute can only be used for isotopes far from the target, where the difference is washed out. Other details are reported in \cite{LEP14}, like the energy range of the experimental data, which reduces the validity of the method, but which is evaluated. Actually this method allows correction of the activity value thanks to measured production cross-sections, rather than determining an uncertainty. Finally the corrected total activity obtained after nine years of cooling was 7.95 10$^{14}$ Bq (8.95 10$^{14}$ Bq with no correction), that is a correction of {\it only} $\sim$ 12\%. This is due to the balance between overestimation and underestimate of the different main contributors. This last point raises the question of the definition of uncertainty. According to the definitions from the Appendix B of GUM \cite{GUM} for a measurement, but applicable to a calculation, of {\it uncertainty} ("parameter, associated with the result of a measurement, that characterizes the dispersion of the values that could reasonably be attributed to the measurand") and {\it error} ("result of a measurement minus a true value of the measurand"), the study developed in \cite{LEP14} is dedicated to the reduction of the error, which is not exactly the estimate of the uncertainty. Nevertheless, whatever the method or the tool used, reduction of the error or assessment of uncertainties, the spallation model developers (and users?) should go beyond the usual validations and should develop this field. For this reason the study developed in \cite{LEP14} must be considered as preliminary and only as one possible way to increase the confidence level of the computational results.

\subsection{Conclusions on the use of spallation models}
\label{conctool}






Developed and improved for several decades, spallation modeling is now reliable enough to be used in numerous applications. The calculated elementary cross-sections can be used to build up databases. Those databases, according to their contents, can be used either in specific purposes as, for example, the meteorites' study via the cosmogenic nuclides, or in particle transport codes, when all the needed information is available. The amount of work is very different in the two cases and the goals also. The main advantage of a database in a transport code is the gain in computation time, compared to the use of a model, whereas the specific databases are most of the time built to fill a lack of experimental data. However, it is worth mentioning also that a database can benefit from the common use of several models and experiments to get the best results in terms of range (projectile, energy, ...) and quality. The other aspect of application is the direct use of spallation models in particle transport codes. In complex geometries models are invoked to give the best results in all types of configuration (projectile, energy, target). These topics have been illustrated through several studies dedicated to three {\it facilities}, Eurisol, MEGAPIE and ESS, and two main conclusions can be drawn. The first one deals with the capabilities of the models. The reliability of most of the spallation models is often high enough for the standard cases and the next steps for the modeling are the particular reactions involving and depending strongly on particular channels. The second one is dedicated to the users. A calculated (and measured) value must be given with its uncertainty. Although validations are one possibility to provide such information, careful estimate of uncertainties will become necessary to make the simulation more and more reliable.

\section{New frontiers}
\label{front}

Modeling of spallation reactions has reached such a level of reliability that the main mechanisms can be considered known and controlled, and makes spallation codes good simulation tools. However some efforts can be made to further improve the modeling, whose future will deal with the details and the extension of the capabilities of the codes. Regarding the details some examples have been mentioned in the previous sections, such as the emisson of Intermediate Mass Fragments, the metastable states or the one-nucleon removal. IMF emission is reasonably well reproduced by some models as far as production cross-sections are concerned, but which could raise problems with the double differential cross-sections \cite{DAV11}. The yield of the metastable nuclei is not always treated or carefully treated, and the one-nucleon removal, where very few channels are open, is badly reproduced by all spallation models, even if a recent study \cite{MAN15} attempted to understand the reasons and gave encouraging results. Some extensions have already been discussed, such as the cluster emission during the INC phase, requiring an ad-hoc coalescence model, or the modeling of reactions with apparently too low-energy projectiles, where surprisingly good results were obtained thanks to careful attention to the needed mechanisms. Another extension can be cited: the heavy ion collision. Basically QMD (or BUU) models are more suitable to simulate heavy-ion reactions than INC models. However, in some cases, an INC model can be competitive with QMD models in simulating light-ion-induced reactions \cite{MAN14}.
In the following subsections, two other topics are considered : the extension to higher energies and the implementation of another degree of freedom, strangeness.


\subsection{Higher energies}
\label{spalhe}

\begin{figure}[hbt]
\begin{center}
\resizebox{.35\textwidth}{!}{
\includegraphics[trim=0cm 0cm 0cm 0cm, clip]{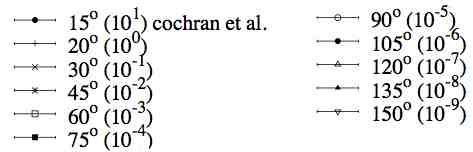}}
\resizebox{.5\textwidth}{!}{
\includegraphics[trim=0cm 0.3cm 0cm 0.5cm, clip]{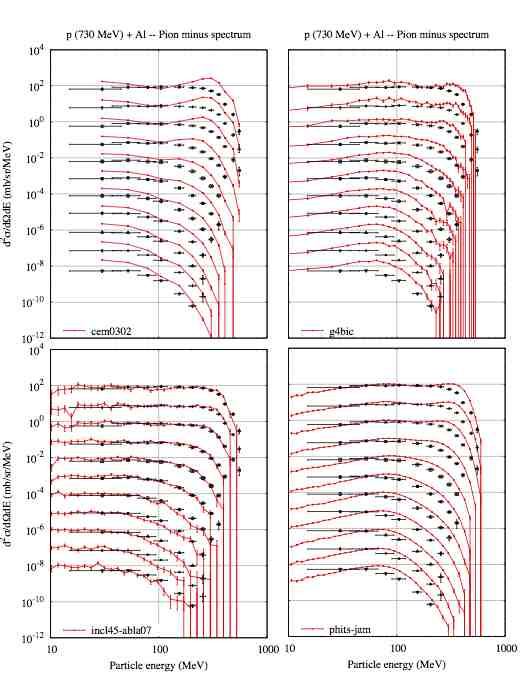}}
\end{center}
\caption{\label{Fig_pion730}
$\pi^-$ double differential cross-sections from the reaction p(730 MeV)+Al. Four models are plotted: CEM03.02 (top left), Binary cascade model from GEANT4 (top right), INCL4.5 (bottom left) and JAM from PHITS (bottom right).  This figure is taken from the website of the IAEA {\it Benchmark on Spallation Models}. Information on the models are given in section \ref{benchiaea}}
\end{figure}

Up to now the modeling of the spallation reactions has been focused on the GeV range. At least two reasons can be put forward. The first one regards the studies on the transmutation of nuclear wastes, which were based on ADS, where the added neutron flux comes from spallation reactions. The optimized proton energy, in terms of neutron multiplicity and cost, was between $\sim$500 MeV and 2-3 GeV (Fig. 16 in \cite{LET00}). The second one is due to galactic cosmic rays' spectra peak between $\sim$100 MeV and $\sim$1 GeV. At these energies only the $\Delta$(1232) resonance seems necessary and, if the results are different according to the models, the production and decay of this resonance are quite well handled. An illustration of this is the pion production and Fig. \ref{Fig_pion730} gives an example of the capability of some models to reproduce $\pi^-$ double differential cross-sections in the collisions p(730 MeV)+Al. The pion production is obviously also driven by other ingredients, in particular the pion potential.
Nevertheless the spallation reaction modeling is also usable with higher energies through the consideration of new open channels. There are two ways to take into account those new channels: either via the production and decay of other resonances or via the direct production of their decay products. A priori the use of all the resonances is more reliable, since that is the necessary step before getting the decay products, which are mainly pions and, in a much smaller amount, kaons. However, the other side of the coin is the needed information related to those resonances : masses, widths, half-lives, decay products, branching ratios, in-medium effects.

\begin{figure}[hbt]
\begin{center}
\resizebox{.5\textwidth}{!}{
\includegraphics{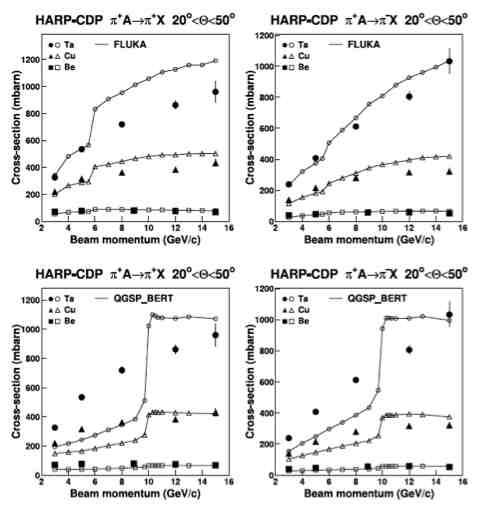}}
\end{center}
\caption{\label{Fig_fluka-geant-harp1}
Comparison of measured inclusive $\pi$ production cross-sections by protons on Be (squares), Cu (triangles) and Ta (circles) with FLUKA on top and GEANT4 employing the QGSP\_BERT below in the intermediate-angle region, as a function of beam momentum. The left panels deal with $\pi^+$ and the right panels with $\pi^-$. The HARP-CDP data are shown with black symbols and the simulations with open symbols. These figures are taken from \cite{BOL10} (Courtesy of J. Wotschack). \href{http://link.springer.com/article/10.1140/epjc/s10052-010-1486-0} {With kind permission of The European Physical Journal (EPJ).}}
\end{figure}

\begin{figure}[hbt]
\begin{center}
\resizebox{.5\textwidth}{!}{
\includegraphics{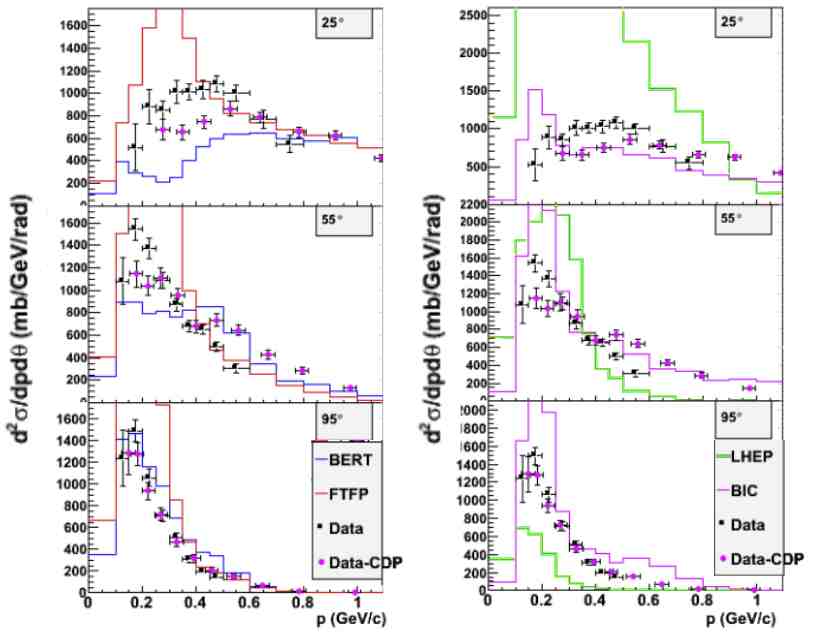}}
\end{center}
\caption{\label{Fig_geant4-harp}
Comparison of measured inclusive $\pi^+$ production cross-sections by 3 GeV/c $\pi^+$ on Pb with GEANT4 employing the bertini cascade (BERT) in blue (left part) and the binary cascade (BIC) in pink (right part) at three angles (25\degree, 55{\degree} and 95\degree) as a function of beam momentum. The two other models plotted (FTFP and LHEP) are dedicated to higher energy beams and are not discussed here. Experimental data are two sets from the HARP experiment \cite{BOL10,CAT08}. Figures are taken from the GEANT4 website.}
\end{figure}

There is unfortunately no study on the reliability of the spallation models at energies above 2-3 GeV, even if one can mention some comparisons between two or three models, mainly using the HARP data, like in \cite{BOL10} testing FLUKA and GEANT4 (bertini cascade) and secondly in the validation pages of the GEANT4 website. From the first study (Fig. \ref{Fig_fluka-geant-harp1})  it is clear that some models are able to reproduce the pion emission reasonably well, but with strong differences according to the target, the type of pions and the projectile energy. Concerning the last point, it is worthwhile to specify that the particle transport codes used different reaction models according to the energy and, for example here, FLUKA uses a spallation model up to 5 GeV (PEANUT) and above a Glauber-Gribov multiple scattering, while in GEANT4 the physics list QGSP\_BERT uses the bertini cascade below 10 GeV and a quark gluon string model above. This explains the sometimes strong discontinuities with energy.
The second case displays the differences between two spallation models available in GEANT4 with a positive pion of (only) 3~GeV/c (Fig. \ref{Fig_geant4-harp}). The bertini model underestimates the pion production with momenta around 300~MeV/c at forward angles, whereas the binary cascade model overestimates it around 200 MeV/c at all angles. On these figures two sets of data of the same HARP experiment are plotted. The reasons for these two published sets and the differences are explained in \cite{BOL09}. Fortunately this doubt on the experimental data does not affect the conclusions on the present modeling. 

\begin{figure}[hbt]
\begin{center}
\resizebox{.5\textwidth}{!}{
\includegraphics{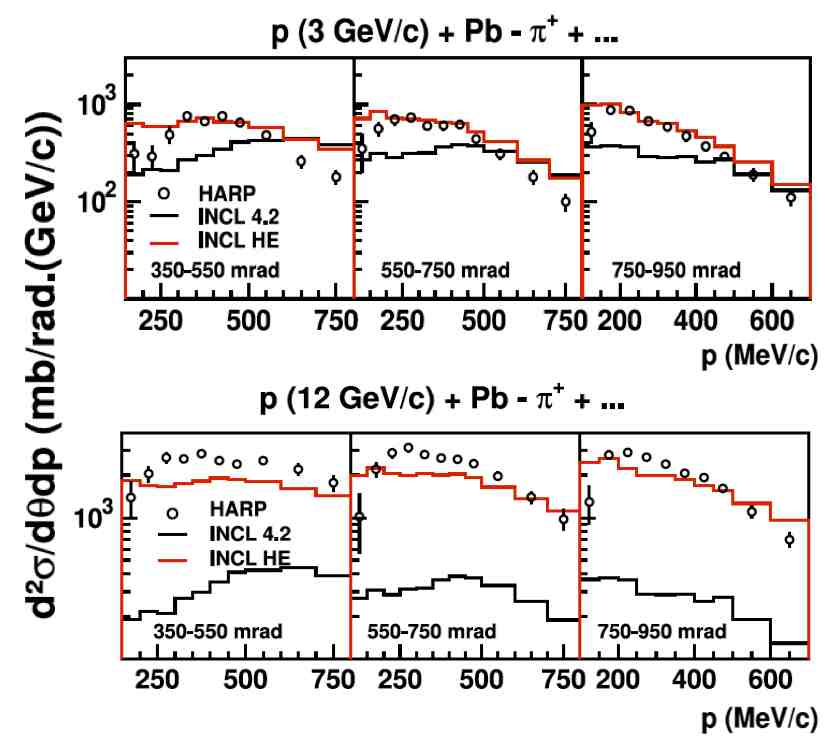}}
\end{center}
\caption{\label{Fig_pedoux}
Double differential-cross sections for the production of $\pi^+$ in proton-induced reactions on Pb at
3 GeV/c (upper part) and at 12 GeV/c (lower part) incident momentum. The multiple pion version of INCL (red line), based on INCL4.2, is compared to the INCL4.2 version (black line) and to the HARP data \cite{CAT08,CAT07a,CAT07b}. Figures taken from \cite{PED11} (Courtesy of J. Cugnon). \href{http://www.sciencedirect.com/science/article/pii/S0375947411005483} {Copyright {\copyright} 2014 Elsevier B.V.}}
\end{figure}
An example of calculation results obtained with a spallation model using the direct multiple pion production instead of the use of the resonances is given in Fig. \ref{Fig_pedoux}. This multiple pion channel was first implemented in INCL4.2 and described in \cite{PED11} and this work has been recently put in the INCL++ version. If the new channels are more than needed at 12 GeV/c, the effects are already clear at 3 GeV/c. Some improvements can still be done, but mainly to treat details or specific domains, such as perhaps the low-energy pions at forward angles. Comparing with the results obtained with other models using the resonances (e.g. FLUKA and binary cascade model in GEANT4), the direct pion production seems justified.
Nevertheless the reliability at energies above $\sim$3 GeV of the spallation models, on pion production, but also other particles and residual nuclei, would deserve a careful study. For example, the calculation results presented here on pion production are rather good, but not perfect and some questions arise: are all the ingredients related to the resonances well-known? Are all the necessary resonances included? The direct multiple pion production  is justified by the large decay widths of the nucleonic resonances, but what about mesonic particles like $\eta$ and $\omega$ with sharp decay widths and which decay into two or three pions? Is the pion absorption on a pair of nucleons or via the $\Delta(1232)$ recombination ($\pi N \rightarrow \Delta$; $\Delta N \rightarrow NN$) well handled?

\subsection{Strangeness}
\label{spalstrange}

Strangeness in the nucleus and obviously hypernuclei can be studied with different goals and achieved by various means. Concerning the goals one can cite the binding energy of the $\Lambda$ (the lightest hyperon and most likely to form a hypernucleus), the value of its potential well, its decay modes and in particular by weak interaction, and more generally the YN and YY interactions, the production rates of the hypernuclei, their lifetimes and decay modes. Four studies using four types of projectile in the energy range of the spallation reactions are mentioned hereafter.

In 1968, A. Lou and D.T. Goodhead studied the production of heavy hypernuclei with a 800 MeV/c $K^-$ beam \cite{LOU68}. They used the simplified cascade model, as one of them, Goodhead, already did for a similar study with 6 GeV/c $K^-$ \cite{EVA67}, and compared their results with experiments. Overall comparisons were encouraging and an analysis of the contributions of the different stages ($K^-$ capture, cascade and evaporation) shows the importance of the cascade. Unfortunately the model was at that time oversimplified. They also tried to extract some information, as the binding energy of $\Lambda$ from the binding energy distribution of the less bound proton, but this latter was too broad and entailed too much uncertainty. The use of $K^+$ beam was much less common, mainly because the $K^+$N interaction is weaker and repulsive. However, this property may be used to probe deeper into the nucleus.

Later, in 1990, a study on the strangeness production in antiproton annihilation on nuclei and based on an precursor of INCL is published \cite{CUG90}. The antiproton beams went up to 4 GeV/c and several targets were considered (from carbon to uranium). The authors compared their calculation results with the available data (particle multiplicities and spectra) and made a detailed analysis. Thus, for example, they showed the importance of secondary reactions in the nucleus   and the various contributions to particle formation (Table IV and V in \cite{CUG90}). In particular the roles of the $\eta$ and $\omega$ were observed and so it is clear that, in addition to the strange particles (kaons, hyperons), other particles are required to correctly model the reaction mechanism. It is worth mentioning here that no resonance was used in this cascade, the outcomes of the decay were directly produced.

In the late 1990s the COSY-13 Collaboration \cite{PYS99} carried out experiments with proton beams with energies around 1-2 GeV on three targets (bismuth, uranium and gold). One of the main objectives was the measurement of the lifetime of the heavy hypernuclei. Within this collaboration Z. Rudy performed calculations \cite{RUD99}  using a BUU model extended to strangeness \cite{WOL93} and showed calculation-measurement comparisons both for the emission spectra of $K^+$, used to estimate the elastic cross-section $K^+N$, and for the delayed fission  cross-sections in the case of two reactions, p(1.5 GeV) + $^{238}$U and p(1.9~GeV) + $^{209}$Bi. In the latter case, where the results were satisfactory, a particular and interesting process is involved: the non-mesonic weak decay of the $\Lambda$ ($\Lambda N \rightarrow NN$). This decay becomes important in the nucleus, since the mesonic weak decay ($\pi N$) is strongly reduced due to the Pauli blocking.

Finally there is the electromagnetic production of strangeness. In their article,  Y. Song and coworkers \cite{SON10} presented their measurements of $^{209}$Bi photofission associated with kaon, pion and proton emission in the forward direction, made at Jefferson Laboratory, and also the comparisons with calculation results obtained using the MCMC (Monte Carlo MultiCollisional) code \cite{GON97}. In the GeV range the photon wavelength is of the order of the nucleon size and, therefore, the individual interactions predominate, which enables the use of an intranuclear cascade model, as in MCMC. Despite the use of assumptions for justifying the non-use of a de-excitation code, the results obtained were correct (e.g. the measured fission probability associated with a kaon was 18\% and the calculations gave 15.4\%). In MCMC extended to the strangeness \cite{PIN98}, the authors use a parameterization for elementary electromagnetic strangeness production based on an article written in 1994 \cite{BOC94}. Since then numerous new data were obtained (e.g. \cite{GLA04}) and progress has been made in the modeling of these elementary interactions \cite{JUL06}. So some improvements might easily be made.

These examples illustrate, over recent decades, the potential capabilities of the spallation models into the strangeness and hypernucleus study. Among the spallation models already cited several times in the previous sections, some of them already include the kaon particles for example, but unfortunately no specific study can be found in the literature. Since various experiments dedicated to hypernuclei are in progress and/or planned in several places and with different types of projectiles (J-PARC (kaon), GSI/FAIR (heavy ion and antiproton), JLab (e$^-$)), the strangeness degree of freedom in spallation modeling should be an interesting and exciting topic.






\section{Conclusion}

In this paper an overview of the spallation reactions, from modeling to applications, is presented. The choices of the topics and examples were driven by the goals to treat the main aspects, to understand how the knowledge and the use of spallation reactions evolved in recent decades, to make an assessment of the reliability of the modeling and to suggest new studies.

Modeling of the spallation reactions is now able to describe the nuclear reactions between a nucleon, or a light particle, and an atomic nucleus in a projectile energy range going from a few tens of MeV up to several GeV, with cross-sections extending over several orders of magnitude. Moreover, the computation time is short enough to allow simulations of complex geometries via particle transport codes, where the models are implemented. This is a good example of the contribution of basics physics, here nuclear physics, to numerous applications where simulations are mandatory. Those applications are as varied as preventing damage to electronic equipment on spacecraft and protection of astronauts, meteorites' study, neutron production to probe the matter or the materials, the transmutation of nuclear wastes, etc. For an example of the impact of the knowledge of the microscopic world on macroscopic effects one can mention the role of pions on the dose rate. Production and emission of pions play a significant role in the excitation energy contained in the remnant nucleus at the end of the intranuclear cascade, so in the way the hot nucleus will de-excite and in the features of the residual nuclei, which give the dose rate. Now production and emission of pions can only be well handled if the elementary production mechanisms are well described (via resonances and their characteristics or directly), if the absorption processes are understood (via resonances and/or on a pair of nucleons), and if the transport is treated correctly (nuclear potential felt by the pion).

To reach such a level of reliability, spallation modeling has continuously made progress, step by step solving the shortcomings and extending its capabilities. This confidence level has been possible only through numerous validations or benchmarks on elementary experimental data. It is worth mentioning that the use of spallation codes to simulate complex geometries urges the model developers to tackle all necessary details, because this comes down to using the models with many nucleus targets and many types of projectiles with broad incident energy spectra. If the particle transport codes need the most reliable spallation models, in return the models benefit from this implementation.

Whatever the quality of the model, improvements in different directions can be made, of course. Regarding the shortcomings, one can cite the metastable isotope and the intermediate mass fragments. The metastable isotopes are important in safety studies, because the decay of the groud-state and of the isomer(s) can be different. The treatment of the metastable state production is either not covered by the spallation model (more precisely de-excitation models) or handled most of the time using tables. This topic is obviously tricky, since it takes place at the end of the de-excitation and depends of the all previous processes. The intermediate mass fragments are produced during the two steps, intranuclear cascade and de-excitation. While now some models are able to give reasonable production rates, the energy spectra are more difficult to achieve, requiring good modeling in each phase. Some experimental sets are available (e.g. PISA collaboration) and a benchmarking study of the modeling would be valuable. Another way to get a better idea of the reliability of the models is the uncertainties' estimate. A precise definition of {\it uncertainty} should first be given to be able to develop the right methodologies. This topic about the spallation reactions remains to be studied. Finally other two sectors deserve careful attention: extension of the spallation modeling up to 10-15 GeV and addition of the strangeness degree of freedom. Some models already include these abilities, but no study has been published on the overall validity, i.e. on different types of observables, especially concerning the strangeness where results obtained with projectile energies below 10 GeV are very rare.

\begin{acknowledgement}
Although only one author put his name on this paper, it is the work of numerous people which is used and presented here. Acknowledging everybody is impossible, but the names of persons who directly or indirectly are at the origin or had an influence on this article must be mentioned. The first one is definitively J. Cugnon, {\it god-father} of INCL and whose contribution to spallation reactions modeling is and will remain essential. Associated with Joseph, the Saclay-Li\`ege team: A. Boudard, S. Leray and D. Mancusi (for the last members). Special thanks to Y. Yariv who spent two years at Saclay. Concerning the implementation of spallation codes in transport codes, J. Hendricks, M. James and L. Waters of LANL must be thanked for their help, and a thought to D. Prael who explained the structure of the LAHET code for the first implementation of INCL in a transport code.  The IAEA benchmark was a collective project with the contribution of the organizers (G. Mank, A. Mengoni, N. Otsuka, S. Leray, D. Filges, Y. Yariv), but of at least five other people (F. Goldenbaum, R. Michel, F. Gallmeier, S. Mashnik and A. Sierk). By using codes some people incited developers to improve the models again and again, which is a good motivation (D. Ene, L. Zanini, D. Schumann, D. Kiselev, M. Wohlmuther and J. Neuhausen). Obviously all experimentalists should be acknowledged for the data they provide, and among them there are Yu. Titarenko, R. Michel and I. Leya for their great contributions to the excitation functions.
\end{acknowledgement}


%


\begin{thebibliography}{00}
%
%

\bibitem{BRO47}
W. M. Brobeck {\it et al.},  Phys. Rev. {\bf 71}, 449-450 (1947).

\bibitem{KAP47}
J. Kaplan, Phys. Rev. {\bf 72}, 738-748 (1947).

\bibitem{SER47}
R. Serber, Phys. Rev. {\bf 72}, 1114-1115 (1947).

\bibitem{OCO48}
P. R. O'connor and G. T. Seaborg, Phys. Rev. {\bf 74}, 1189-1190 (1948).

\bibitem{SEA48}
G. T. Seaborg and I. Perlman, Rev. Mod. Phys. {\bf 20}, 585-667 (1948).

\bibitem{MAN11}
D. Mancusi {\it et al.}, Rev. Mod. Phys. {\bf 20}, 585-667 (2011).

\bibitem{MAS00}
S. G. Mashnik, arXiv:astro-ph/0008382v1 (2000).

\bibitem{AMM09}
K. Ammon {\it et al.}, Meteoritics \& Planetary Science {\bf 44}, 485-503 (2009).

\bibitem{SINQ}
SINQ - http://www.psi.ch/sinq/

\bibitem{ISIS}
ISIS - http://www.isis.stfc.ac.uk

\bibitem{SNS}
SNS - http://neutrons.ornl.gov

\bibitem{JSNS}
JSNS - http://j-parc.jp/MatLife/en/index.html - M. Futakawa {\it et al.}, Neutron News {\bf 22}, 15-19 (2011).

\bibitem{ESS}
ESS - http://europeanspallationsource.se

\bibitem{CSNS}
CSNS - http://csns.ihep.ac.cn/english/index.htm

\bibitem{FIL09}
D. Filges and F. Goldenbaum, "Handbook of Spallation Research: Theory, Experimentsand Applications", John Wiley \& Sons, 2010.

\bibitem{KOR10}
Yu. A. Korovin {\it et al.},  Nucl. Instrum. Meth. {\bf A 624}, 20-26 (2010).

\bibitem{WAT11}
Y. Watanabe {\it et al.}, J. Korean Phys. Soc. {\bf 59}, 1040, doi: 10.3938/jkps.59.1040, (2011).

\bibitem{AIC91}
J. Aichelin, Phys. Rep. {\bf 202}, 233-360 (1991).

\bibitem{BUN85}
V. E. Bunakov and G. V. Matvejev, Z. Phys. A {\bf 322}, 511-521 (1985).

\bibitem{CUG12}
J. Cugnon, Few-Body Systems {\bf 53}, 143-149 (2012).

\bibitem{UU33}
E. A. Uehling and G. E. Uhlenbeck, Phys. Rev.  {\bf 43}, 552--561 (1933).

\bibitem{MET58a}
N. Metropolis  {\it et al.}, Phys. Rev. {\bf 110}, 185 (1958) .

\bibitem{MET58b}
N. Metropolis  {\it et al.}, Phys. Rev. {\bf 110}, 204 (1958) .

\bibitem{GOL48}
M. L. Goldberger, Phys. Rev. {\bf 74}, 1269 (1948) .

\bibitem{BER63}
H. W. Bertini, Phys. Rev. {\bf 131}, 1801 (1963) .

\bibitem{CUG97}
J. Cugnon  {\it et al.}, Phys. Rev. {\bf C 56}, 2431 (1997) .

\bibitem{BOU02}
A. Boudard {\it et al.}, Phys. Rev. {\bf C 66}, 044615 (2002).

\bibitem{BOU13}
A. Boudard  {\it et al.}, Phys. Rev. C {\bf 87} 014606 (2013).

\bibitem{MAS08}
S. G. Mashnik {\it et al.}, "CEM03.03 and LAQGSM03.03 Event Generators for the MCNP6, MCNPX, and MARS15 Transport Codes", LANL Report LA-UR-08-2931; arXiv:0805.0751v2 [nucl-th] (2008).

\bibitem{GRI66}
J. J. Griffin, Phys. Rev. Lett. {\bf 17}, 478 (1966).

\bibitem{CUG05}
J. Cugnon  and P. Henrotte, Eur. Phys. J. A. {\bf 16} 393-407 (2005).

\bibitem{HAS14}
S. Hashimoto {\it et al.}, Nucl. Instrum. Meth. {\bf B 333}, 27-41 (2014).

\bibitem{WEA73}
A.K. Weaver {\it et al.}, Phys. Med. Biol. 18, 64 (1973).

\bibitem{LER02}
S. Leray {\it et al.}, Phys. Rev. {\bf C 65}, 044621 (2002).

\bibitem{VIL07}
C. Villagrasa-Canton {\it et al.}, Phys. Rev. {\bf C 75}, 044603 (2007).

\bibitem{NAP04}
P. Napolitani {\it et al.}, Phys. Rev. {\bf C 70}, 054607 (2004).

\bibitem{ENQ01}
T. Enqvist {\it et al.},  Nucl. Phys. {\bf A 686}, 481 (2001).	

\bibitem{WEI37}
V. Weisskopf, Phys. Rev. {\bf 52}: 295-303 (1937).

\bibitem{GIL65}
A. Gilbert and A.G.W. Cameron, Can. J. Phys. {\bf 43}: 1446-1496 (1965).

\bibitem{IGN75}
A.V.Ignatyuk {\it et al.}, Sov. J. Nucl. Phys. {\bf 21}, 255 (1975).

\bibitem{SHM82}
K. H. Schmidt {\it et al.}, Zeit. f\"ur Physik A Hadrons and Nuclei, 308, 215-225 (1982).

\bibitem{IGN76}
A.V.Ignatyuk {\it et al.}, Sov. J. Nucl. Phys. {\bf 21}(6), 612 (1976).

\bibitem{TOK81}
J. Toke and W.J. Swiatecki, Nucl. Phys. {\bf A 372}, 141-150 (1981).

\bibitem{GUE88}
C. Guet {\it et al.}, Phys. Lett. {\bf B 205}(4), 427-431 (1988).

\bibitem{SHL91}
S. Shlomo et J. B. Natowitz, Phys. Rev. {\bf C 44}, 2878-2880 (1991).

\bibitem{JUN98}
R. Junghans {\it et al.}, Nucl. Phys. {\bf A 629}, 635-655 (1998).

\bibitem{WEI40}
V. Weisskopf and D. H. Ewing, Phys. Rev. {\bf 57}, 472-485 (1940).

\bibitem{HAU52}
W.Hauser and H. Feshbach, Phys. Rev. {\bf 87}, 366-373 (1952).

\bibitem{HUI62}
J. R. Huizenga and G. Igo, Nucl. Phys. {\bf 29}, 462-473 (1962).

\bibitem{BAS74}
R. Bass, Nucl. Phys. {\bf A 23}, 45-63 (1974).

\bibitem{FIL08}
D. Filges {\it et al.}, Joint ICTP-IAEA Advanced Workshop on Model Codes for Spallation Reactions (2008);\\
www-nds.iaea.org/publications/indc/indc-nds-0530/

\bibitem{ENSDF}
The Evaluated Structure Data File (ENSDF) maintained by the National Nuclear Data Center (NNDC), Brookhaven National Laboratory, http://www.nndc.bnl.gov/.

\bibitem{PHI13}
T. Ogawa {\it et al.}, Nucl. Instrum. Meth. {\bf B 325}, 35 (2014).

\bibitem{SIE86}
A.J. Sierk, Phys. Rev.  {\bf C 33}, 2039-2053 (1986).

\bibitem{ILJ92}
A.S. Iljinov and M.V. Mebel, Nucl. Phys. {\bf A 543}, 517-557 (1992).

\bibitem{JUR05}
B. Jurado {\it et al.}, Nucl. Phys. {\bf A  747}, 14-43 (2005).

\bibitem{BEN98}
J. Benlliure {\it et al.}, Nucl. Phys. {\bf A  628}, 458-478 (1998).

\bibitem{ATC80}
F. Atchison, "Spallation and fission in heavy metal nuclei under medium energy proton bombardment Meeting on Targets for neutron beam spallation sources", Ed. G. Bauer, KFA J\"ulich Germany, J\"ul-conf-34 (1980)

\bibitem{FUR01a}
S. Furihata {\it et al.}, “The Gem Code - A simulation Program for the Evaporation and Fission Process of an Excited Nucleus -” JAERI-Data/Code 2001-015 (2001).

\bibitem{MOR88}
L.G. Moretto and G.J. Wozniak. The Decay of Hot Nuclei. LBL-26207. DE89 J06609 (1988). 

\bibitem{BON76}
J.P. Bondorf, Journal de Physique {\bf 37}, C5 - 195 (1976).

\bibitem{SU88}
R. K. Su {\it et al.}, Phys. Rev. {\bf C  37}, 1770-1773 (1988).

\bibitem{BON95}
J.P. Bondorf {\it et al.}, Phys. Rep. {\bf 257}, 133-221 (1995).

\bibitem{BOT00}
A. S. Botvina {\it et al.}, Phys. Rev.  {\bf E 62}, R64-R67 (2000).

\bibitem{SCH02}
K.-H. Schmidt {\it et al.}, Nucl. Phys. {\bf A 710}, 157-179 (2002).

\bibitem{NAT02}
J. B. Natowitz {\it et al.}, Phys. Rev. {\bf C 65}, 034618 (2002).

\bibitem{ENQ99}
T. Enqvist {\it et al.},  Nucl. Phys. {\bf A 658}, 47-66 (1999).

\bibitem{BUY05}
N. Buyukcizmeci {\it et al.}, Eur. Phys. J. {\bf A 25}, 57 (2005).

\bibitem{SOU07}
G.A. Souliotis {\it et al.}, Phys. Rev. {\bf C 75}, 011601R (2007). 

\bibitem{BOT06}
A.S. Botvina {\it et al.}, Phys. Rev. {\bf C 74}, 044609 (2006). 

\bibitem{FER50}
E. Fermi, Prog. of Theo. Phys. {\bf 5}, 570-583 (1950).

\bibitem{EPH67}
M. \'Epherre and \'E. Gradsztajn, LE JOURNAL DE PHYSIQUE {\bf 28}, 745-751 (1967).

\bibitem{CAR11}
B. V. Carlson {\it et al.}, J. Phys. : Conf. Ser., {\bf 312}, 082017 (2011).

\bibitem{SAL01}
M. Salvatores and \'E. Fort, CLEFS CEA N. 45, pages 22-29 (2001) - In french. 

\bibitem{PRA89}
R.E. Prael. User guide to LCS: the LAHET code system. LA-UR-89-3014 (1989).

\bibitem{PRA01}
R.E. Prael. Release Notes for LAHET Code System with {LAHET}\texttrademark Version 3.16 (2001).

\bibitem{REE94}
R.C. Reedy and J. Masarik, Lunar. Planet. Sci., XXV, 1119-1120 (1994)

\bibitem{MAS99}
J. Masarik and J. Beer, J. Geo. Res. {\bf 104}, 12099-12111 (1999)

\bibitem{ARM72}
T. W. Armstrong and K. C. Chandler, Nucl. Sci. Eng. {\bf 49}, 110 (1972).

\bibitem{CHA72}
K. C. Chandler and T. W. Armstrong. OPERATING INSTRUCTIONS FOR THE HIGH-ENERGY NUCLEON-MESON TRANSPORT CODE HETC. ORNL- 4744 (1972).

\bibitem{GAB85}
T. A. Gabriel. The high energy transport code HETC. ORNL/TM-9727 (1985). 

\bibitem{COL71}
W. A. Coleman and T. W. Armstrong, Nucl. Sci. Eng. {\bf 43}, 353 (1971).  

\bibitem{BRI86}
J. F. Briesmeister. MCNP - A general Monte Carlo Code for Neutron and Photon Transport. LA-7396-M Revision 2 (1986).

\bibitem{DAV03}
J.-C. David {\it et al.}, Spallation Neutron Production on Thick Target at Saturne. Proceedings of the International Workshop on Nuclear Data for the Transmutation of Nuclear Waste - GSI-Darmstadt, Germany - September 1-5, 2003. ISSN 3-00-012276-1.

\bibitem{HEN05}
J. S. Hendricks {\it et al.}, MCNPX EXTENSIONS VERSION 2.5.0. LA-UR-05-2675 (2005).

\bibitem{GEA03}
Geant4 collaboration, Geant4 - a simulation toolkit, Nucl. Instr. Meth. Phys. Res. {\bf A506}, 250-303 (2003).

\bibitem{KAI11}
P. Kaitaniemi {\it et al.}, Prog. Nucl. Sci. Tech. {\bf 2}, 788-793  (2011).

\bibitem{MAN14}
D. Mancusi {\it et al.},  Phys. Rev. {\bf C 90}, 054602 (2014).

\bibitem{SAT13}
T. Sato {\it et al.}, Particle and Heavy Ion Transport code System, PHITS, version 2.52, Journal of
Nuclear Science and Technology {\bf 50}, 913-923 (2013).

\bibitem{GEA87}
B. Brun {\it et al.}, GEANT3 User's guide, Rep. DD/EE/84-1, 584 pp., Eur. Org. for Nucl. Res., Geneva (1987).

\bibitem{BRI93}
J. F. Briesmeister, MCNP - A general Monte Carlo N-particle transport code version 4A, LA-12625-M (1993).

\bibitem{FLU07a}
G. Battistoni {\it et al.}, "The FLUKA code: Description and benchmarking", Proceedings of the Hadronic Shower Simulation Workshop 2006, Fermilab 6--8 September 2006, M. Albrow, R. Raja eds., AIP Conference Proceeding 896, 31-49, (2007).

\bibitem{FLU07b}
A. Ferrari {\it et al.}, "FLUKA: a multi-particle transport code", CERN-2005-10 (2005), INFN/TC$\_$05/11, SLAC-R-773.

\bibitem{MOK04a}
N.V. Mokhov, "The Mars Code System User's Guide", Fermilab-FN-628 (1995).

\bibitem{MOK04b}
O.E. Krivosheev, N.V. Mokhov, "MARS Code Status", Proc. Monte Carlo 2000 Conf., p. 943, Lisbon, October 23-26, 2000, Fermilab-Conf-00/181 (2000).

\bibitem{MOK04c}
N.V. Mokhov, "Status of MARS Code", Fermilab-Conf-03/053 (2003).

\bibitem{MOK04d}
N.V. Mokhov {\it et al.}, "Recent Enhancements to the MARS15 Code", Fermilab-Conf-04/053 (2004); http://www-ap.fnal.gov/MARS/.

\bibitem{PRA88}
R.E. Prael and M. Bozoian, Adaptation of the Multistage Pre-equilibrium Model for the Monte Carlo method. LA-UR-88-238 (1988).

\bibitem{FLU87}
P. A. Aarnio {\it et al.},  Enhancements to the FLUKA86 program (FLUKA87), CERN/TIS-RP/190 (1987).

\bibitem{BAR81}
J. Barish {\it et al.}, HETFIS: High-Energy Nucleon-Meson Transport Code with Fission, ORNL/TM-7882 (1981)

\bibitem{MAS06}
S. G. Mashnik {\it et al.}, CEM03 and LAQGSM03 - new modeling tools for nuclear applications,  Journal of Physics-Conference Series, Institute of Physics, 41(1), pp. 340-351 (2006).

\bibitem{MAS05}
S. G. Mashnik {\it et al.}, CEM03.01 User Manual, LANL Report LA-UR-05-7321 (2005); RSICC Code Package PSR-532; http://www-rsicc.ornl.gov/codes/psr/psr5/psr532.html; http://www.nea.fr/abs/html/psr-0532.html.

\bibitem{FUR00}
S. Furihata, Nucl. Instr. and Meth. in Phys. Res. {\bf B171}, 251 (2000).

\bibitem{JAM00}
Y. Nara {\it et al.},  Phys. Rev. {\bf C 61}, 024901 (2000).

\bibitem{JQM95}
K. Niita {\it et al.}, Phys. Rev. {\bf C 52}, 2620-2635 (1995).

\bibitem{DOS59}
I. Dostrovsky {\it et al.}, Phys. Rev. {\bf 116}, 683 (1959).

\bibitem{GUD83}
K. K. Gudima {\it et al.}, Nucl. Phys. {\bf A 401} 329 (1983).

\bibitem{BOT87}
A. S. Botvina A. S. {\it et al.}, Nucl. Phys. {\bf A 475} 663 (1987).

\bibitem{BLA72}
M. Blann, Phys. Rev. Lett. {\bf 28}, 757 (1972).

\bibitem{BER72}
L. Bertocchi, Nuovo Cimento {\bf 11A}, 45 (1972); and references therein.

\bibitem{SOR95}
H. Sorge, Phys. Rev. {\bf C 52}, 3291 (1995).

\bibitem{RAN95}
J. Ranft, Phys. Rev. {\bf D 51}, 54 (1995).

\bibitem{KOD06}
I. Kodeli {\it et al.}, SINBAD Shielding Benchmark Experiments Status and Planned Activities,  {\it The American Nuclear Society's 14th Biennial Topical Meeting of the Radiation Protection and Shielding Division, Carlsbad New Mexico, USA. April 3-6, 2006}; https://www.oecd-nea.org/science/wprs/shielding/sinbad/.

\bibitem{BLA93}
M. Blann {\it et al.}, International Code Comparison for Intermediate Energy Nuclear Data. NEA/OECD, NSC/DOC(94)-2, Paris, 1993.

\bibitem{MIC97}
R. Michel et P. Nagel. International Codes and Model Intercomparison for Intermediate Energy Activation Yields. NEA/OECD, NSC/DOC(97)-1, Paris, 1997.

\bibitem{FIL95}
D. Filges {\it et al.}, Thick Target Benchmark for Lead and Tungsten. NEA/OECD, NSC/DOC(95) 2, Paris, 1995.

\bibitem{MEU05}
J.-P. Meulders {\it et al.}, High and Intermediate energy Nuclear Data for Accelerator-driven Systems. HINDAS final report, 2005. http://www.theo.phys.ulg.ac.be/$\sim$cugnon/\\
Final\_Scientific\_Report\_HINDAS.pdf 

\bibitem{MEU00}
J.-P. Meulders {\it et al.},  Physical Aspects of Lead as a Neutron Producing Target for Accelerator Transmutation Devices 
Final Report on the Concerted Action: Physical Aspects of Lead as a Neutron Producing Target for Accelerator Transmutation Devices, 2001, European Atomic Energy Commission, Luxembourg, 142p.

\bibitem{NUD11}
E. Gonz{\' a}lez-Romero {\it et al.}, NUDATRA/EUROTRANS nuclear data for nuclear waste transmutation, "Proceedings of the Workshop on Technology and Components of Accelerator-driven Systems", Karlsruhe, Germany, 15-17 March 2010, ISBN 978-92-64-11727-3 \copyright OECD 2011.

\bibitem{AND13}
E. Gonz{a}lez, Final Report of the project ANDES (Accurate Nuclear Data for nuclear Energy Sustainability), http://www.andes-nd.eu/system/files/docs/ANDES\_FinalReport\_v8x.pdf.

\bibitem{LET02}
A. Letourneau {\it et al.}, Nucl. Phys. {\bf A 712} 133 (2002).

\bibitem{BOU03}
A. Boudard {\it et al.}, Comparison of INCL4 with Experiments, "Proceedings of the International Workshop on Nuclear Data for the Transmutation of Nuclear Waste", GSI-Darmstadt, Germany, September 1-5, 2003, ISBN 3-00-012276-1, Editors: Aleksandra Kelic and Karl-Heinz Schmidt

\bibitem{LER02}
S. Leray {\it et al.}, Phys. Rev. {\bf C 65} 044621 (2002).

\bibitem{HER01}
C.M. Herbach {\it et al.}, FZJ Juelich annual report (2001)

\bibitem{HER00}
C.M.Herbach {\it et al.}, Proc. of the SARE-5meeting, OECD, Paris, July 2000

\bibitem{TAI03}
J. Taieb  {\it et al.}, Nucl. Phys. {\bf A 724} 413 (2003).

\bibitem{BER03}
M. Bernas  {\it et al.}, Nucl. Phys. {\bf A 725} 213 (2003).

\bibitem{AOU04}
Th. Aoust and J. Cugnon, Eur. Phys. J. {\bf A 21} 79 (2004).

\bibitem{AOU06}
Th. Aoust and J. Cugnon, Phys. Rev. {\bf C 74} 064607 (2006).

\bibitem{CUG11}
J. Cugnon  {\it et al.}, Journal of Physics: Conference Series {\bf 312} (2011) 082019.

\bibitem{BUB07}
A.Bubak  {\it et al.}, Phys. Rev. {\bf C 76} 014618 (2007).

\bibitem{BUD08}
A.Budzanowski {\it et al.}, Phys. Rev. {\bf C 78} 024603 (2008).

\bibitem{AYY14}
Y. Ayyad  {\it et al.}, Phys. Rev. {\bf C 89} 054610 (2014).

\bibitem{DAV11}
J.-C. David {\it et al.},  Report on the predicting capabilities of the standard simulation tools in the 150-600 MeV energy range, FP7-ANDES, WP4 Deliverable D4.1.

\bibitem{GRE80}
R. E. L. Green and R. G. Korteling, Phys. Rev. {\bf C 22} 1594 (1980).

\bibitem{CUG13}
J. Cugnon {\it et al.},  Report on the validation of the simulation tools developed in Task 4.4 and assessment of the expected reduction of uncertainty on key parameters of the ADS, FP7-ANDES, WP4 Deliverable D4.6.

\bibitem{DAV13}
J.-C. David {\it et al.},  Eur. Phys. J. {\bf A 49} 29 (2013).

\bibitem{MAN15}
D. Mancusi {\it et al.}, Phys. Rev. {\bf C 91} 034602 (2015)

\bibitem{RAP06}
B. Rapp {\it et al.}, "Benchmark calculations on particle production within the EURISOL DS project". EURISOL DS/Task5/TN-06-04, 2006.

\bibitem{DAV07a}
J.-C. David {\it et al.}, "Benchmark calculations on residue production within the EURISOL-DS project ; Part I : thin targets". Internal Rapport, DAPNIA-07-04, 2007.

\bibitem{DAV07b}
J.-C. David {\it et al.}, "Benchmark calculations on residue production within the EURISOL-DS project ; Part II : thick targets". Internal Rapport, DAPNIA-07-59, 2007.

\bibitem{FEL06}
M. Felcini and A. Ferrari, Validation of FLUKA calculated cross-sections for radioisotope production in proton-on-target collisions at proton energies around 1GeV, EURISOL DS/Task 5/TN-06-01.

\bibitem{PIE06}
L. Pienkowski {\it et al.}, FLUKA simulation for NESSI experiment, EURISOL DS/Task5/TN-06-05.

\bibitem{ENE06}
D. Ene, "High-energy neutron attenuation in iron and concrete: Verification of FLUKA Monte Carlo code by comparison with HIMAC experimental results", EURISOL DS/Task5/TN-06-14.

\bibitem{ENE09}
D. Ene {\it et al.}, Layout of the EURISOL post-accelerator, CEA Saclay Internal report: IRFU-09-109 - (2009); EURISOL DS Technical Report 05-25-2009-0036.

\bibitem{EXFOR}
http://www-nds.iaea.org/exfor/; N. Otuka {\it et al.}, Nuclear Data Sheets {\bf 120}, 272-276 (2014).

\bibitem{KON11}
A. Koning and A. Mengoni, Quality Improvement of the EXFOR database, NEA/NSC/WPEC/DOC(2010)428, OECD 2011.

\bibitem{MCN08}
MCNPX User's Manual Version 2.6.0, April 2008, Denise B. Pelowitz, editor, LA-CP-07-1473

\bibitem{PHIXX}
http://phits.jaea.go.jp/Reference.html

\bibitem{YOU92}
P. G. Young {\it et al.}, LA-12343-MS, Los Alamos National Laboratory, 1992.

\bibitem{IWA12}
Y. Iwamoto {\it et al.}, Nucl. Instrum. Meth. {\bf A 690} 10-16 (2012). 

\bibitem{YAS13}
H. Yashima {\it et al.}, Radiation Protection Dosimetry (2013), pp. 1-5, doi: 10.1093/rpd/nct334.

\bibitem{LEY11}
I. Leya and R. Michel,  Nucl. Instrum. Meth. {\bf B 269}, 2487-2503 (2011).

\bibitem{IWA04}
Y. Iwamoto {\it et al.}, Phys. Rev. {\bf C 70} 024602 (2004).

\bibitem{MAS95}
S. G. Mashnik, User Manual for the Code CEM95, (1995), http://www.nea.fr/abs/html/iaea1247.html.

\bibitem{MAS00b}
S. G. Mashnik and A. J. Sierk, Proc. AccApp00, November 12-16, 2000, Washington, DC (USA), ANS, La Grange Park, IL, 2001, pp. 328-341; E-print: nucl-th/0011064.

\bibitem{MCNXX}
https://laws.lanl.gov/vhosts/mcnp.lanl.gov/\\
references.shtml\#mcnp6\_refs\_h

\bibitem{MAS11}
S. Mashnik, "Validation and Verification of MCNP6 Against High-Energy Experimental Data and Calculations by Other Codes. I. The CEM Testing Primer", LA-UR-11-05129 (2011).

\bibitem{MAS11b}
S. Mashnik, "Validation and Verification of MCNP6 Against High-Energy Experimental Data and Calculations by Other Codes. II. The LAQGSM Testing Primer", LA-UR-11-05627 (2011).

\bibitem{BAN11}
S. Banerjee {\it et al.}, Journal of Physics: Conference series {\bf 331}  032034 (2011).

\bibitem{SIH08}
L. Sihver {\it et al.}, Acta Astronautica {\bf 63}  865-877 (2008).

\bibitem{MOK07}
N.V. Mokhov and S.I. Striganov, FERMILAB-CONF-07-009-AD January 2007.

\bibitem{LER11}
S. Leray {\it et al.}, J. Korean Phys. Soc. {\bf 59}, 791, doi: 10.3938/jkps.59.791, (2011).

\bibitem{DAV11b}
J.-C. David {\it et al.}, Prog. Nucl. Sci. Tech. {\bf 2}, 942-947 (2011).

\bibitem{HJO96}
E.L. Hjort {\it et al.}, Phys. Rev. {\bf C 53}, 237 (1996).	

\bibitem{GUE05}
A. Guertin {\it et al.}, Eur. Phys. J.  {\bf A 23}, 49 (2005).	

\bibitem{MEI92}
M.M. Meier {\it et al.}, Nucl. Sci. Eng. {\bf 110}, 289 (1992).

\bibitem{AMI92}
W.B. Amian {\it et al.}, Nucl. Sci. Eng. {\bf 112}, 78 (1992).

\bibitem{ISH97}
K. Ishibashi {\it et al.}, J. Nucl. Sci. Tech.  {\bf 34}, 529 (1997).	

\bibitem{LET00}
A. Letourneau {\it et al.}, Nucl. Instrum. Meth. {\bf B 170}, 299 (2000).

\bibitem{FRA90}
J. Franz {\it et al.}, Nucl. Phys.  {\bf A 510}, 774 (1990).	

\bibitem{BER73}
F.E. Bertrand {\it et al.}, Phys. Rev.  {\bf C 8}, 1045 (1973).

\bibitem{COW96}
A.A. Cowley {\it et al.}, Phys. Rev.  {\bf C 54}, 778 (1996).	

\bibitem{BUD09}
A. Budzanowski {\it et al.}, Phys. Rev.  {\bf C 80}, 054604 (2009).	

\bibitem{FOR91}
S.V. Fortsch {\it et al.}, Phys. Rev.  {\bf C 43}, 691 (1991).	

\bibitem{CHR80}
R.E. Chrien {\it et al.}, Phys. Rev.  {\bf C 21}, 1014 (1980).	

\bibitem{MCG84}
J.A. McGill {\it et al.}, Phys. Rev.  {\bf C 29}, 204 (1984).	

\bibitem{HER06}
C.-M. Herbach {\it et al.}, Nucl. Phys.  {\bf A 765}, 426 (2006).	

\bibitem{COC72}
D.R.F.Cochran {\it et al.}, Phys. Rev. {\bf D 6}, 3085 (1972).

\bibitem{ENY85}
H.Enyo {\it et al.}, Phys. Lett. {\bf B 159}, 1 (1985).

\bibitem{VIL07}
C. Villagrasa-Canton {\it et al.},  Phys. Rev. {\bf C 75}, 044603 (2007).	

\bibitem{NAP04}
P. Napolitani {\it et al.},  Phys. Rev. {\bf C 70}, 054607 (2004).	

\bibitem{AUD06}
L. Audouin {\it et al.},  Nucl. Phys. {\bf A 768}, 1 (2006).	

\bibitem{RIC06}
M.V. Ricciardi {\it et al.}, Phys. Rev. {\bf C 73}, 014607 (2006).	

\bibitem{BER06}
M. Bernas {\it et al.},  Nucl. Phys. {\bf A 765}, 197 (2006).	

\bibitem{MIC95}
R.Michel {\it et al.}, Nucl. Instrum. Meth. {\bf B 103}, 183 (1995).		

\bibitem{SCH96}
Th.Schiekel {\it et al.}, Nucl. Instrum. Meth. {\bf B 114}, 91 (1996).		

\bibitem{MIC02}
R.Michel {\it et al.}, J. Nucl. Sci. Technol. Suppl.2, 242 (2002).		

\bibitem{AMM08}
K.Ammon {\it et al.}, Nucl. Instrum. Meth. {\bf B 266}, 2 (2008).		

\bibitem{TIT08}
Yu.E.Titarenko {\it et al.}, Phys. Rev. {\bf C 78}, 034615 (2008).		

\bibitem{GLO01}
M.Gloris {\it et al.}, Nucl. Instrum. Meth. {\bf A 463}, 593 (2001).		

\bibitem{LEY05}
I.Leya {\it et al.}, Nucl. Instrum. Meth. {\bf B 229}, 1 (2005).		

\bibitem{TIT06}
Yu.E.Titarenko {\it et al.}, Nucl. Instrum. Meth. {\bf A}, 562 (2006).

\bibitem{SCH11}
D. Schumann {\it et al.},  Journal of Physics {\bf G 38}, 065103 (2011).

\bibitem{MIC09}
R. Michel, "The Concept of Intrinsic Discrepancy Applied to the Comparison of Experimental and Theoretical Data for Benchmarking Spallation Models." IAEA - Consultants' Meeting on Spallation Reactions (6-7 October), 2009. https://www-nds.iaea.org/spallations/2009cm/

\bibitem{SAW12}
Y. Sawada {\it et al.}, Nucl. Instrum. Meth. {\bf B 291}, 38 (2012).		

\bibitem{ULM95}
J.L. Ulmann {\it et al.}, "APT PADIONUCLIDE PRODUCTION EXPERIMENT TECHNICAL REPORT", LA-UR-95-3327 (1995).

\bibitem{JAN11}
J. Janczyszyn {\it et al.}, "Benchmark on radionuclides production and heat generation rates in lead target exposed to 660 MeV protons." AGH - University of Science and Technology (Krakow) - EC EUROTRANS-NUDATRA and IAEA CRP Report, 2011.

\bibitem{POH07}
W. Pohorecki {\it et al.}, "Thick lead target exposed to 660 MeV protons : benchmark model on radioactive nuclides production and heat generation, and beyond.", in Proceedings of the International Conference on Nuclear Data for Science and Technology, April 22-27, 2007, Nice, France, editors O.Bersillon, F.Gunsing, E.Bauge, R.Jacqmin, and S.Leray, EDP Sciences, 2008, pp 1225-1228 - DOI: 10.1051/ndata:07745

\bibitem{MCN05}
MCNPX User's Manual Version 2.5.0, April 2005, Denise B. Pelowitz, editor, LA-CP-05-0369

\bibitem{FAS03}
A. Fasso {\it et al.}, "The physics models of FLUKA : status and recent developments". Computing in High Energy and Nuclear Physics, 24-28 March 2003, La Jolla, California, ePrint hep-ph/0306267, 2003.

\bibitem{FER05}
A. Ferrari {\it et al.}, "FLUKA : A multi-particle transport code (Program version 2005)". CERN-2005-010, SLAC- R-773, INFN-TC-05-11, 2005. 

\bibitem{ADA05}
J. Adam {\it et al.}, Eur. Phys. J.  {\bf A 23}, 61-68 (2005). DOI 10.1140/epja/i2004-10031-y

\bibitem{MAJ06}
M. Majerle {\it et al.}, Journal of Physics: Conference Series 41 (2006) 331-339. doi:10.1088/1742-6596/41/1/036

\bibitem{MAJ07}
M. Majerle {\it et al.}, Nucl. Instrum. Meth. {\bf A 580}, 110-113 (2007).

\bibitem{KRA10}
A. Kr\'asa {\it et al.}, Nucl. Instrum. Meth. {\bf A 615}, 70-77 (2010).

\bibitem{OH11}
J. Oh {\it et al.}, Prog. Nucl. Sci. Tech. {\bf 1}, 85-88 (2011).

\bibitem{MEI89}
M. M. Meier {\it et al.},  Nucl. Sci. and Eng. {\bf 102}, 310 (1989).

\bibitem{MEI90}
M. M. Meier {\it et al.},  Nucl. Sci. and Eng. {\bf 104}, 339 (1990).

\bibitem{IWA10}
Y. Iwamoto {\it et al.},  Nucl. Instrum. Meth. {\bf A 620}, 484-489 (2010).

\bibitem{MAJ09}
M. Majerle, PhD Thesis, Czech Technical University in Prague (2009).

\bibitem{MEN98}
S. M\' enard, PhD Thesis, Universit\' e d'Orsay (1998).

\bibitem{VAR99}
C. Varignon, PhD Thesis, Universit\' e de Caen (1999).

\bibitem{MEI99}
S. Meigo {\it et al.},  Nucl. Instrum. Meth. {\bf A 431}, 521-530 (1999).

\bibitem{KON14}
A. YU Konobeyev and U. Fischer, "Reference data for evaluation of gas production cross-sections in proton induced reactions at intermediate energies", KIT scientific reports 7660, doi: 10.5445/KSP/1000038463, (2014).

\bibitem{KON12}
A.J. Koning and D. Rochman, "Modern Nuclear Data Evaluation With The TALYS Code System", Nuclear Data Sheets 113 (2012) 2841; www.talys.eu/tendl-2013.html

\bibitem{BAR99}
V. S. Barashenkov {\it et al.}, Atomnaya Energiya 87, 283 (1999).

\bibitem{ENDF6}
N.M. Larson, Compact Covariance Matrix, in ENDF-102 Data Formats and Procedures for the Evaluated Nuclear Data File ENDF-6, BNL-NCS-4494505-Rev., cross-section Evaluation Working Group, Brookhaven National Laboratory, NY, USA, (2005); https://www.oecd-nea.org/dbdata/data/endf102.htm

\bibitem{WAT04}
Y. Watanabe {\it et al.}, in Proceedings of Inter. Conf. on Nucl. Data for Sci. and Techn. (ND2004), (Santa Fe, USA, 2004), p. 326 (2004).

\bibitem{NII95}
K. Niita {\it et al.},  Phys. Rev. {\bf C 52}, 2620 (1995); JAERIData/Code 99-042 (1999). 	

\bibitem{NAR01}
Y. Nara {\it et al.},  Phys. Rev. {\bf C 61}, 024901 (2001). 	

\bibitem{CHA06}
M. B. Chadwick {\it et al.},  Nucl. Data. Sheets {\bf C 107}, 2931 (2006). 	

\bibitem{KON09}
A.J. Koning and D. Rochman, TENDL-2009: "TALYS-based Evaluated Nuclear Data Library"; http://www.talys.eu/tendl-2009/

\bibitem{TAK09}
H. Takada {\it et al.}, J. Nucl. Sci. Technol. {\bf 46}, 589-598 (2009).		

\bibitem{TRA13}
R. Trappitsch and I. Leya, Meteoritics \& Planetary Science, {\bf 48}, 195-210 (2013). doi: 10.1111/maps.12051

\bibitem{SIM83}
J. A. Simpson, Ann. Rev. Nucl. Part. Sci. {\bf 33}, 323-381 (1983).

\bibitem{REE94}
R.C. Reedy and J. Masarik, Lunar. Planet. Sci., XXV, 1119-1120 (1994).

\bibitem{LEY09}
I. Leya and J. Masarik, Meteoritics \& Planetary Science {\bf  44}, 1061-1086 (2009).

\bibitem{REE13}
R. C. Reedy,  Nucl. Instrum. Meth. {\bf B 294}, 470-474 (2013).

\bibitem{CAR96}
R. F. Carlson, At. Data Nucl. Data Tables {\bf 63}, 93 (1996).

\bibitem{PRA97}
R. E. Prael and M. B. Chadwick, "Applications of Evaluated Nuclear Data in the LAHET\texttrademark CODE"LA-UR-97-1744 (1997).

\bibitem{BAR93}
 B. C. Barashenkov, cross-sections of Interactions of Particles and Nuclei with Nuclei (OlYal, Dubna, 1993) (in Russian).

\bibitem{DAV11c}
J.-C. David {\it et al.}, Memorie della Societ\`a Astronomica Italiana {\bf 82}, 909-912 (2011).

\bibitem{MAS07}
S. G. Mashnik, R. E. Prael, and K. K. Gudima, LANL Report LA-UR-06-8652 (2007).

\bibitem{MAS98}
S. G. Mashnik {\it et al.}, LANL Report LA-UR-98-6000 (1998); Eprint: nucl-th/9812071; Proc. SARE-4, Knoxville, TN, September 13-16, 1998, edited by T. A. Gabriel, ORNL, 1999, pp. 151-162.

\bibitem{MAS14}
S. G. Mashnik and L. M. Kerby,  Nucl. Instrum. Meth. {\bf A 764}, 59-81 (2014).

\bibitem{KIM98}
E. Kim {\it et al.},  Nucl. Sci. Eng. {\bf 129}, 209 (1998).

\bibitem{KIM99}
E. Kim {\it et al.},  Nucl. Sci. Tech. {\bf 36}, 29 (1999).

\bibitem{SIS07}
J. M. Sisterson, Nucl. Instrum. Meth. {\bf B 261}, 993-995 (2007).

\bibitem{SEK11}
S. Sekimoto {\it et al.}, J. Korean Phys. Soc. {\bf 59}, 1916-1919 (2011).

\bibitem{YAS11}
H. Yashima {\it et al.}, Proc. Radiochim. Acta {\bf 1}, 135-139 (2011).

\bibitem{NIN11}
K. Ninomiya {\it et al.}, Proc. Radiochim. Acta {\bf 1}, 123-126 (2011).

\bibitem{VEE77}
L. R. Veeser {\it et al.}, Phys. Rev. {\bf C 16}, 1792 (1977).
 
\bibitem{UNO96}
Y. Uno {\it et al.}, Nucl. Sci. Eng. {\bf 122}, 247 (1996).

\bibitem{KON10}
A.J. Koning and D. Rochman, TENDL-2010: "TALYS-based Evaluated Nuclear Data Library"; ftp://ftp.nrg.eu/pub/www/talys/tendl2010/\\
tendl2010.html

\bibitem{LEY00}
I. Leya {\it et al.}, Meteoritics \& Planetary Science {\bf  35}, 259-286 (2000).

\bibitem{ING00}
A. Ingemarsson {\it et al.}, Nucl. Phys. {\bf A 676}, 3 (2000).

\bibitem{IGO63}
G. Igo and B. D. Wilkins, Phys. Rev. {\bf 131}, 1251 (1963).

\bibitem{STE64}
P. H. Stelson and F. K. McGowan, Phys. Rev. {\bf B 133}, 911 (1964).

\bibitem{LEV91}
V. N. Levkovskij, "Activation cross-sections for Nuclides of Average Masses (A = 40-100) by Protons and Alpha-Particles with Average Energies (E = 10-50 MeV)" (Inter Vesi, Moscow, 1991).

\bibitem{YAD08}
A. Yadav {\it et al.}, Phys. Rev. {\bf C 78}, 044606 (2008).

\bibitem{TAN60}
S. Tanaka, J. Phys. Soc. Jpn. {\bf 15}, 2159 (1960).

\bibitem{TIT02}
Yu E. Titarenko {\it et al.}, Phys. Rev. {\bf C 65} 064610 (2002).

\bibitem{EUR09}
J. Cornell, Y. Blumenfeld, and G. Fortuna, "Final report of the eurisol design study (2005-2009)". Published by GANIL, 2009.

\bibitem{ENE09b}
D. Ene {\it et al.}, In-target yields for RIB production: Part II: two stage target configuration, CEA Saclay Internal report: IRFU-09-87 (2009)

\bibitem{CHA10}
S. Chabod {\it et al.},  Eur. Phys. J.  {\bf A 45}, 131-145 (2010).

\bibitem{STO05}
T. Stora {\it et al.}, "The EURISOL Facility: Feasibility study for the 100-kW direct targets" (2005), CERN EDMS Doc. \#758813, http://edms.cern.ch/document/758813/2; and on the http://www.eurisol.org/site02/ website in direct\_target/publications\_internal\_task\_note.php

\bibitem{COR03}
J.Cornell, "The EURISOL report - Feasibility study for the EURopean Isotope-Separation-On-Line radioactive beam facility", (Ganil, Caen, 2003) appendix C; available on http://pro.ganil-spiral2.eu/eurisol/feasibility-study-reports/feasibility-study-appendix-c

\bibitem{PAG07}
R. Page {\it et al.}, "Selection of key experiments with the associated instrumentation" (2007); http://www.eurisol.org/site02/physics\_and\_instrumentation/

\bibitem{CHA08}
S. Chabod {\it et al.}, "Optimization of in-target yields for RIB production : Part I : direct targets", Internal Report, Irfu-08-21 (2008).

\bibitem{WIL98}
W. B. Wilson {\it et al.}, "Status of CINDER'90 Codes and Data,", LA-UR-98-361 (1998).

\bibitem{RAP10}
B. Rapp {\it et al.}, Shielding Aspects of Accelerators, Targets and Irradiation Facilities Eighth Meeting (SATIF-8). NEA/NSC/DOC(2010)6, page 251 (2010).

\bibitem{ENE09c}
D. Ene {\it et al.}, Radiation protection aspects of the EURISOL Multi-MW target shielding, CEA Saclay Internal report: IRFU-09-15 (2009).

\bibitem{MOO03}
R. Moormann. "Ess-target inventories and their radiotoxic and toxic potential", Document ESS-R-1205-R.Moormann-1-02 (2003).

\bibitem{MOO08}
R. Moormann {\it et al.}, "Safety aspects of high power targets for European spallation sources", International Conference on the Physics of Reactors - Nuclear Power : A Sustainable Resource; Interlaken, Switzerland, September 14-19 (2008).

\bibitem{NNDC}
National Nuclear data Centre On-Line Service, http://www.nndc.bnl.gov/ 

\bibitem{IWA02}
H. Iwase {\it et al.}, J. Nucl. Sci. Tech. {\bf 39}, 1142 (2002).

\bibitem{FUR01}
S. Furihata and H. Nakashima, "Analysis of activation yields by INC/GEM", JAERI-Conf-2001-006 (2006) 

\bibitem{ZAN08}
L. Zanini {\it et al.}, "Neutronic and Nuclear Post-test Analysis of MEGAPIE", PSI Report 08-04 (2008). ISSN 1019-0643.

\bibitem{BAU01}
G.S Bauer  {\it et al.}, "MEGAPIE, a 1MW pilot experiment for a liquid metal spallation target", Journal of Nuclear Materials {\bf 296}, 17-33 (2001). ISSN 0022-3115. 

\bibitem{ZAN11}
L. Zanini {\it et al.}, J. Nucl. Mat. {\bf 415}, 367-377 (2011).

\bibitem{SEN10}
F. Michel-Sendis {\it et al.}, Nucl. Instrum. Meth. {\bf B 268}, 2257-2271 (2010).

\bibitem{RID07}
R. Ridikas {\it et al.},  Eur. Phys. J.  {\bf A 32}, 1-4 (2007).

\bibitem{DAV08}
J.-C. David  {\it et al.}, "NUDATRA - WP 5.4 - Task T5.4.3 Activation calculations for the MEGAPIE target with INCL4 and ABLA, and comparison
with other codes", Internal report, Irfu-08-453 (2008).

\bibitem{MCN11}
Denise B. Pelowitz {\it et al.}, "MCNPX 2.7.0 Extensions", LA-UR-11-02295 (2011).

\bibitem{THI11}
N. Thiolli\`ere  {\it et al.}, Nucl. Sci. and Eng. {\bf 169}, 178-187 (2011).

\bibitem{DAV11d}
J.-C. David, "INCL4.5-Abla07: What's new for the assessment of spallation target activation?", 4th. HPTW, Malm\"o, Sweden, May 2-6, 2011; www.hep.princeton.edu/\~{}mcdonald/mumu/target/ in David/david\_050411.pdf

\bibitem{DAV12}
J.-C. David, "Spallation: understanding for predicting ! ?", report of Habilitation \`a Diriger des Recherches (in french), Irfu-12-259; http://tel.archives-ouvertes.fr/tel-00811587

\bibitem{TAL08}
Y. Tall {\it et al.}, "Volatile elements production rates in a proton-irradiated molten lead-bismuth target", in Proceedings of the International Conference on Nuclear Data for Science and Technology, April 22-27, 2007, Nice, France, editors O.Bersillon, F.Gunsing, E.Bauge, R.Jacqmin, and S.Leray, EDP Sciences, 2008, pp 1069-1072 - DOI: 10.1051/ndata:07762.

\bibitem{ZAN14}
L. Zanini {\it et al.}, Nuclear Data Sheets {\bf 119}, 292-295 (2014).

\bibitem{KOR92}
Yu. A. Korovin {\it et al.}, The Code for the Calculation of Nuclide Composition and Activity of Irradiated Materials,  J. Yadernye Konstanty (Nucl. Constants) {\bf 117}, 3-4 (1992).

\bibitem{RAM59}
W.J.Ramler {\it et al.}, Phys. Rev. {\bf 114}, 154 (1959).

\bibitem{LAM85a}
R.M.Lambrecht and S.Mirzadeh, J.of Labelled Compounds and Radiopharmaceut. {\bf 21}, 1288 (1984).

\bibitem{LAM85b}
R.M.Lambrecht and S.Mirzadeh, Applied Radiation and Isotopes {\bf 36}, 443 (1985).

\bibitem{BAR74}
A.R.Barnett and J.S.Lilley, Phys. Rev. {\bf C 9}, 2010 (1974).

\bibitem{DEC74}
G.Deconninck and M.Longree, Annales de la Soci\'et\'e Scientifique de Bruxelles {\bf 88}, 341 (1974).

\bibitem{KEL49}
E.L.Kelly and E.Segre, Phys. Rev. {\bf 75}, 999 (1949).

\bibitem{PAT99}
H.B.Patel, {\it et al.}, Nuovo Cimento {\bf A 112}, 1439 (1999).

\bibitem{RIZ90}
I.A.Rizvi {\it et al.}, Applied Radiation and Isotopes {\bf 41}, 215 (1990)

\bibitem{STI74}
J.D.Stickler and K.J.Hofstetter, Phys. Rev. {\bf C 9}, 1064 (1974).

\bibitem{SIN94}
N.L.Singh {\it et al.},  Nuovo Cimento {\bf A 107}, 1635 (1994)

\bibitem{HER05}
A.Hermanne {\it et al.}, "Experimental Study of the cross sections of alpha-Particle Induced Reactions on 209Bi", Conf.on Nucl.Data for Sci.and Techn., Santa Fe 2004, p.957 (2004); A.Hermanne {\it et al.}, Applied Radiation and Isotopes {\bf 63}, 1 (2005).

\bibitem{RAT92}
S.S.Rattan {\it et al.}, Radiochimica Acta {\bf 57}, 7 (1992).

\bibitem{CAR13}
J. M. Carpenter, National School on Neutron and X-ray Scattering, Oak Ridge 11-24 August (2013); Carpenter-NeutronGeneration2013.pdf at http://neutrons.ornl.gov/conf/nxs2013/lecture/pdf/

\bibitem{ESSTDR}
ESS Technical Design Report; http://eval.esss.lu.se/cgi-bin/public/DocDB/ShowDocument?docid=274

\bibitem{LEP14}
A. Leprince {\it et al.}, Reliability and use of the INCL4.6-Abla07 spallation model in the frame of the European Spallation Source (ESS) target design, CEA Saclay Internal report: IRFU-14-25 (2014).

\bibitem{LER10}
S. Leray {\it et al.}, Nucl. Instrum. Meth. {\bf B  268}, 581-586 (2010).

\bibitem{GUM}
GUM (Guide to the expression of Uncertainty in Measurement), JCGM 100:2008; http://www.bipm.org/fr/publications/guides/gum.html

\bibitem{BOL10}
A. Bolshakova {\it et al.}, Eur. Phys. J.  {\bf C 70}, 543-553 (2010); and references therein.

\bibitem{CAT08}
M. G. Catanesi  {\it et al.}, Phys. Rev. {\bf C 77}, 055207 (2008).

\bibitem{BOL09}
A. Bolshakova {\it et al.}, Eur. Phys. J.  {\bf C 62}, 293 (2009).

\bibitem{PED11}
S. Pedoux and J. Cugnon {\it et al.}, Nucl. Phys. {\bf A 866}, 16-36 (2011).

\bibitem{CAT07a}
M.G. Catanesi {\it et al.}, Eur. Phys. J.  {\bf C 54}, 37 (2008).

\bibitem{CAT07b}
M.G. Catanesi {\it et al.}, Eur. Phys. J.  {\bf C 31}, 787 (2007).

\bibitem{LOU68}
A. Lou and D. T. Goodhead,  Phys. Rev. {\bf 168}, 1214-1223 (1968).

\bibitem{EVA67}
D.A. Evans and D.T. Goodhead, Nucl. Phys. {\bf B 3}, 441-463 (1967).

\bibitem{CUG90}
J. Cugnon {\it et al.}, Phys. Rev. {\bf C 41}, 1701-1718 (1990).

\bibitem{PYS99}
K. Pysz {\it et al.}, Nucl. Instrum. Meth. {\bf A  420}, 356-365 (1999).

\bibitem{RUD99}
Z. Rudy,  {\it Non-mesonic hyperon decay in heavy hypernuclei}, Habilitation - Report No. 1811/PH, 1999, Institute of Physics, Jagellonian University.

\bibitem{WOL93}
Gy. Wolf {\it et al.}, Nucl. Phys. {\bf A 552}, 549-570 (1993).

\bibitem{SON10}
Y. Song {\it et al.}, Physics of Atomic Nuclei  {\bf 73}, 1707-1712 (2010).

\bibitem{GON97}
M. Goncalves {\it et al.},  Phys. Lett. {\bf B 406}, 1-6 (1997).

\bibitem{PIN98}
S. de Pina {\it et al.}, Phys. Lett. {\bf B 434}, 1-6 (1998).

\bibitem{BOC94}
M. Bockhorst {\it et al.}, Zeitschrift f\"ur Physik {\bf C 63}, 37-47 (1994).

\bibitem{GLA04}
K.-H. Glander {\it et al.}, Eur. Phys. J.  {\bf  A 19}, 251-273 (2004).

\bibitem{JUL06}
B. Juli\' a-D\' iaz,{\it et al.}, Phys. Rev. {\bf  C 73}, 055204 (2006).

\end{thebibliography}
\end{document}